\documentclass[longauth]{aaEC}

\usepackage{lastpage}
\usepackage{graphicx}
\usepackage{natbib}
\usepackage{scalerel}
\usepackage{fontawesome}
\usepackage{tabularx}
\usepackage{booktabs}  % for better looking tables
\usepackage{siunitx}  % for better number alignment
\usepackage[normalem]{ulem}
\usepackage[table]{xcolor}
\usepackage{ragged2e}
\usepackage{txfonts}
\usepackage[pdfencoding=auto,psdextra]{hyperref}
\hypersetup{
    colorlinks=true,
    linkcolor=blue,
    filecolor=magenta,      
    urlcolor=blue,
    citecolor=blue
}
\makeatletter
\renewcommand*\aa@pageof{, page \thepage{} of \pageref*{LastPage}}
\makeatother

\usepackage[utf8]{inputenc}

\usepackage[switch, modulo]{lineno}
\usepackage{amsmath,amssymb}
\usepackage{siunitx}
\usepackage{xspace}
\usepackage[nameinlink,capitalise]{cleveref}
\usepackage{euclid}
\usepackage{ISTL_macros}
\usepackage{glossaries}
\glsdisablehyper
\usepackage{orcidlink} %this has to come after hyperref
\newcommand{\orcid}[1]{\orcidlink{#1}}
\newcommand{\corr}[1]{{#1}}

\crefname{section}{Sect.}{Sects.} % Singular and plural forms
\Crefname{section}{Sect.}{Sects.} % For capitalized references

\newacronym{eft}{EFTofLSS}{effective field theory of large-scale structure}
\newacronym{lss}{LSS}{large-scale structure}
\newacronym{drone}{DR1}{Data Release 1}
\newacronym{drthree}{DR3}{Data Release 3}
\newacronym{bbn}{BBN}{Big Bang Nucleosynthesis}
\newcommand{\bao}{BAO\xspace}
\newcommand{\rsd}{RSD\xspace}

\newcommand{\eft}{\gls{eft}\xspace}

\newcommand{\drthree}{\gls{drthree}\xspace}
\newcommand{\bbn}{\gls{bbn}\xspace}

\AtBeginDocument{\nolinenumbers}

\begin{document}
%%%% Version Friday 11th of July 2025 11:11:35 AM UT								\title{\Euclid preparation}
\title{\Euclid preparation}
\subtitle{XCVI. Cosmology Likelihood for Observables in Euclid (CLOE). 3. Inference and forecasts\thanks{Dedicated to our colleague Karim Benabed, whose insight helped shape this manuscript and elevate its scientific quality.}}         				
%%%% Version Saturday 26th of July 2025 07:51:27 AM UT												
%%%% Please do not edit the author list -- contact ECEB Bureau for changes
\author{Euclid Collaboration: G.~Ca\~nas-Herrera\orcid{0000-0003-2796-2149}\thanks{\email{canasherrera@strw.leidenuniv.nl}}\inst{\ref{aff1},\ref{aff2},\ref{aff3}}
\and L.~W.~K.~Goh\orcid{0000-0002-0104-8132}\inst{\ref{aff4}}
\and L.~Blot\orcid{0000-0002-9622-7167}\inst{\ref{aff5},\ref{aff6}}
\and M.~Bonici\orcid{0000-0002-8430-126X}\inst{\ref{aff7},\ref{aff8}}
\and S.~Camera\orcid{0000-0003-3399-3574}\inst{\ref{aff9},\ref{aff10},\ref{aff11}}
\and V.~F.~Cardone\inst{\ref{aff12},\ref{aff13}}
\and P.~Carrilho\orcid{0000-0003-1339-0194}\inst{\ref{aff14}}
\and S.~Casas\orcid{0000-0002-4751-5138}\inst{\ref{aff15}}
\and S.~Davini\orcid{0000-0003-3269-1718}\inst{\ref{aff16}}
\and S.~Di~Domizio\orcid{0000-0003-2863-5895}\inst{\ref{aff17},\ref{aff16}}
\and S.~Farrens\orcid{0000-0002-9594-9387}\inst{\ref{aff4}}
\and S.~Gouyou~Beauchamps\inst{\ref{aff18},\ref{aff19}}
\and S.~Ili\'c\orcid{0000-0003-4285-9086}\inst{\ref{aff20},\ref{aff21}}
\and S.~Joudaki\orcid{0000-0001-8820-673X}\inst{\ref{aff22},\ref{aff23}}
\and F.~Keil\orcid{0000-0002-8108-1679}\inst{\ref{aff21}}
\and A.~M.~C.~Le~Brun\orcid{0000-0002-0936-4594}\inst{\ref{aff24}}
\and M.~Martinelli\orcid{0000-0002-6943-7732}\inst{\ref{aff12},\ref{aff13}}
\and C.~Moretti\orcid{0000-0003-3314-8936}\inst{\ref{aff25},\ref{aff26},\ref{aff27},\ref{aff28},\ref{aff29}}
\and V.~Pettorino\inst{\ref{aff1}}
\and A.~Pezzotta\orcid{0000-0003-0726-2268}\inst{\ref{aff30},\ref{aff31}}
\and Z.~Sakr\orcid{0000-0002-4823-3757}\inst{\ref{aff32},\ref{aff21},\ref{aff33}}
\and A.~G.~S\'anchez\orcid{0000-0003-1198-831X}\inst{\ref{aff31}}
\and D.~Sciotti\orcid{0009-0008-4519-2620}\inst{\ref{aff12},\ref{aff13}}
\and K.~Tanidis\inst{\ref{aff34}}
\and I.~Tutusaus\orcid{0000-0002-3199-0399}\inst{\ref{aff21}}
\and V.~Ajani\orcid{0000-0001-9442-2527}\inst{\ref{aff4},\ref{aff35},\ref{aff36}}
\and M.~Crocce\orcid{0000-0002-9745-6228}\inst{\ref{aff19},\ref{aff18}}
\and A.~Fumagalli\orcid{0009-0004-0300-2535}\inst{\ref{aff37},\ref{aff28}}
\and C.~Giocoli\orcid{0000-0002-9590-7961}\inst{\ref{aff38},\ref{aff39}}
\and L.~Legrand\orcid{0000-0003-0610-5252}\inst{\ref{aff40},\ref{aff41}}
\and M.~Lembo\orcid{0000-0002-5271-5070}\inst{\ref{aff42},\ref{aff43}}
\and G.~F.~Lesci\orcid{0000-0002-4607-2830}\inst{\ref{aff44},\ref{aff38}}
\and D.~Navarro-Gironés\orcid{0000-0003-0507-372X}\inst{\ref{aff3}}
\and A.~Nouri-Zonoz\orcid{0009-0006-6164-8670}\inst{\ref{aff45}}
\and S.~Pamuk\orcid{0009-0004-0852-8624}\inst{\ref{aff46}}
\and A.~Pourtsidou\orcid{0000-0001-9110-5550}\inst{\ref{aff14},\ref{aff47}}
\and M.~Tsedrik\orcid{0000-0002-0020-5343}\inst{\ref{aff14},\ref{aff47}}
\and J.~Bel\inst{\ref{aff48}}
\and C.~Carbone\orcid{0000-0003-0125-3563}\inst{\ref{aff7}}
\and K.~Benabed\inst{\ref{aff96}}
\and J.~Claramunt~Gonzalez\orcid{0009-0009-5387-6341}\inst{\ref{aff49}}
\and C.~A.~J.~Duncan\orcid{0009-0003-3573-0791}\inst{\ref{aff50}}
\and M.~Kilbinger\orcid{0000-0001-9513-7138}\inst{\ref{aff4}}
\and A.~Porredon\orcid{0000-0002-2762-2024}\inst{\ref{aff22},\ref{aff51}}
\and D.~Sapone\orcid{0000-0001-7089-4503}\inst{\ref{aff52}}
\and E.~Sellentin\inst{\ref{aff53},\ref{aff3}}
\and P.~L.~Taylor\orcid{0000-0001-6999-4718}\inst{\ref{aff54},\ref{aff55}}
\and N.~Tessore\orcid{0000-0002-9696-7931}\inst{\ref{aff56}}
\and B.~Altieri\orcid{0000-0003-3936-0284}\inst{\ref{aff57}}
\and A.~Amara\inst{\ref{aff58}}
\and L.~Amendola\orcid{0000-0002-0835-233X}\inst{\ref{aff32}}
\and S.~Andreon\orcid{0000-0002-2041-8784}\inst{\ref{aff30}}
\and N.~Auricchio\orcid{0000-0003-4444-8651}\inst{\ref{aff38}}
\and C.~Baccigalupi\orcid{0000-0002-8211-1630}\inst{\ref{aff28},\ref{aff27},\ref{aff29},\ref{aff25}}
\and M.~Baldi\orcid{0000-0003-4145-1943}\inst{\ref{aff59},\ref{aff38},\ref{aff39}}
\and S.~Bardelli\orcid{0000-0002-8900-0298}\inst{\ref{aff38}}
\and R.~Bender\orcid{0000-0001-7179-0626}\inst{\ref{aff31},\ref{aff60}}
\and A.~Biviano\orcid{0000-0002-0857-0732}\inst{\ref{aff27},\ref{aff28}}
\and D.~Bonino\orcid{0000-0002-3336-9977}\inst{\ref{aff11}}
\and E.~Branchini\orcid{0000-0002-0808-6908}\inst{\ref{aff17},\ref{aff16},\ref{aff30}}
\and M.~Brescia\orcid{0000-0001-9506-5680}\inst{\ref{aff61},\ref{aff62}}
\and J.~Brinchmann\orcid{0000-0003-4359-8797}\inst{\ref{aff63},\ref{aff64}}
\and V.~Capobianco\orcid{0000-0002-3309-7692}\inst{\ref{aff11}}
\and J.~Carretero\orcid{0000-0002-3130-0204}\inst{\ref{aff22},\ref{aff65}}
\and M.~Castellano\orcid{0000-0001-9875-8263}\inst{\ref{aff12}}
\and G.~Castignani\orcid{0000-0001-6831-0687}\inst{\ref{aff38}}
\and S.~Cavuoti\orcid{0000-0002-3787-4196}\inst{\ref{aff62},\ref{aff66}}
\and K.~C.~Chambers\orcid{0000-0001-6965-7789}\inst{\ref{aff67}}
\and A.~Cimatti\inst{\ref{aff68}}
\and C.~Colodro-Conde\inst{\ref{aff69}}
\and G.~Congedo\orcid{0000-0003-2508-0046}\inst{\ref{aff14}}
\and C.~J.~Conselice\orcid{0000-0003-1949-7638}\inst{\ref{aff50}}
\and L.~Conversi\orcid{0000-0002-6710-8476}\inst{\ref{aff70},\ref{aff57}}
\and Y.~Copin\orcid{0000-0002-5317-7518}\inst{\ref{aff71}}
\and F.~Courbin\orcid{0000-0003-0758-6510}\inst{\ref{aff72},\ref{aff73}}
\and H.~M.~Courtois\orcid{0000-0003-0509-1776}\inst{\ref{aff74}}
\and M.~Cropper\orcid{0000-0003-4571-9468}\inst{\ref{aff75}}
\and A.~Da~Silva\orcid{0000-0002-6385-1609}\inst{\ref{aff76},\ref{aff77}}
\and H.~Degaudenzi\orcid{0000-0002-5887-6799}\inst{\ref{aff78}}
\and S.~de~la~Torre\inst{\ref{aff79}}
\and G.~De~Lucia\orcid{0000-0002-6220-9104}\inst{\ref{aff27}}
\and A.~M.~Di~Giorgio\orcid{0000-0002-4767-2360}\inst{\ref{aff80}}
\and H.~Dole\orcid{0000-0002-9767-3839}\inst{\ref{aff81}}
\and F.~Dubath\orcid{0000-0002-6533-2810}\inst{\ref{aff78}}
\and X.~Dupac\inst{\ref{aff57}}
\and S.~Dusini\orcid{0000-0002-1128-0664}\inst{\ref{aff82}}
\and S.~Escoffier\orcid{0000-0002-2847-7498}\inst{\ref{aff83}}
\and M.~Farina\orcid{0000-0002-3089-7846}\inst{\ref{aff80}}
\and F.~Faustini\orcid{0000-0001-6274-5145}\inst{\ref{aff84},\ref{aff12}}
\and S.~Ferriol\inst{\ref{aff71}}
\and F.~Finelli\orcid{0000-0002-6694-3269}\inst{\ref{aff38},\ref{aff85}}
\and P.~Fosalba\orcid{0000-0002-1510-5214}\inst{\ref{aff18},\ref{aff19}}
\and S.~Fotopoulou\orcid{0000-0002-9686-254X}\inst{\ref{aff86}}
\and N.~Fourmanoit\orcid{0009-0005-6816-6925}\inst{\ref{aff83}}
\and M.~Frailis\orcid{0000-0002-7400-2135}\inst{\ref{aff27}}
\and E.~Franceschi\orcid{0000-0002-0585-6591}\inst{\ref{aff38}}
\and S.~Galeotta\orcid{0000-0002-3748-5115}\inst{\ref{aff27}}
\and K.~George\orcid{0000-0002-1734-8455}\inst{\ref{aff60}}
\and W.~Gillard\orcid{0000-0003-4744-9748}\inst{\ref{aff83}}
\and B.~Gillis\orcid{0000-0002-4478-1270}\inst{\ref{aff14}}
\and P.~G\'omez-Alvarez\orcid{0000-0002-8594-5358}\inst{\ref{aff87},\ref{aff57}}
\and J.~Gracia-Carpio\inst{\ref{aff31}}
\and B.~R.~Granett\orcid{0000-0003-2694-9284}\inst{\ref{aff30}}
\and A.~Grazian\orcid{0000-0002-5688-0663}\inst{\ref{aff88}}
\and F.~Grupp\inst{\ref{aff31},\ref{aff60}}
\and L.~Guzzo\orcid{0000-0001-8264-5192}\inst{\ref{aff89},\ref{aff30}}
\and S.~V.~H.~Haugan\orcid{0000-0001-9648-7260}\inst{\ref{aff90}}
\and H.~Hoekstra\orcid{0000-0002-0641-3231}\inst{\ref{aff3}}
\and W.~Holmes\inst{\ref{aff91}}
\and I.~Hook\orcid{0000-0002-2960-978X}\inst{\ref{aff92}}
\and F.~Hormuth\inst{\ref{aff93}}
\and A.~Hornstrup\orcid{0000-0002-3363-0936}\inst{\ref{aff94},\ref{aff95}}
\and P.~Hudelot\inst{\ref{aff96}}
\and K.~Jahnke\orcid{0000-0003-3804-2137}\inst{\ref{aff97}}
\and M.~Jhabvala\inst{\ref{aff98}}
\and B.~Joachimi\orcid{0000-0001-7494-1303}\inst{\ref{aff56}}
\and E.~Keih\"anen\orcid{0000-0003-1804-7715}\inst{\ref{aff99}}
\and S.~Kermiche\orcid{0000-0002-0302-5735}\inst{\ref{aff83}}
\and A.~Kiessling\orcid{0000-0002-2590-1273}\inst{\ref{aff91}}
\and B.~Kubik\orcid{0009-0006-5823-4880}\inst{\ref{aff71}}
\and K.~Kuijken\orcid{0000-0002-3827-0175}\inst{\ref{aff3}}
\and M.~K\"ummel\orcid{0000-0003-2791-2117}\inst{\ref{aff60}}
\and M.~Kunz\orcid{0000-0002-3052-7394}\inst{\ref{aff45}}
\and H.~Kurki-Suonio\orcid{0000-0002-4618-3063}\inst{\ref{aff100},\ref{aff101}}
\and O.~Lahav\orcid{0000-0002-1134-9035}\inst{\ref{aff56}}
\and R.~Laureijs\inst{\ref{aff1},\ref{aff102}}
\and S.~Ligori\orcid{0000-0003-4172-4606}\inst{\ref{aff11}}
\and P.~B.~Lilje\orcid{0000-0003-4324-7794}\inst{\ref{aff90}}
\and V.~Lindholm\orcid{0000-0003-2317-5471}\inst{\ref{aff100},\ref{aff101}}
\and I.~Lloro\orcid{0000-0001-5966-1434}\inst{\ref{aff103}}
\and G.~Mainetti\orcid{0000-0003-2384-2377}\inst{\ref{aff104}}
\and D.~Maino\inst{\ref{aff89},\ref{aff7},\ref{aff105}}
\and E.~Maiorano\orcid{0000-0003-2593-4355}\inst{\ref{aff38}}
\and O.~Mansutti\orcid{0000-0001-5758-4658}\inst{\ref{aff27}}
\and S.~Marcin\inst{\ref{aff106}}
\and O.~Marggraf\orcid{0000-0001-7242-3852}\inst{\ref{aff107}}
\and K.~Markovic\orcid{0000-0001-6764-073X}\inst{\ref{aff91}}
\and N.~Martinet\orcid{0000-0003-2786-7790}\inst{\ref{aff79}}
\and F.~Marulli\orcid{0000-0002-8850-0303}\inst{\ref{aff44},\ref{aff38},\ref{aff39}}
\and R.~Massey\orcid{0000-0002-6085-3780}\inst{\ref{aff108}}
\and H.~J.~McCracken\orcid{0000-0002-9489-7765}\inst{\ref{aff96}}
\and E.~Medinaceli\orcid{0000-0002-4040-7783}\inst{\ref{aff38}}
\and M.~Melchior\inst{\ref{aff106}}
\and Y.~Mellier\inst{\ref{aff109},\ref{aff96}}
\and M.~Meneghetti\orcid{0000-0003-1225-7084}\inst{\ref{aff38},\ref{aff39}}
\and E.~Merlin\orcid{0000-0001-6870-8900}\inst{\ref{aff12}}
\and G.~Meylan\inst{\ref{aff110}}
\and A.~Mora\orcid{0000-0002-1922-8529}\inst{\ref{aff111}}
\and M.~Moresco\orcid{0000-0002-7616-7136}\inst{\ref{aff44},\ref{aff38}}
\and L.~Moscardini\orcid{0000-0002-3473-6716}\inst{\ref{aff44},\ref{aff38},\ref{aff39}}
\and C.~Neissner\orcid{0000-0001-8524-4968}\inst{\ref{aff112},\ref{aff65}}
\and S.-M.~Niemi\inst{\ref{aff1}}
\and J.~W.~Nightingale\orcid{0000-0002-8987-7401}\inst{\ref{aff113}}
\and C.~Padilla\orcid{0000-0001-7951-0166}\inst{\ref{aff112}}
\and S.~Paltani\orcid{0000-0002-8108-9179}\inst{\ref{aff78}}
\and F.~Pasian\orcid{0000-0002-4869-3227}\inst{\ref{aff27}}
\and K.~Pedersen\inst{\ref{aff114}}
\and W.~J.~Percival\orcid{0000-0002-0644-5727}\inst{\ref{aff8},\ref{aff115},\ref{aff116}}
\and S.~Pires\orcid{0000-0002-0249-2104}\inst{\ref{aff4}}
\and G.~Polenta\orcid{0000-0003-4067-9196}\inst{\ref{aff84}}
\and M.~Poncet\inst{\ref{aff117}}
\and L.~A.~Popa\inst{\ref{aff118}}
\and L.~Pozzetti\orcid{0000-0001-7085-0412}\inst{\ref{aff38}}
\and F.~Raison\orcid{0000-0002-7819-6918}\inst{\ref{aff31}}
\and R.~Rebolo\inst{\ref{aff69},\ref{aff119},\ref{aff120}}
\and A.~Renzi\orcid{0000-0001-9856-1970}\inst{\ref{aff121},\ref{aff82}}
\and J.~Rhodes\orcid{0000-0002-4485-8549}\inst{\ref{aff91}}
\and G.~Riccio\inst{\ref{aff62}}
\and E.~Romelli\orcid{0000-0003-3069-9222}\inst{\ref{aff27}}
\and M.~Roncarelli\orcid{0000-0001-9587-7822}\inst{\ref{aff38}}
\and R.~Saglia\orcid{0000-0003-0378-7032}\inst{\ref{aff60},\ref{aff31}}
\and B.~Sartoris\orcid{0000-0003-1337-5269}\inst{\ref{aff60},\ref{aff27}}
\and J.~A.~Schewtschenko\orcid{0000-0002-4913-6393}\inst{\ref{aff14}}
\and P.~Schneider\orcid{0000-0001-8561-2679}\inst{\ref{aff107}}
\and T.~Schrabback\orcid{0000-0002-6987-7834}\inst{\ref{aff122}}
\and A.~Secroun\orcid{0000-0003-0505-3710}\inst{\ref{aff83}}
\and E.~Sefusatti\orcid{0000-0003-0473-1567}\inst{\ref{aff27},\ref{aff28},\ref{aff29}}
\and G.~Seidel\orcid{0000-0003-2907-353X}\inst{\ref{aff97}}
\and M.~Seiffert\orcid{0000-0002-7536-9393}\inst{\ref{aff91}}
\and S.~Serrano\orcid{0000-0002-0211-2861}\inst{\ref{aff18},\ref{aff123},\ref{aff19}}
\and P.~Simon\inst{\ref{aff107}}
\and C.~Sirignano\orcid{0000-0002-0995-7146}\inst{\ref{aff121},\ref{aff82}}
\and G.~Sirri\orcid{0000-0003-2626-2853}\inst{\ref{aff39}}
\and A.~Spurio~Mancini\orcid{0000-0001-5698-0990}\inst{\ref{aff124}}
\and L.~Stanco\orcid{0000-0002-9706-5104}\inst{\ref{aff82}}
\and J.~Steinwagner\orcid{0000-0001-7443-1047}\inst{\ref{aff31}}
\and P.~Tallada-Cresp\'{i}\orcid{0000-0002-1336-8328}\inst{\ref{aff22},\ref{aff65}}
\and D.~Tavagnacco\orcid{0000-0001-7475-9894}\inst{\ref{aff27}}
\and A.~N.~Taylor\inst{\ref{aff14}}
\and I.~Tereno\inst{\ref{aff76},\ref{aff125}}
\and S.~Toft\orcid{0000-0003-3631-7176}\inst{\ref{aff126},\ref{aff127}}
\and R.~Toledo-Moreo\orcid{0000-0002-2997-4859}\inst{\ref{aff128}}
\and F.~Torradeflot\orcid{0000-0003-1160-1517}\inst{\ref{aff65},\ref{aff22}}
\and L.~Valenziano\orcid{0000-0002-1170-0104}\inst{\ref{aff38},\ref{aff85}}
\and J.~Valiviita\orcid{0000-0001-6225-3693}\inst{\ref{aff100},\ref{aff101}}
\and T.~Vassallo\orcid{0000-0001-6512-6358}\inst{\ref{aff60},\ref{aff27}}
\and G.~Verdoes~Kleijn\orcid{0000-0001-5803-2580}\inst{\ref{aff102}}
\and A.~Veropalumbo\orcid{0000-0003-2387-1194}\inst{\ref{aff30},\ref{aff16},\ref{aff17}}
\and Y.~Wang\orcid{0000-0002-4749-2984}\inst{\ref{aff129}}
\and J.~Weller\orcid{0000-0002-8282-2010}\inst{\ref{aff60},\ref{aff31}}
\and G.~Zamorani\orcid{0000-0002-2318-301X}\inst{\ref{aff38}}
\and F.~M.~Zerbi\inst{\ref{aff30}}
\and E.~Zucca\orcid{0000-0002-5845-8132}\inst{\ref{aff38}}
\and M.~Ballardini\orcid{0000-0003-4481-3559}\inst{\ref{aff42},\ref{aff43},\ref{aff38}}
\and M.~Bolzonella\orcid{0000-0003-3278-4607}\inst{\ref{aff38}}
\and A.~Boucaud\orcid{0000-0001-7387-2633}\inst{\ref{aff130}}
\and E.~Bozzo\orcid{0000-0002-8201-1525}\inst{\ref{aff78}}
\and C.~Burigana\orcid{0000-0002-3005-5796}\inst{\ref{aff131},\ref{aff85}}
\and R.~Cabanac\orcid{0000-0001-6679-2600}\inst{\ref{aff21}}
\and M.~Calabrese\orcid{0000-0002-2637-2422}\inst{\ref{aff132},\ref{aff7}}
\and P.~Casenove\inst{\ref{aff117}}
\and D.~Di~Ferdinando\inst{\ref{aff39}}
\and J.~A.~Escartin~Vigo\inst{\ref{aff31}}
\and L.~Gabarra\orcid{0000-0002-8486-8856}\inst{\ref{aff34}}
\and S.~Matthew\orcid{0000-0001-8448-1697}\inst{\ref{aff14}}
\and N.~Mauri\orcid{0000-0001-8196-1548}\inst{\ref{aff68},\ref{aff39}}
\and R.~B.~Metcalf\orcid{0000-0003-3167-2574}\inst{\ref{aff44},\ref{aff38}}
\and M.~P\"ontinen\orcid{0000-0001-5442-2530}\inst{\ref{aff100}}
\and C.~Porciani\orcid{0000-0002-7797-2508}\inst{\ref{aff107}}
\and V.~Scottez\inst{\ref{aff109},\ref{aff133}}
\and M.~Tenti\orcid{0000-0002-4254-5901}\inst{\ref{aff39}}
\and M.~Viel\orcid{0000-0002-2642-5707}\inst{\ref{aff28},\ref{aff27},\ref{aff25},\ref{aff29},\ref{aff26}}
\and M.~Wiesmann\orcid{0009-0000-8199-5860}\inst{\ref{aff90}}
\and Y.~Akrami\orcid{0000-0002-2407-7956}\inst{\ref{aff134},\ref{aff135}}
\and S.~Alvi\orcid{0000-0001-5779-8568}\inst{\ref{aff42}}
\and I.~T.~Andika\orcid{0000-0001-6102-9526}\inst{\ref{aff136},\ref{aff137}}
\and R.~E.~Angulo\orcid{0000-0003-2953-3970}\inst{\ref{aff138},\ref{aff139}}
\and S.~Anselmi\orcid{0000-0002-3579-9583}\inst{\ref{aff82},\ref{aff121},\ref{aff6}}
\and M.~Archidiacono\orcid{0000-0003-4952-9012}\inst{\ref{aff89},\ref{aff105}}
\and F.~Atrio-Barandela\orcid{0000-0002-2130-2513}\inst{\ref{aff140}}
\and A.~Balaguera-Antolinez\orcid{0000-0001-5028-3035}\inst{\ref{aff69}}
\and M.~Bethermin\orcid{0000-0002-3915-2015}\inst{\ref{aff141}}
\and A.~Blanchard\orcid{0000-0001-8555-9003}\inst{\ref{aff21}}
\and S.~Borgani\orcid{0000-0001-6151-6439}\inst{\ref{aff142},\ref{aff28},\ref{aff27},\ref{aff29},\ref{aff26}}
\and M.~L.~Brown\orcid{0000-0002-0370-8077}\inst{\ref{aff50}}
\and S.~Bruton\orcid{0000-0002-6503-5218}\inst{\ref{aff143}}
\and A.~Calabro\orcid{0000-0003-2536-1614}\inst{\ref{aff12}}
\and B.~Camacho~Quevedo\orcid{0000-0002-8789-4232}\inst{\ref{aff18},\ref{aff19}}
\and A.~Cappi\inst{\ref{aff38},\ref{aff144}}
\and F.~Caro\inst{\ref{aff12}}
\and C.~S.~Carvalho\inst{\ref{aff125}}
\and T.~Castro\orcid{0000-0002-6292-3228}\inst{\ref{aff27},\ref{aff29},\ref{aff28},\ref{aff26}}
\and F.~Cogato\orcid{0000-0003-4632-6113}\inst{\ref{aff44},\ref{aff38}}
\and S.~Conseil\orcid{0000-0002-3657-4191}\inst{\ref{aff71}}
\and S.~Contarini\orcid{0000-0002-9843-723X}\inst{\ref{aff31}}
\and A.~R.~Cooray\orcid{0000-0002-3892-0190}\inst{\ref{aff145}}
\and O.~Cucciati\orcid{0000-0002-9336-7551}\inst{\ref{aff38}}
\and F.~De~Paolis\orcid{0000-0001-6460-7563}\inst{\ref{aff146},\ref{aff147},\ref{aff148}}
\and G.~Desprez\orcid{0000-0001-8325-1742}\inst{\ref{aff102}}
\and A.~D\'iaz-S\'anchez\orcid{0000-0003-0748-4768}\inst{\ref{aff149}}
\and J.~M.~Diego\orcid{0000-0001-9065-3926}\inst{\ref{aff46}}
\and P.~Dimauro\orcid{0000-0001-7399-2854}\inst{\ref{aff12},\ref{aff150}}
\and A.~Enia\orcid{0000-0002-0200-2857}\inst{\ref{aff59},\ref{aff38}}
\and Y.~Fang\inst{\ref{aff60}}
\and A.~G.~Ferrari\orcid{0009-0005-5266-4110}\inst{\ref{aff39}}
\and P.~G.~Ferreira\orcid{0000-0002-3021-2851}\inst{\ref{aff34}}
\and A.~Finoguenov\orcid{0000-0002-4606-5403}\inst{\ref{aff100}}
\and A.~Franco\orcid{0000-0002-4761-366X}\inst{\ref{aff147},\ref{aff146},\ref{aff148}}
\and K.~Ganga\orcid{0000-0001-8159-8208}\inst{\ref{aff130}}
\and J.~Garc\'ia-Bellido\orcid{0000-0002-9370-8360}\inst{\ref{aff134}}
\and T.~Gasparetto\orcid{0000-0002-7913-4866}\inst{\ref{aff27}}
\and V.~Gautard\inst{\ref{aff151}}
\and R.~Gavazzi\orcid{0000-0002-5540-6935}\inst{\ref{aff79},\ref{aff96}}
\and E.~Gaztanaga\orcid{0000-0001-9632-0815}\inst{\ref{aff19},\ref{aff18},\ref{aff23}}
\and F.~Giacomini\orcid{0000-0002-3129-2814}\inst{\ref{aff39}}
\and G.~Gozaliasl\orcid{0000-0002-0236-919X}\inst{\ref{aff152},\ref{aff100}}
\and M.~Guidi\orcid{0000-0001-9408-1101}\inst{\ref{aff59},\ref{aff38}}
\and C.~M.~Gutierrez\orcid{0000-0001-7854-783X}\inst{\ref{aff153}}
\and A.~Hall\orcid{0000-0002-3139-8651}\inst{\ref{aff14}}
\and S.~Hemmati\orcid{0000-0003-2226-5395}\inst{\ref{aff154}}
\and C.~Hern\'andez-Monteagudo\orcid{0000-0001-5471-9166}\inst{\ref{aff120},\ref{aff69}}
\and H.~Hildebrandt\orcid{0000-0002-9814-3338}\inst{\ref{aff51}}
\and J.~Hjorth\orcid{0000-0002-4571-2306}\inst{\ref{aff114}}
\and J.~J.~E.~Kajava\orcid{0000-0002-3010-8333}\inst{\ref{aff155},\ref{aff156}}
\and Y.~Kang\orcid{0009-0000-8588-7250}\inst{\ref{aff78}}
\and V.~Kansal\orcid{0000-0002-4008-6078}\inst{\ref{aff157},\ref{aff158}}
\and D.~Karagiannis\orcid{0000-0002-4927-0816}\inst{\ref{aff42},\ref{aff159}}
\and K.~Kiiveri\inst{\ref{aff99}}
\and C.~C.~Kirkpatrick\inst{\ref{aff99}}
\and S.~Kruk\orcid{0000-0001-8010-8879}\inst{\ref{aff57}}
\and F.~Lacasa\orcid{0000-0002-7268-3440}\inst{\ref{aff160},\ref{aff81}}
\and M.~Lattanzi\orcid{0000-0003-1059-2532}\inst{\ref{aff43}}
\and J.~Le~Graet\orcid{0000-0001-6523-7971}\inst{\ref{aff83}}
\and F.~Lepori\orcid{0009-0000-5061-7138}\inst{\ref{aff161}}
\and G.~Leroy\orcid{0009-0004-2523-4425}\inst{\ref{aff162},\ref{aff108}}
\and J.~Lesgourgues\orcid{0000-0001-7627-353X}\inst{\ref{aff15}}
\and L.~Leuzzi\orcid{0009-0006-4479-7017}\inst{\ref{aff44},\ref{aff38}}
\and T.~I.~Liaudat\orcid{0000-0002-9104-314X}\inst{\ref{aff163}}
\and S.~J.~Liu\orcid{0000-0001-7680-2139}\inst{\ref{aff80}}
\and A.~Loureiro\orcid{0000-0002-4371-0876}\inst{\ref{aff164},\ref{aff165}}
\and J.~Macias-Perez\orcid{0000-0002-5385-2763}\inst{\ref{aff166}}
\and G.~Maggio\orcid{0000-0003-4020-4836}\inst{\ref{aff27}}
\and M.~Magliocchetti\orcid{0000-0001-9158-4838}\inst{\ref{aff80}}
\and F.~Mannucci\orcid{0000-0002-4803-2381}\inst{\ref{aff167}}
\and R.~Maoli\orcid{0000-0002-6065-3025}\inst{\ref{aff168},\ref{aff12}}
\and J.~Mart\'{i}n-Fleitas\orcid{0000-0002-8594-569X}\inst{\ref{aff111}}
\and C.~J.~A.~P.~Martins\orcid{0000-0002-4886-9261}\inst{\ref{aff169},\ref{aff63}}
\and L.~Maurin\orcid{0000-0002-8406-0857}\inst{\ref{aff81}}
\and M.~Migliaccio\inst{\ref{aff170},\ref{aff171}}
\and M.~Miluzio\inst{\ref{aff57},\ref{aff172}}
\and P.~Monaco\orcid{0000-0003-2083-7564}\inst{\ref{aff142},\ref{aff27},\ref{aff29},\ref{aff28}}
\and A.~Montoro\orcid{0000-0003-4730-8590}\inst{\ref{aff19},\ref{aff18}}
\and G.~Morgante\inst{\ref{aff38}}
\and C.~Murray\inst{\ref{aff130}}
\and S.~Nadathur\orcid{0000-0001-9070-3102}\inst{\ref{aff23}}
\and K.~Naidoo\orcid{0000-0002-9182-1802}\inst{\ref{aff23}}
\and A.~Navarro-Alsina\orcid{0000-0002-3173-2592}\inst{\ref{aff107}}
\and S.~Nesseris\orcid{0000-0002-0567-0324}\inst{\ref{aff134}}
\and L.~Pagano\orcid{0000-0003-1820-5998}\inst{\ref{aff42},\ref{aff43}}
\and F.~Passalacqua\orcid{0000-0002-8606-4093}\inst{\ref{aff121},\ref{aff82}}
\and K.~Paterson\orcid{0000-0001-8340-3486}\inst{\ref{aff97}}
\and L.~Patrizii\inst{\ref{aff39}}
\and A.~Pisani\orcid{0000-0002-6146-4437}\inst{\ref{aff83},\ref{aff173}}
\and D.~Potter\orcid{0000-0002-0757-5195}\inst{\ref{aff161}}
\and S.~Quai\orcid{0000-0002-0449-8163}\inst{\ref{aff44},\ref{aff38}}
\and M.~Radovich\orcid{0000-0002-3585-866X}\inst{\ref{aff88}}
\and P.~Reimberg\orcid{0000-0003-3410-0280}\inst{\ref{aff109}}
\and I.~Risso\orcid{0000-0003-2525-7761}\inst{\ref{aff174}}
\and G.~Rodighiero\orcid{0000-0002-9415-2296}\inst{\ref{aff121},\ref{aff88}}
\and S.~Sacquegna\orcid{0000-0002-8433-6630}\inst{\ref{aff146},\ref{aff147},\ref{aff148}}
\and M.~Sahl\'en\orcid{0000-0003-0973-4804}\inst{\ref{aff175}}
\and E.~Sarpa\orcid{0000-0002-1256-655X}\inst{\ref{aff25},\ref{aff26},\ref{aff29}}
\and J.~Schaye\orcid{0000-0002-0668-5560}\inst{\ref{aff3}}
\and A.~Schneider\orcid{0000-0001-7055-8104}\inst{\ref{aff161}}
\and M.~Sereno\orcid{0000-0003-0302-0325}\inst{\ref{aff38},\ref{aff39}}
\and A.~Silvestri\orcid{0000-0001-6904-5061}\inst{\ref{aff2}}
\and L.~C.~Smith\orcid{0000-0002-3259-2771}\inst{\ref{aff176}}
\and J.~Stadel\orcid{0000-0001-7565-8622}\inst{\ref{aff161}}
\and C.~Tao\orcid{0000-0001-7961-8177}\inst{\ref{aff83}}
\and G.~Testera\inst{\ref{aff16}}
\and R.~Teyssier\orcid{0000-0001-7689-0933}\inst{\ref{aff173}}
\and S.~Tosi\orcid{0000-0002-7275-9193}\inst{\ref{aff17},\ref{aff174}}
\and A.~Troja\orcid{0000-0003-0239-4595}\inst{\ref{aff121},\ref{aff82}}
\and M.~Tucci\inst{\ref{aff78}}
\and C.~Valieri\inst{\ref{aff39}}
\and A.~Venhola\orcid{0000-0001-6071-4564}\inst{\ref{aff177}}
\and D.~Vergani\orcid{0000-0003-0898-2216}\inst{\ref{aff38}}
\and F.~Vernizzi\orcid{0000-0003-3426-2802}\inst{\ref{aff178}}
\and G.~Verza\orcid{0000-0002-1886-8348}\inst{\ref{aff179}}
\and N.~A.~Walton\orcid{0000-0003-3983-8778}\inst{\ref{aff176}}}
										   
%%%% please do not edit the affiliation list -- contact ECEB Bureau for changes
\institute{European Space Agency/ESTEC, Keplerlaan 1, 2201 AZ Noordwijk, The Netherlands\label{aff1}
\and
Institute Lorentz, Leiden University, Niels Bohrweg 2, 2333 CA Leiden, The Netherlands\label{aff2}
\and
Leiden Observatory, Leiden University, Einsteinweg 55, 2333 CC Leiden, The Netherlands\label{aff3}
\and
Universit\'e Paris-Saclay, Universit\'e Paris Cit\'e, CEA, CNRS, AIM, 91191, Gif-sur-Yvette, France\label{aff4}
\and
Center for Data-Driven Discovery, Kavli IPMU (WPI), UTIAS, The University of Tokyo, Kashiwa, Chiba 277-8583, Japan\label{aff5}
\and
Laboratoire Univers et Th\'eorie, Observatoire de Paris, Universit\'e PSL, Universit\'e Paris Cit\'e, CNRS, 92190 Meudon, France\label{aff6}
\and
INAF-IASF Milano, Via Alfonso Corti 12, 20133 Milano, Italy\label{aff7}
\and
Waterloo Centre for Astrophysics, University of Waterloo, Waterloo, Ontario N2L 3G1, Canada\label{aff8}
\and
Dipartimento di Fisica, Universit\`a degli Studi di Torino, Via P. Giuria 1, 10125 Torino, Italy\label{aff9}
\and
INFN-Sezione di Torino, Via P. Giuria 1, 10125 Torino, Italy\label{aff10}
\and
INAF-Osservatorio Astrofisico di Torino, Via Osservatorio 20, 10025 Pino Torinese (TO), Italy\label{aff11}
\and
INAF-Osservatorio Astronomico di Roma, Via Frascati 33, 00078 Monteporzio Catone, Italy\label{aff12}
\and
INFN-Sezione di Roma, Piazzale Aldo Moro, 2 - c/o Dipartimento di Fisica, Edificio G. Marconi, 00185 Roma, Italy\label{aff13}
\and
Institute for Astronomy, University of Edinburgh, Royal Observatory, Blackford Hill, Edinburgh EH9 3HJ, UK\label{aff14}
\and
Institute for Theoretical Particle Physics and Cosmology (TTK), RWTH Aachen University, 52056 Aachen, Germany\label{aff15}
\and
INFN-Sezione di Genova, Via Dodecaneso 33, 16146, Genova, Italy\label{aff16}
\and
Dipartimento di Fisica, Universit\`a di Genova, Via Dodecaneso 33, 16146, Genova, Italy\label{aff17}
\and
Institut d'Estudis Espacials de Catalunya (IEEC),  Edifici RDIT, Campus UPC, 08860 Castelldefels, Barcelona, Spain\label{aff18}
\and
Institute of Space Sciences (ICE, CSIC), Campus UAB, Carrer de Can Magrans, s/n, 08193 Barcelona, Spain\label{aff19}
\and
Universit\'e Paris-Saclay, CNRS/IN2P3, IJCLab, 91405 Orsay, France\label{aff20}
\and
Institut de Recherche en Astrophysique et Plan\'etologie (IRAP), Universit\'e de Toulouse, CNRS, UPS, CNES, 14 Av. Edouard Belin, 31400 Toulouse, France\label{aff21}
\and
Centro de Investigaciones Energ\'eticas, Medioambientales y Tecnol\'ogicas (CIEMAT), Avenida Complutense 40, 28040 Madrid, Spain\label{aff22}
\and
Institute of Cosmology and Gravitation, University of Portsmouth, Portsmouth PO1 3FX, UK\label{aff23}
\and
Laboratoire d'etude de l'Univers et des phenomenes eXtremes, Observatoire de Paris, Universit\'e PSL, Sorbonne Universit\'e, CNRS, 92190 Meudon, France\label{aff24}
\and
SISSA, International School for Advanced Studies, Via Bonomea 265, 34136 Trieste TS, Italy\label{aff25}
\and
ICSC - Centro Nazionale di Ricerca in High Performance Computing, Big Data e Quantum Computing, Via Magnanelli 2, Bologna, Italy\label{aff26}
\and
INAF-Osservatorio Astronomico di Trieste, Via G. B. Tiepolo 11, 34143 Trieste, Italy\label{aff27}
\and
IFPU, Institute for Fundamental Physics of the Universe, via Beirut 2, 34151 Trieste, Italy\label{aff28}
\and
INFN, Sezione di Trieste, Via Valerio 2, 34127 Trieste TS, Italy\label{aff29}
\and
INAF-Osservatorio Astronomico di Brera, Via Brera 28, 20122 Milano, Italy\label{aff30}
\and
Max Planck Institute for Extraterrestrial Physics, Giessenbachstr. 1, 85748 Garching, Germany\label{aff31}
\and
Institut f\"ur Theoretische Physik, University of Heidelberg, Philosophenweg 16, 69120 Heidelberg, Germany\label{aff32}
\and
Universit\'e St Joseph; Faculty of Sciences, Beirut, Lebanon\label{aff33}
\and
Department of Physics, Oxford University, Keble Road, Oxford OX1 3RH, UK\label{aff34}
\and
Institute for Particle Physics and Astrophysics, Dept. of Physics, ETH Zurich, Wolfgang-Pauli-Strasse 27, 8093 Zurich, Switzerland\label{aff35}
\and
LINKS Foundation, Via Pier Carlo Boggio, 61 10138 Torino, Italy\label{aff36}
\and
Ludwig-Maximilians-University, Schellingstrasse 4, 80799 Munich, Germany\label{aff37}
\and
INAF-Osservatorio di Astrofisica e Scienza dello Spazio di Bologna, Via Piero Gobetti 93/3, 40129 Bologna, Italy\label{aff38}
\and
INFN-Sezione di Bologna, Viale Berti Pichat 6/2, 40127 Bologna, Italy\label{aff39}
\and
DAMTP, Centre for Mathematical Sciences, Wilberforce Road, Cambridge CB3 0WA, UK\label{aff40}
\and
Kavli Institute for Cosmology Cambridge, Madingley Road, Cambridge, CB3 0HA, UK\label{aff41}
\and
Dipartimento di Fisica e Scienze della Terra, Universit\`a degli Studi di Ferrara, Via Giuseppe Saragat 1, 44122 Ferrara, Italy\label{aff42}
\and
Istituto Nazionale di Fisica Nucleare, Sezione di Ferrara, Via Giuseppe Saragat 1, 44122 Ferrara, Italy\label{aff43}
\and
Dipartimento di Fisica e Astronomia "Augusto Righi" - Alma Mater Studiorum Universit\`a di Bologna, via Piero Gobetti 93/2, 40129 Bologna, Italy\label{aff44}
\and
Universit\'e de Gen\`eve, D\'epartement de Physique Th\'eorique and Centre for Astroparticle Physics, 24 quai Ernest-Ansermet, CH-1211 Gen\`eve 4, Switzerland\label{aff45}
\and
Instituto de F\'isica de Cantabria, Edificio Juan Jord\'a, Avenida de los Castros, 39005 Santander, Spain\label{aff46}
\and
Higgs Centre for Theoretical Physics, School of Physics and Astronomy, The University of Edinburgh, Edinburgh EH9 3FD, UK\label{aff47}
\and
Aix-Marseille Universit\'e, Universit\'e de Toulon, CNRS, CPT, Marseille, France\label{aff48}
\and
Methodology and Statistics Unit, Institute of Psychology, Leiden University, Wassenaarseweg 52, 2333 AK Leiden, The Netherlands\label{aff49}
\and
Jodrell Bank Centre for Astrophysics, Department of Physics and Astronomy, University of Manchester, Oxford Road, Manchester M13 9PL, UK\label{aff50}
\and
Ruhr University Bochum, Faculty of Physics and Astronomy, Astronomical Institute (AIRUB), German Centre for Cosmological Lensing (GCCL), 44780 Bochum, Germany\label{aff51}
\and
Departamento de F\'isica, FCFM, Universidad de Chile, Blanco Encalada 2008, Santiago, Chile\label{aff52}
\and
Mathematical Institute, University of Leiden, Einsteinweg 55, 2333 CA Leiden, The Netherlands\label{aff53}
\and
Center for Cosmology and AstroParticle Physics, The Ohio State University, 191 West Woodruff Avenue, Columbus, OH 43210, USA\label{aff54}
\and
Department of Physics, The Ohio State University, Columbus, OH 43210, USA\label{aff55}
\and
Department of Physics and Astronomy, University College London, Gower Street, London WC1E 6BT, UK\label{aff56}
\and
ESAC/ESA, Camino Bajo del Castillo, s/n., Urb. Villafranca del Castillo, 28692 Villanueva de la Ca\~nada, Madrid, Spain\label{aff57}
\and
School of Mathematics and Physics, University of Surrey, Guildford, Surrey, GU2 7XH, UK\label{aff58}
\and
Dipartimento di Fisica e Astronomia, Universit\`a di Bologna, Via Gobetti 93/2, 40129 Bologna, Italy\label{aff59}
\and
Universit\"ats-Sternwarte M\"unchen, Fakult\"at f\"ur Physik, Ludwig-Maximilians-Universit\"at M\"unchen, Scheinerstrasse 1, 81679 M\"unchen, Germany\label{aff60}
\and
Department of Physics "E. Pancini", University Federico II, Via Cinthia 6, 80126, Napoli, Italy\label{aff61}
\and
INAF-Osservatorio Astronomico di Capodimonte, Via Moiariello 16, 80131 Napoli, Italy\label{aff62}
\and
Instituto de Astrof\'isica e Ci\^encias do Espa\c{c}o, Universidade do Porto, CAUP, Rua das Estrelas, PT4150-762 Porto, Portugal\label{aff63}
\and
Faculdade de Ci\^encias da Universidade do Porto, Rua do Campo de Alegre, 4150-007 Porto, Portugal\label{aff64}
\and
Port d'Informaci\'{o} Cient\'{i}fica, Campus UAB, C. Albareda s/n, 08193 Bellaterra (Barcelona), Spain\label{aff65}
\and
INFN section of Naples, Via Cinthia 6, 80126, Napoli, Italy\label{aff66}
\and
Institute for Astronomy, University of Hawaii, 2680 Woodlawn Drive, Honolulu, HI 96822, USA\label{aff67}
\and
Dipartimento di Fisica e Astronomia "Augusto Righi" - Alma Mater Studiorum Universit\`a di Bologna, Viale Berti Pichat 6/2, 40127 Bologna, Italy\label{aff68}
\and
Instituto de Astrof\'{\i}sica de Canarias, V\'{\i}a L\'actea, 38205 La Laguna, Tenerife, Spain\label{aff69}
\and
European Space Agency/ESRIN, Largo Galileo Galilei 1, 00044 Frascati, Roma, Italy\label{aff70}
\and
Universit\'e Claude Bernard Lyon 1, CNRS/IN2P3, IP2I Lyon, UMR 5822, Villeurbanne, F-69100, France\label{aff71}
\and
Institut de Ci\`{e}ncies del Cosmos (ICCUB), Universitat de Barcelona (IEEC-UB), Mart\'{i} i Franqu\`{e}s 1, 08028 Barcelona, Spain\label{aff72}
\and
Instituci\'o Catalana de Recerca i Estudis Avan\c{c}ats (ICREA), Passeig de Llu\'{\i}s Companys 23, 08010 Barcelona, Spain\label{aff73}
\and
UCB Lyon 1, CNRS/IN2P3, IUF, IP2I Lyon, 4 rue Enrico Fermi, 69622 Villeurbanne, France\label{aff74}
\and
Mullard Space Science Laboratory, University College London, Holmbury St Mary, Dorking, Surrey RH5 6NT, UK\label{aff75}
\and
Departamento de F\'isica, Faculdade de Ci\^encias, Universidade de Lisboa, Edif\'icio C8, Campo Grande, PT1749-016 Lisboa, Portugal\label{aff76}
\and
Instituto de Astrof\'isica e Ci\^encias do Espa\c{c}o, Faculdade de Ci\^encias, Universidade de Lisboa, Campo Grande, 1749-016 Lisboa, Portugal\label{aff77}
\and
Department of Astronomy, University of Geneva, ch. d'Ecogia 16, 1290 Versoix, Switzerland\label{aff78}
\and
Aix-Marseille Universit\'e, CNRS, CNES, LAM, Marseille, France\label{aff79}
\and
INAF-Istituto di Astrofisica e Planetologia Spaziali, via del Fosso del Cavaliere, 100, 00100 Roma, Italy\label{aff80}
\and
Universit\'e Paris-Saclay, CNRS, Institut d'astrophysique spatiale, 91405, Orsay, France\label{aff81}
\and
INFN-Padova, Via Marzolo 8, 35131 Padova, Italy\label{aff82}
\and
Aix-Marseille Universit\'e, CNRS/IN2P3, CPPM, Marseille, France\label{aff83}
\and
Space Science Data Center, Italian Space Agency, via del Politecnico snc, 00133 Roma, Italy\label{aff84}
\and
INFN-Bologna, Via Irnerio 46, 40126 Bologna, Italy\label{aff85}
\and
School of Physics, HH Wills Physics Laboratory, University of Bristol, Tyndall Avenue, Bristol, BS8 1TL, UK\label{aff86}
\and
FRACTAL S.L.N.E., calle Tulip\'an 2, Portal 13 1A, 28231, Las Rozas de Madrid, Spain\label{aff87}
\and
INAF-Osservatorio Astronomico di Padova, Via dell'Osservatorio 5, 35122 Padova, Italy\label{aff88}
\and
Dipartimento di Fisica "Aldo Pontremoli", Universit\`a degli Studi di Milano, Via Celoria 16, 20133 Milano, Italy\label{aff89}
\and
Institute of Theoretical Astrophysics, University of Oslo, P.O. Box 1029 Blindern, 0315 Oslo, Norway\label{aff90}
\and
Jet Propulsion Laboratory, California Institute of Technology, 4800 Oak Grove Drive, Pasadena, CA, 91109, USA\label{aff91}
\and
Department of Physics, Lancaster University, Lancaster, LA1 4YB, UK\label{aff92}
\and
Felix Hormuth Engineering, Goethestr. 17, 69181 Leimen, Germany\label{aff93}
\and
Technical University of Denmark, Elektrovej 327, 2800 Kgs. Lyngby, Denmark\label{aff94}
\and
Cosmic Dawn Center (DAWN), Denmark\label{aff95}
\and
Institut d'Astrophysique de Paris, UMR 7095, CNRS, and Sorbonne Universit\'e, 98 bis boulevard Arago, 75014 Paris, France\label{aff96}
\and
Max-Planck-Institut f\"ur Astronomie, K\"onigstuhl 17, 69117 Heidelberg, Germany\label{aff97}
\and
NASA Goddard Space Flight Center, Greenbelt, MD 20771, USA\label{aff98}
\and
Department of Physics and Helsinki Institute of Physics, Gustaf H\"allstr\"omin katu 2, 00014 University of Helsinki, Finland\label{aff99}
\and
Department of Physics, P.O. Box 64, 00014 University of Helsinki, Finland\label{aff100}
\and
Helsinki Institute of Physics, Gustaf H{\"a}llstr{\"o}min katu 2, University of Helsinki, Helsinki, Finland\label{aff101}
\and
Kapteyn Astronomical Institute, University of Groningen, PO Box 800, 9700 AV Groningen, The Netherlands\label{aff102}
\and
SKA Observatory, Jodrell Bank, Lower Withington, Macclesfield, Cheshire SK11 9FT, UK\label{aff103}
\and
Centre de Calcul de l'IN2P3/CNRS, 21 avenue Pierre de Coubertin 69627 Villeurbanne Cedex, France\label{aff104}
\and
INFN-Sezione di Milano, Via Celoria 16, 20133 Milano, Italy\label{aff105}
\and
University of Applied Sciences and Arts of Northwestern Switzerland, School of Engineering, 5210 Windisch, Switzerland\label{aff106}
\and
Universit\"at Bonn, Argelander-Institut f\"ur Astronomie, Auf dem H\"ugel 71, 53121 Bonn, Germany\label{aff107}
\and
Department of Physics, Institute for Computational Cosmology, Durham University, South Road, Durham, DH1 3LE, UK\label{aff108}
\and
Institut d'Astrophysique de Paris, 98bis Boulevard Arago, 75014, Paris, France\label{aff109}
\and
Institute of Physics, Laboratory of Astrophysics, Ecole Polytechnique F\'ed\'erale de Lausanne (EPFL), Observatoire de Sauverny, 1290 Versoix, Switzerland\label{aff110}
\and
Aurora Technology for European Space Agency (ESA), Camino bajo del Castillo, s/n, Urbanizacion Villafranca del Castillo, Villanueva de la Ca\~nada, 28692 Madrid, Spain\label{aff111}
\and
Institut de F\'{i}sica d'Altes Energies (IFAE), The Barcelona Institute of Science and Technology, Campus UAB, 08193 Bellaterra (Barcelona), Spain\label{aff112}
\and
School of Mathematics, Statistics and Physics, Newcastle University, Herschel Building, Newcastle-upon-Tyne, NE1 7RU, UK\label{aff113}
\and
DARK, Niels Bohr Institute, University of Copenhagen, Jagtvej 155, 2200 Copenhagen, Denmark\label{aff114}
\and
Department of Physics and Astronomy, University of Waterloo, Waterloo, Ontario N2L 3G1, Canada\label{aff115}
\and
Perimeter Institute for Theoretical Physics, Waterloo, Ontario N2L 2Y5, Canada\label{aff116}
\and
Centre National d'Etudes Spatiales -- Centre spatial de Toulouse, 18 avenue Edouard Belin, 31401 Toulouse Cedex 9, France\label{aff117}
\and
Institute of Space Science, Str. Atomistilor, nr. 409 M\u{a}gurele, Ilfov, 077125, Romania\label{aff118}
\and
Consejo Superior de Investigaciones Cientificas, Calle Serrano 117, 28006 Madrid, Spain\label{aff119}
\and
Universidad de La Laguna, Departamento de Astrof\'{\i}sica, 38206 La Laguna, Tenerife, Spain\label{aff120}
\and
Dipartimento di Fisica e Astronomia "G. Galilei", Universit\`a di Padova, Via Marzolo 8, 35131 Padova, Italy\label{aff121}
\and
Universit\"at Innsbruck, Institut f\"ur Astro- und Teilchenphysik, Technikerstr. 25/8, 6020 Innsbruck, Austria\label{aff122}
\and
Satlantis, University Science Park, Sede Bld 48940, Leioa-Bilbao, Spain\label{aff123}
\and
Department of Physics, Royal Holloway, University of London, TW20 0EX, UK\label{aff124}
\and
Instituto de Astrof\'isica e Ci\^encias do Espa\c{c}o, Faculdade de Ci\^encias, Universidade de Lisboa, Tapada da Ajuda, 1349-018 Lisboa, Portugal\label{aff125}
\and
Cosmic Dawn Center (DAWN)\label{aff126}
\and
Niels Bohr Institute, University of Copenhagen, Jagtvej 128, 2200 Copenhagen, Denmark\label{aff127}
\and
Universidad Polit\'ecnica de Cartagena, Departamento de Electr\'onica y Tecnolog\'ia de Computadoras,  Plaza del Hospital 1, 30202 Cartagena, Spain\label{aff128}
\and
Infrared Processing and Analysis Center, California Institute of Technology, Pasadena, CA 91125, USA\label{aff129}
\and
Universit\'e Paris Cit\'e, CNRS, Astroparticule et Cosmologie, 75013 Paris, France\label{aff130}
\and
INAF, Istituto di Radioastronomia, Via Piero Gobetti 101, 40129 Bologna, Italy\label{aff131}
\and
Astronomical Observatory of the Autonomous Region of the Aosta Valley (OAVdA), Loc. Lignan 39, I-11020, Nus (Aosta Valley), Italy\label{aff132}
\and
ICL, Junia, Universit\'e Catholique de Lille, LITL, 59000 Lille, France\label{aff133}
\and
Instituto de F\'isica Te\'orica UAM-CSIC, Campus de Cantoblanco, 28049 Madrid, Spain\label{aff134}
\and
CERCA/ISO, Department of Physics, Case Western Reserve University, 10900 Euclid Avenue, Cleveland, OH 44106, USA\label{aff135}
\and
Technical University of Munich, TUM School of Natural Sciences, Physics Department, James-Franck-Str.~1, 85748 Garching, Germany\label{aff136}
\and
Max-Planck-Institut f\"ur Astrophysik, Karl-Schwarzschild-Str.~1, 85748 Garching, Germany\label{aff137}
\and
Donostia International Physics Center (DIPC), Paseo Manuel de Lardizabal, 4, 20018, Donostia-San Sebasti\'an, Guipuzkoa, Spain\label{aff138}
\and
IKERBASQUE, Basque Foundation for Science, 48013, Bilbao, Spain\label{aff139}
\and
Departamento de F{\'\i}sica Fundamental. Universidad de Salamanca. Plaza de la Merced s/n. 37008 Salamanca, Spain\label{aff140}
\and
Universit\'e de Strasbourg, CNRS, Observatoire astronomique de Strasbourg, UMR 7550, 67000 Strasbourg, France\label{aff141}
\and
Dipartimento di Fisica - Sezione di Astronomia, Universit\`a di Trieste, Via Tiepolo 11, 34131 Trieste, Italy\label{aff142}
\and
California Institute of Technology, 1200 E California Blvd, Pasadena, CA 91125, USA\label{aff143}
\and
Universit\'e C\^{o}te d'Azur, Observatoire de la C\^{o}te d'Azur, CNRS, Laboratoire Lagrange, Bd de l'Observatoire, CS 34229, 06304 Nice cedex 4, France\label{aff144}
\and
Department of Physics \& Astronomy, University of California Irvine, Irvine CA 92697, USA\label{aff145}
\and
Department of Mathematics and Physics E. De Giorgi, University of Salento, Via per Arnesano, CP-I93, 73100, Lecce, Italy\label{aff146}
\and
INFN, Sezione di Lecce, Via per Arnesano, CP-193, 73100, Lecce, Italy\label{aff147}
\and
INAF-Sezione di Lecce, c/o Dipartimento Matematica e Fisica, Via per Arnesano, 73100, Lecce, Italy\label{aff148}
\and
Departamento F\'isica Aplicada, Universidad Polit\'ecnica de Cartagena, Campus Muralla del Mar, 30202 Cartagena, Murcia, Spain\label{aff149}
\and
Observatorio Nacional, Rua General Jose Cristino, 77-Bairro Imperial de Sao Cristovao, Rio de Janeiro, 20921-400, Brazil\label{aff150}
\and
CEA Saclay, DFR/IRFU, Service d'Astrophysique, Bat. 709, 91191 Gif-sur-Yvette, France\label{aff151}
\and
Department of Computer Science, Aalto University, PO Box 15400, Espoo, FI-00 076, Finland\label{aff152}
\and
Instituto de Astrof\'\i sica de Canarias, c/ Via Lactea s/n, La Laguna 38200, Spain. Departamento de Astrof\'\i sica de la Universidad de La Laguna, Avda. Francisco Sanchez, La Laguna, 38200, Spain\label{aff153}
\and
Caltech/IPAC, 1200 E. California Blvd., Pasadena, CA 91125, USA\label{aff154}
\and
Department of Physics and Astronomy, Vesilinnantie 5, 20014 University of Turku, Finland\label{aff155}
\and
Serco for European Space Agency (ESA), Camino bajo del Castillo, s/n, Urbanizacion Villafranca del Castillo, Villanueva de la Ca\~nada, 28692 Madrid, Spain\label{aff156}
\and
ARC Centre of Excellence for Dark Matter Particle Physics, Melbourne, Australia\label{aff157}
\and
Centre for Astrophysics \& Supercomputing, Swinburne University of Technology,  Hawthorn, Victoria 3122, Australia\label{aff158}
\and
Department of Physics and Astronomy, University of the Western Cape, Bellville, Cape Town, 7535, South Africa\label{aff159}
\and
Universit\'e Libre de Bruxelles (ULB), Service de Physique Th\'eorique CP225, Boulevard du Triophe, 1050 Bruxelles, Belgium\label{aff160}
\and
Department of Astrophysics, University of Zurich, Winterthurerstrasse 190, 8057 Zurich, Switzerland\label{aff161}
\and
Department of Physics, Centre for Extragalactic Astronomy, Durham University, South Road, Durham, DH1 3LE, UK\label{aff162}
\and
IRFU, CEA, Universit\'e Paris-Saclay 91191 Gif-sur-Yvette Cedex, France\label{aff163}
\and
Oskar Klein Centre for Cosmoparticle Physics, Department of Physics, Stockholm University, Stockholm, SE-106 91, Sweden\label{aff164}
\and
Astrophysics Group, Blackett Laboratory, Imperial College London, London SW7 2AZ, UK\label{aff165}
\and
Univ. Grenoble Alpes, CNRS, Grenoble INP, LPSC-IN2P3, 53, Avenue des Martyrs, 38000, Grenoble, France\label{aff166}
\and
INAF-Osservatorio Astrofisico di Arcetri, Largo E. Fermi 5, 50125, Firenze, Italy\label{aff167}
\and
Dipartimento di Fisica, Sapienza Universit\`a di Roma, Piazzale Aldo Moro 2, 00185 Roma, Italy\label{aff168}
\and
Centro de Astrof\'{\i}sica da Universidade do Porto, Rua das Estrelas, 4150-762 Porto, Portugal\label{aff169}
\and
Dipartimento di Fisica, Universit\`a di Roma Tor Vergata, Via della Ricerca Scientifica 1, Roma, Italy\label{aff170}
\and
INFN, Sezione di Roma 2, Via della Ricerca Scientifica 1, Roma, Italy\label{aff171}
\and
HE Space for European Space Agency (ESA), Camino bajo del Castillo, s/n, Urbanizacion Villafranca del Castillo, Villanueva de la Ca\~nada, 28692 Madrid, Spain\label{aff172}
\and
Department of Astrophysical Sciences, Peyton Hall, Princeton University, Princeton, NJ 08544, USA\label{aff173}
\and
INAF-Osservatorio Astronomico di Brera, Via Brera 28, 20122 Milano, Italy, and INFN-Sezione di Genova, Via Dodecaneso 33, 16146, Genova, Italy\label{aff174}
\and
Theoretical astrophysics, Department of Physics and Astronomy, Uppsala University, Box 516, 751 37 Uppsala, Sweden\label{aff175}
\and
Institute of Astronomy, University of Cambridge, Madingley Road, Cambridge CB3 0HA, UK\label{aff176}
\and
Space physics and astronomy research unit, University of Oulu, Pentti Kaiteran katu 1, FI-90014 Oulu, Finland\label{aff177}
\and
Institut de Physique Th\'eorique, CEA, CNRS, Universit\'e Paris-Saclay 91191 Gif-sur-Yvette Cedex, France\label{aff178}
\and
Center for Computational Astrophysics, Flatiron Institute, 162 5th Avenue, 10010, New York, NY, USA\label{aff179}}

\date{\today}

\titlerunning{CLOE. 3. Inference and forecasts}
   \authorrunning{Euclid Collaboration: G.~Cañas-Herrera et al.}

  \abstract
   {The \Euclid mission aims to measure the positions, shapes, and redshifts of over a billion galaxies to provide unprecedented constraints on the nature of dark matter and dark energy. Achieving this goal requires a continuous reassessment of the mission's scientific performance, particularly in terms of its ability to constrain cosmological parameters, as our understanding of how to model large-scale structure observables {improves}. In this study, we present the first scientific forecasts using the Cosmology Likelihood for Observables in Euclid (\CLOE), a dedicated \Euclid cosmological pipeline developed to support this endeavour. Using advanced Bayesian inference techniques applied to synthetic \Euclid-like data, we sampled the posterior distribution of cosmological and nuisance parameters across a variety of cosmological models and \Euclid primary probes: cosmic shear, angular photometric galaxy clustering, galaxy-galaxy lensing, and spectroscopic galaxy clustering. We validated the capability of \CLOE to produce reliable cosmological forecasts, showcasing \Euclid's potential to achieve a figure of merit for the dark energy parameters \wo and \wa, which exceed 400 when all primary probes are combined. Furthermore, we illustrate the behaviour of the posterior probability distribution of the parameters of interest given different priors and scale cuts. Finally, we emphasise the importance of addressing computational challenges, proposing further exploration of innovative data science techniques to efficiently navigate the \Euclid high-dimensional parameter space in upcoming cosmological data releases.}


   \maketitle
   
\section{Introduction}\label{sec:introduction}
\Euclid, a medium-class mission of the European Space Agency (ESA) under the \textit{Cosmic Vision 2015--2025} programme, is designed to investigate the accelerated expansion of the Universe, attributed to dark energy \citep{Perlmutter, SN1, SN2}, and to explore the nature of dark matter \citep{Feng2010}. It also aims to probe the initial conditions that seeded cosmic structure and to test the limits of general relativity \citep{Laureijs11}. Following its successful launch on July $1$ 2023, \Euclid has begun mapping the cosmos in unprecedented detail, focusing on the \gls{lss} and compiling one of the most extensive galaxy catalogues to date.

As a cornerstone of contemporary cosmology, \Euclid will complement and extend results from current surveys to test the concordance cosmological model, also known as \LCDM, where cosmic acceleration is driven by a cosmological constant $\varLambda$ \citep{Heymans:2020gsg, DESY3, DESIDR1} and the major matter component is Cold Dark Matter. The mission also targets the \wowaCDM\ extension, where dark energy is treated as a barotropic fluid with a redshift-dependent equation of state (EoS) $w(z) \equiv p / \rho c^2$. In particular, we adopted the Chevallier--Polarski--Linder (CPL) parametrisation, a widely used Maclaurin expansion for $w(z)$ EoS \citep{Chevallier:2000qy,PhysRevLett.90.091301},
\begin{equation}\label{eq:weos}
w(z) = w_0 + w_a \,\frac{z}{1+z}\;,
\end{equation}
where \wo is the present day ($z=0$) value of the EoS, while \wa measures how fast it evolves with redshift. In \LCDM, $\wo=-1$ and $\wa=0$. 
In particular, we aimed to quantify the performance of \Euclid in discerning the nature of the dark Universe by comparing the so-called dark energy figure of merit (\fom), which is defined as the inverse square root of the covariance-matrix determinant for the dark energy parameters \wo and \wa \citep{Wang2008},
\begin{equation}
{\rm FoM}=\frac{1}{\sqrt{\det {\rm Cov}(w_{\rm 0},w_{a})}}\;.
\label{eq:FOM}
\end{equation}
A larger \fom indicates a more precise measurement of the dark energy properties. For this reason, it is crucial to increase, as much as possible, the available data used for the analysis to decrease the statistical uncertainty associated with these parameters.

%Euclid is a 1.2-meter telescope equipped with two cutting-edge scientific instruments. This allow \Euclid to measure the positions, shapes, and redshifts of galaxies across the optical and near-infrared spectrum. The payload includes a high-fidelity panoramic visible instrument (VIS), equipped with one filter (\IE) \citep{EuclidSkyVIS}, a Near-Infrared Spectrometer and Photometer (NISP) equipped with three filters (\YE, \JE, and \HE) and a slitless spectrograph, featuring a set of `blue' and `red' grisms \citep{EuclidSkyNISP, EuclidSkyNISPCU}. With a common field-of-view of 0.53 square degrees, \Euclid will survey approximately 14\,000 square degrees of the sky, reaching redshifts nearing $z\approx2$, forming the Euclid Wide Survey \citep{Scaramella-EP1}. \Euclid will generate a comprehensive galaxy redshift catalogue spanning magnitudes of up to 24 in the \YE, \JE, and \HE bands. This treasure trove of data promises to furnish invaluable insights, supported by 30 million spectroscopic redshifts for galaxy-clustering analysis and a staggering 1.5 billion photometric galaxy images, crucial for weak lensing and photometric galaxy-clustering investigations. See \citet{EuclidSkyOverview} for a comprehensive review of the science that \Euclid will enable.

As the \Euclid\ survey progresses and the covered area expands, it becomes essential to refine cosmological forecasts to ensure the readiness of Bayesian analysis pipelines for the mission’s scientific exploitation. The initial forecasts outlined in the \Euclid\ Definition Study Report \citep{Laureijs11} set the stage for the mission’s cosmological objectives, but recent developments call for a reassessment. First, our improved understanding of the theoretical complexity involved in modelling large-scale structure observables -- such as cosmic shear, angular galaxy clustering, galaxy-galaxy lensing, and spectroscopic clustering -- now demands more accurate treatments. These models must incorporate systematic effects, which introduce a large number of nuisance parameters and substantially enlarge the parameter space.

Second, the field has advanced beyond traditional approximation methods such as Fisher forecasts. There is growing consensus within the Euclid Consortium \citep{Blanchard-EP7} that robust cosmological inference requires sampling the full posterior probability distribution. Bayesian statistics offer a rigorous framework for model testing and parameter estimation, driving the need for software capable of performing \corr{exploration of full posterior probability distribution using different sampling methods (e.g.\ Metropolis-Hastings Monte Carlo and nested sampling)}. Such tools must compute likelihoods based on theoretical predictions and observational data while modelling a broad range of probes and their combinations.

This paper presents cosmological forecasts and Bayesian analysis results derived using the \Euclid\ cosmological pipeline. It reflects a coordinated effort across the consortium, including theoretical modelling \citep[][referred to as Paper 1]{2025arXiv251009118E}, software development \citep[][Paper 2]{2026arXiv260322475E}, review, and validation \citep[][Paper 4]{2025arXiv251009141E}. The result of this work is the development of the Cosmological Likelihood for Observables in Euclid\ (\CLOE), a tool that produces theoretical predictions, evaluates likelihoods, and constrains cosmological parameters.

The paper is structured as follows: \cref{sec:methodology} introduces the Bayesian inference framework underlying \CLOE\ and the sampling techniques used. In \cref{sec:theory}, we describe the main \Euclid\ probes and the theoretical modelling, including non-linear treatments and alternative cosmological scenarios (\cref{sec:models}). Section \ref{sec:synthetic_data} and \cref{sec:covmat} detail the construction of synthetic data vectors and their associated covariance matrices. Forecast results are presented in \cref{sec:results}, followed by our conclusions in \cref{sec:conclusions}.

\section{Methodology}\label{sec:methodology}
Constraints on cosmological parameters are best obtained assuming a Bayesian statistical framework,\footnote{See \citet{Ivezic:2014:SDM:2578955} and \citet{trotta2017bayesian} for extensive reviews of Bayesian statistics in astronomy.} whereby the parameters of interest are assumed to be random variables following probability distributions. Bayesian statistical analyses rely on Bayes' Theorem \citep{bayes1763}, which yields the probability distribution of the parameters $\vec{\theta}$ given a model $M$ and the observed data $\vec{d}$. 
%\stefc{Given that both parameterS and datA are plural, we should use vector or set notation for them.} 
This probability distribution, $\post$, called a posterior distribution, is defined as
\begin{equation}\label{eq:BayesTh}
\post=\frac{\like\,\prior}{\evid}\;,
\end{equation}
where $\like$ is the probability of the data given the parameters of the assumed model; and, in the case of fixed data as a multivalued function of the parameters, it is known as the likelihood. $\prior$ is the prior distribution, which is the probability distribution of the parameters $\vec{\theta}$ taking into account all available external (i.e.\ \textit{a priori}) information; and $\evid$ is the evidence, which gives the probability of observing the data given the external information as well as the chosen model $M$.

%The likelihood, $\mathcal{L}$, which incorporates the underlying assumptions of the model $M$, serves as a probability distribution, assessing the agreement between that model, characterised by a specific set of parameter values $\vec{\theta}$, and the observed experimental data $\vec{d}$. A primary focus in cosmology is determining the degree of compatibility of a given cosmological model, such as the concordance model or its variants, with the available data. Hence, establishing a link between the observed data $\vec{d}$ and its theoretical predictions derived from the model becomes imperative. 

\corr{In cosmology, the likelihood $\mathcal{L}$ is commonly modelled as a multivariate Gaussian distribution. This approximation is justified by the large number of independent Fourier modes of the underlying cosmological fields that contribute to the measured observables. According to the Central Limit Theorem, the distribution of the observable estimator constructed from a sufficiently large number of independent random variables converges towards a Gaussian, regardless of the distribution of the individual variables. Since cosmological surveys typically probe an enormous number of modes, the resulting sampling distributions of the measured quantities are well captured by a multivariate Gaussian, such as}
\begin{equation}\label{eq:likelihood_gaussian}
-2\ln{\mathcal{L}} = [\vec{d}-\vec{T}(\vec{\theta})]^{\sf T}\, \tens{C}^{-1}\,[\vec{d}-\vec{T}(\vec{\theta})]+\mathrm{const.}\;,
\end{equation}
where $\vec{d}$ is the data vector, $\vec{T}(\vec{\theta})$ is the theory vector, and $\tens{C}$ is the covariance matrix of the data $\vec{d}$.\footnote{For illustration purposes, we are using a multivariate Gaussian distribution. However, for an exhaustive description of the current likelihood implementation in \CLOE, the reader is kindly referred to the \CLOE code implementation article Paper 2.} Indeed, \citet{JB2026}, and \citet{2026arXiv260401309E} studied the distribution of two-point statistics and their non-Gaussianity{, showing} that the assumption of a Gaussian likelihood is sufficient in the context of \Euclid. 

Assuming that the posterior distribution $P$ is Gaussian, we can obtain its multidimensional covariance matrix $\tens{C}$ using Fisher analysis. The Fisher matrix \citep{1995PhDT........19B,1996ApJ...465...34V,TTH97} is defined as the Hessian of the logarithmic likelihood function
\begin{equation}
  F_{\alpha\beta} = \left\langle- \left. \frac{\partial^2 \ln \mathcal{L}}{\partial\theta_\alpha \partial\theta_\beta}\right|_{\vec{\theta}_{\rm ref}} \right\rangle\;, 
  \label{eq:fisher-general}
\end{equation}
where $\alpha$ and $\beta$ denote the elements of the parameter set and the derivatives are evaluated at the point ${\mathbf \theta_{\rm ref}}$ of the parameter space. This point should agree with the maximum of the likelihood distribution, and in practical terms, corresponds to the fiducial value assumed in the analysis. The covariance matrix of the posterior distribution is the inverse of the obtained Fisher matrix. Fisher analysis is a fast way of exploring the posterior distribution and has been widely used in the literature \citep{Martinelli20, Bonici23, EP-Dournac, Ilic-EP15, Frusciante23, Casas23a, Nesseris22}. The Fisher matrix approach assumes Gaussian posteriors, an assumption that can fail not only in large parameter spaces but also in cases with poorly constrained parameters or non-linear parametrisations, commonly found in large-scale structure analyses. For example, the \Om--$\sigma_8$ degeneracy in weak lensing led to non-Gaussian posteriors even in low-dimensional spaces, motivating the use of combinations like of parameters for a more Gaussian behaviour.

\subsection{Sampling the posterior distribution}
When the primary objective is to estimate the optimal values for the parameter set $\vec{\theta}$ that best fit a model $M$, this goal translates into obtaining the corresponding posterior distribution $\post$. Thus, determining the most suitable parameter values $\vec{\theta}$ entails exploring the parameter space and assessing the quality of the model-data fit across a broad range of parameter values within the space allowed by the prior $\Pi$. 

In cosmology, evaluating the posterior distributions $\post$ of numerous parameters with non-conjugate prior distributions poses considerable challenges. Hence, numerical techniques are indispensable for sampling the posterior distribution. Typically, this involves assessing potential parameter values via a \corr{sampling-based} approach. These techniques rely on random sampling from the actual posterior distribution. The prevalent methods include Markov chain Monte Carlo (MCMC) techniques, which enhance sampling efficiency by iteratively refining the parameter-space exploration, and other alternative algorithms, such as `nested sampling', which directly estimate the evidence while exploring the parameter space, rather than just sampling from the posterior distribution. 

Choosing the appropriate sampling technique is crucial. In a high-dimensional space, simultaneously drawing samples from the posterior distributions of multiple parameters can be a slow and computationally intensive task, especially in light of the \Euclid survey analysis requirements. While MCMC Metropolis--Hastings, MH, \citep{MH} has been successfully used in \citet{DESIDR1} and \citet{DESIDR1_full_shape}, it struggles to properly sample the posterior distribution if it is multi-modal or non-Gaussian. In contrast, nested sampling \citep{nested_sampling} efficiently samples such posterior distributions while simultaneously computing the evidence $\mathcal{Z}(\vec{d}|M)$, given by the integral
\begin{equation}\label{eq:evidence}
    \mathcal{Z}=\int \mathcal{L}(\vec{\theta})\,\Pi(\vec{\theta})\, \diff \vec{\theta}\;.
\end{equation}
To do this efficiently, the technique works by first drawing a number of 
$n_{\rm{live}}$ `live points' from the prior, and in each subsequent iteration $i$, replacing the point with the lowest likelihood value $\mathcal{L}_i$ (now denoted as a dead point) with a new point with a greater likelihood. Denoting the prior volume $X(\mathcal{L})$ as the fraction of the prior contained within an \textit{isocurvature likelihood contour}, it is given by 
\begin{equation}
    X(\mathcal{L})=\int_{\mathcal{L}(\vec{\theta})>\mathcal{L}} \Pi(\vec{\theta})\, \diff\vec{\theta}\;.
\end{equation}
Then the evidence can be easily approximated as a sum of the area under the $\mathcal{L}(X)$ curve
\begin{equation}
    \mathcal{Z}\approx\,\sum_{i=1}(X_{i-1}-X_i)\,\mathcal{L}_i\;.
\end{equation}

When the posterior mass $\mathcal{Z}_{\rm live}\approx \left<\mathcal{L}\right>_{\rm live}X_{\rm live}$ contained by the current set of live points is a small enough fraction of the total $\mathcal{Z}$, the posterior distribution is considered to be converged \citep{2011MNRAS.414.1418K}. 

There are currently several state-of-the-art implementations of nested sampling for Bayesian inference, notably \multinest\footnote{\url{https://github.com/JohannesBuchner/MultiNest}}\faGithub\hspace{1pt} \citep{MultiNest} and \polychord\footnote{\url{https://github.com/PolyChord/PolyChordLite}}\faGithub\hspace{1pt} \citep{Polychord,Polychord2}. \polychord has been used in the latest survey analyses \citep{DESY3,Li:2023tui} and has been proven to give robust and accurate posteriors \citep{lemos:2022}.

Although nested sampling algorithms scale better than MCMC MH with an increasing number of dimensions, they are still computationally expensive. In particular, \polychord parallelisation scales optimally when one core is assigned to each live point. Typically, $25\,D$ live points are needed for robust posterior sampling, where $D$ refers to the number of dimensions of the parameter space. In a typical cosmological run where LSS data is used, this requirement implies approximately an order of $\mathcal{O}(10^3)$ number of live points. For this reason, novel nested sampling algorithms have been developed, aiming to speed up the calculation of the evidence in \cref{eq:evidence}, or allowing for more efficient inference on how the boundary of the live points should be drawn, for instance, by using machine learning.  In particular, the \nautilus\footnote{\url{https://github.com/johannesulf/nautilus}}\faGithub\hspace{1pt} code uses a neural network-based algorithm to determine efficient boundaries instead of calculating the corresponding integral \citep{nautilus}. This software is currently growing in popularity as it is fast and less computationally demanding.\\

\subsection{Analysis setup: Sampling and samples}
In this work, we present the main forecasting results using \nautilus as the sampling software, which we have interfaced with \href{https://github.com/CobayaSampler/cobaya}{\cobaya}\footnote{\url{https://github.com/CobayaSampler/cobaya}}\faGithub\hspace{1pt} \citep{2019ascl.soft10019T} using its \texttt{get\_model} wrapper. We run \nautilus using 4000 live points, 16 neural networks, and 512 likelihood evaluations at each step. We keep default values for both the maximum fraction of the evidence contained in the posterior live mass, and the minimum effective sample size. \corr{For validation, we confirmed that both \nautilus and \polychord yield consistent posterior distributions and compatible evidence values for the \Euclid target cosmological model.\footnote{Since \nautilus performs optimally for $N_\mathrm{dim}<50$, we relied on \polychord to validate our analysis given the large parameter space explored. Nevertheless, due to its superior computational efficiency, \nautilus was our preferred choice for this study.}, and has been used in \citet{kids-legacy}}.

Posterior samples obtained from \nautilus are further processed using \href{https://github.com/cmbant/getdist}{\getdist}\footnote{\url{https://github.com/cmbant/getdist}}\faGithub\hspace{1pt} \citep{getdist} to extract summary statistics and visualise posterior distributions. When evaluating the \fom, we consistently marginalise over both cosmological and nuisance parameters. To accurately account for potential non-Gaussian features in the posteriors, we compute the 68\% and 95\% confidence interval areas using the \texttt{Polygon} routine from \texttt{matplotlib}, instead of relying solely on \cref{eq:FOM}. Both methods have been cross-validated for Gaussian posteriors in \citet{Casas23}, showing consistent results. All posterior distributions are analysed within the \href{https://datalabs.esa.int}{\texttt{ESA datalabs}\footnote{\url{https://datalabs.esa.int}}} environment \citep{Navarro2024}, which also serves as the reference framework for generating the \CLOE-related figures presented in this paper.

\section{Theory vectors}\label{sec:theory} 
In this section, we describe the primary observational probes used by \Euclid to derive cosmological forecasts. Specifically, we describe how we construct the theory vectors given the synthetic data to constrain the underlying cosmological parameters (see \cref{sec:synthetic_data} for more details). Since our goal is to assess the constraining power of \Euclid as a multi-probe experiment, we focus on both the photometric and spectroscopic observables, as well as on their combination, in which case the probes are treated independently.

\subsection{Euclid's primary observables and their combination}\label{subsec:probes}
To explore the nature of the dark sector of the Universe, we use two complementary observational probes: weak lensing and galaxy clustering. For an overview of \Euclid's primary probes as well as their mathematical description, see \citet{EuclidSkyOverview} and Paper 1. They play a crucial role in constraining the underlying parameters that describe the main components of the Universe, i.e.\ dark energy and dark matter. More specifically, it has been widely shown \citep{2023MNRAS.518..477A, Asgari-kids1000, DESY3, Yan:2024zsz, More:2023knf}  that cosmic shear and angular clustering are able to constrain the underlying matter distribution in the Universe, encoded in the mean matter density related to the critical density $\Omega_{\rm m}$, as well as the amplitude of the linear matter power spectrum on scales of $8\,\hMpc$, \sotto, along with the derived parameter $S_8$, defined as
\begin{equation}\label{eq:S8}
S_8 = \sotto \sqrt{\frac{\Om}{0.3}}\; . 
\end{equation}
Their combination, the so-called 3\texttimes2pt probe, which also includes the cross-correlation between cosmic shear and angular clustering (`galaxy-galaxy lensing', XC), has the potential to constrain most of the underlying cosmological parameters \citep{2017arXiv170609359K, 2017MNRAS.465.1454H}. Furthermore, spectroscopic galaxy clustering can provide additional information by constraining two main effects, the growth of cosmic structures (via redshift-space distortions, \rsd), and the background expansion history and geometry of the Universe \citep[via baryon acoustic oscillations, \bao,][]{eBOSS}.

Hence, in this work, we produce forecasts using the following probes:

\begin{enumerate}
    \item Weak lensing (WL);
    \item 3\texttimes2pt: combination of weak lensing, angular clustering (GCph) and galaxy-galaxy lensing (XC);
    \item 3D spectroscopic galaxy clustering (GCsp);
    \item Full \Euclid analysis: combination of the 3\texttimes2pt joint with spectroscopic galaxy clustering (3\texttimes2pt + GCsp).
\end{enumerate}
\begin{table*}[htpb!]
\centering
\caption{Summary of the specifications used to produce the theoretical predictions for both photometric and spectroscopic probes.}
\begin{tabular}{>{\raggedright\arraybackslash}
m{4cm}
>{\raggedright\arraybackslash}m{7.5cm}
>{\raggedright\arraybackslash}m{5.5cm}}
\textbf{Name and Specification} & \textbf{Photometric Probe} & \textbf{Spectroscopic Probe} \\ \hline \hline
Boltzmann Solver  & \camb \citep{Lewis:1999bs} & \camb \citep{Lewis:1999bs} \\ \hline
Non-linear Scales  & \hmcode \citep{Mead:2020vgs} & EFTofLSS %\citep{Moretti24}
\\ \hline
High GCph scale cut  & $\ell_{\rm max}$(WL) = $5000$, \par $\ell_{\rm max}$(GCph) = $\ell_{\rm max}$(XC) = $3000$ & $k_{\rm max} = 0.3\,\kMpc$ for all redshift bins  \\ \hline
Low GCph scale cut  & $\ell_{\rm max}$(WL) = $5000$, \par $\ell_{\rm max}$(GCph) = $\ell_{\rm max}$(XC) = 750 & $k_{\rm max} = 0.3\,\kMpc$ for all redshift bins  \\ \hline
Intrinsic Alignment model & zNLA & $-$ \\ \hline
Galaxy bias  & Linear only, polynomial fitting up to third-order across all redshift bins &  Linear and quadratic, one parameter each per redshift bin \\ \hline
Magnification galaxy bias  & Polynomial fitting up to third-order across all redshift bins & $-$\\ \hline
Systematic nuisance effects & Multiplicative bias, error in the redshift-bin mean distribution & Per-bin purity factors, Poissonian shot noise for extra-stochastic parameters \\ \hline
\end{tabular}
\label{tab:theoretical_probes}
\end{table*}
We focus our forecasting and validation efforts on two-point statistics in harmonic space for photometric probes, and in Fourier space for the spectroscopic one. Specifically, we consider the angular power spectra $\cl{\ell}$ for WL and 3\texttimes2pt, defined, in the Limber approximation, as
\begin{equation}
\cl{\ell}[ij][AB] = \int\de z\,
\frac{c W_i^{A}(z) \, W_j^{B}(z)}
{H(z) \, f_{K}^{2}[r(z)]} \, P_{AB}\left [\frac{\ell + 1/2}{f_K[r(z)]}, \, z \right ] \; ,
\label{eq: cijabgen}
\end{equation}
where $W_i^{A}(z)$ is the radial weight function for the tracer $A$, and $P_{AB}(k, z)$ is the 3D power spectrum for the $(A, B)$ {probe} combination, {with} $A$ and $B$ {being} lensing or photometric galaxy clustering. Further {details on}  { \cref{eq: cijabgen}} {can be} found in Paper 1, Eqs.\ (28--62).\\

For spectroscopic galaxy clustering, we focus on the Legendre multipoles $P_\ell(k)$ in Fourier space,
\begin{equation}
P_{\ell}(k, z) =  \frac{2 \ell + 1}{2} 
\int_{-1}^{-1}{\de \mu_k \,L_{\ell}(\mu_k) \,P_{\rm gg}^{\rm spectro}(k, \mu_k, z)} \, ,
\label{eq: pellgg}
\end{equation}
where $P_{\rm gg}^{\rm spectro}(k, \mu_k, z)$ is the galaxy power spectrum, and $L_\ell(\mu_k)$ is the Legendre polynomial of order $\ell$. Additional details are provided in {Eqs.\ (83--100) of} Paper 1, as well as in Euclid Preparation: Crocce et al. (in prep.) and Euclid Preparation: Moretti et al. (in prep.).

\subsection{Cosmological models}\label{sec:models}
We evaluate \CLOE's capability to constrain parameters for various cosmological models, including the standard \LCDM model and its extensions, as described in \citet{Blanchard-EP7}. A brief overview of these models is listed below.

Assuming a flat universe with no curvature, the \LCDM model is fully specified by five parameters: the baryon density parameter \Ob, the cold dark matter density parameter \Oc, the Hubble constant $H_0$, the primordial spectral index \ns, and the primordial amplitude of scalar perturbations \As. For all the models investigated, we also assume the presence of one massive neutrino species.%., with effective relativistic degrees of freedom $N_{\mathrm{eff}}=3.046$, and total mass $\sum m_{\nu}=0.06$\,eV.

We further consider minimal extensions to \LCDM, by allowing variations in (1) curvature, (2) the dark energy EoS with parameters \wo and \wa, and (3) deviations from general relativity. In the first case, we relax the assumption of $\OK=0$ and sample the curvature energy density \OK as an additional cosmological parameter. In the second case, we consider a $\wowaCDM$ model where we vary the EoS of dark energy, $w(z)$, following \cref{eq:weos}.

Finally, we consider modifications to general relativity by allowing variations in the growth of structures through the parameter \gammag, which governs the scaling relation between the growth rate $f(z)$ and the matter energy density \Omz\,,
\begin{equation}
    f(z)=[\Omz]^{\gammag}.
\end{equation}
A value inconsistent with the fiducial of $\gammag\approx0.545$ would point to a deviation in growth history \citep{1991MNRAS.251..128L,PhysRevD.72.043529}, and hence hint at a gravitational theory different from general relativity. 

In conclusion, we explicitly list the six models that we constrain in this paper using the probes described in \cref{subsec:probes}:
\begin{enumerate}
    \item \LCDM (flat);
    \item \LCDM + \gammag (flat);
    \item \LCDM (non-flat);
    \item \wowaCDM (flat);
    \item \wowaCDM + \gammag (flat);
    \item \wowaCDM (non-flat).\footnote{We do not perform forecasts for the cosmological models \LCDM + \gammag (non-flat) and \wowaCDM + \gammag (non-flat), as the current parametrisation for \gammag in \CLOE does not hold for non-flat geometrical cosmologies. The reader is kindly referred to Paper 1 for more details.}
\end{enumerate}
In this work, we adopt the flat \wowaCDM\ model as the baseline \Euclid\ cosmology, following \citet{EuclidSkyOverview}. Detailed descriptions of the theoretical modelling for each observational probe within the cosmological models considered are provided in Paper 1. For reference, the cosmology used in the \Euclid\ Flagship Simulation 2 corresponds to a flat \LCDM\ model \citep{EuclidSkyFlagship}.

\subsection{Theoretical description of the probes}\label{subsec:theoretical_specs}

To generate the theory vectors, we define a specific setup for the calculation of the background quantities, the recipe used to describe the evolution of the observables over non-linear scales, and the inclusion of multiple systematic effects. In addition, we also identify different sets of scale cuts to explore the constraining power of the data vectors, motivated by the pessimistic and optimistic cases presented in \citet{Blanchard-EP7}. An exhaustive summary of these specifications can be found in \cref{tab:theoretical_probes}.

To compute the corresponding two-point statistics for the different probes, we start by making a call to the Boltzmann solver  \href{https://github.com/cmbant/CAMB}{\camb}\footnote{\url{https://github.com/cmbant/CAMB}}\faGithub\hspace{1pt} \citep{Lewis:1999bs} to obtain the cosmological background quantities and the leading-order density perturbations. The list includes the Hubble parameter $H(z)$, the comoving distance $r(z)$ and angular diameter $D_{\rm A}(z)$ distances, the growth factor $D_+(z)$ and growth rate $f(z)$, as {well} as the linear matter power spectrum $P_{\rm m}(k,z)$.

Despite the substantial amount of information encoded {in the} large scales {of the matter power spectrum}, {a significant fraction of it can also be recovered at the small scales.} However, our linear predictions are {only accurate} on large scales{, hence we} rely on non-linear prescriptions for each observable to model smaller scales and access this additional information. In this analysis, we adopt a baseline approach that is based on a single non-linear recipe, as described in Euclid Preparation: Crocce et al. (in prep.). {A} comprehensive description and comparison of different theoretical frameworks will be explored in future work (Euclid Preparation: Carriltho et al., in prep.; Euclid Preparation: Moretti et al., in prep.). 

\vspace{15pt}

In terms of the photometric probes, the non-linear matter power spectrum is modelled using prescriptions from \hmcode \citep{Mead:2020vgs}, which is a state-of-the-art recipe commonly adopted in Stage-III experiments \citep{Heymans:2020gsg, DESY3}. This is based on an extension of the original halo model formalism \citep{CooShe2002} that also includes the modelling of baryonic effects on small scales. In the version of \hmcode used for this analysis, these {effects} are captured {by} a single parameter, $\log_{10}({T_{\rm AGN}}/$K). {It quantifies} the feedback from active galactic nuclei {(AGN)} in the subgrid prescription for thermal AGN feedback {based on} \citet{2009MNRAS.398...53B}, which was {then} used in the BAHAMAS simulations \citep{2017MNRAS.465.2936M}, {to fit to this version of \hmcode}. Baryonic feedback and its impact {on} cosmological {parameter inference} is further studied in Euclid Preparation: Carrilho et al. (in prep.).

We model the linear galaxy bias $b_{\rm G}(z)$ and magnification bias $b_{\rm mag}(z)$ for the galaxy clustering and galaxy-galaxy lensing probes respectively with a third-order polynomial expansion,
\begin{align}
    & b_{\rm G}(z) = b_{\rm G,0} + b_{\rm G,1}\,z + b_{\rm G,2}\,z^2 + b_{\rm G,3}\,z^3\;,\\
    & b_{\rm mag}(z) = b_{\rm mag,0} + b_{\rm mag,1}\,z + b_{\rm mag,2}\,z^2 + b_{\rm mag,3}\,z^3\;,
\end{align}
which is multiplied to the matter power spectrum $P_{\rm m}$ following Eqs. (47) and (48) of Paper 1 and integrated to obtain the harmonic space galaxy-galaxy correlation {power} spectrum $C_{ij}^{gg}(\ell)$ (see Eq.\ 37 of Paper 1). 
For the weak lensing probe, {galaxy intrinsic alignment (IA) is} modelled {by} the non-linear alignment (NLA) framework \citep{2007NJPh....9..444B}, including a redshift-dependent intrinsic alignment kernel as a proof of concept. This choice is adopted as a placeholder, with the recognition that more physically motivated and accurate models will be required for future analyses aiming at higher precision. We model the galaxy bias using a low-order polynomial to reflect its expected smooth evolution with redshift, given the simple magnitude-based selection of the sample. This choice ensures consistency with the IA model, which is described using a few parameters. While the polynomial form is not theoretically motivated, it provides a good fit to the measured bias values and offers a practical balance between accuracy and model complexity. Systematic effects such as the shear multiplicative bias due to imperfect shear calibration $m_{\rm L}^i$ and the photometric redshift uncertainty $\Delta z_{\rm L}^i$ are included using specific extra parameters per tomographic bin. Finally, the effect from \rsd is included in the modelling of the galaxy density kernel $W_i^{\rm g}(\ell,z)$ as described in Paper 1.

\vspace{15pt}

For the spectroscopic probe, the final non-linear recipe is based on the recently developed formalism of the \eft; see Paper 1 and Euclid Preparation: Moretti et al. (in prep.), and \citet{IvaSimZal2020} for a detailed description of this model. This framework provides a state-of-the-art description of the clustering of biased tracers in redshift space, accounting for the non-linear evolution of the matter density field, galaxy bias, and \rsd. The complete model includes 11 free parameters per spectroscopic bin, including the linear bias $b^1_{{\rm G},\, i}$, local quadratic bias $b^2_{{\rm G},\, i}$, and non-local quadratic and cubic bias, $b_{{\rm G_2},\, i}$ and $\bGthree$ (Euclid Preparation: Moretti et al., in prep.). The small-scale damping of the clustering signal caused by the \rsd smearing is modelled using a set of \eft counter-terms, which, at leading order, are limited to three extra parameters, $(c_0,\,c_2,\,c_4)$, each one scaling with a different power of $\mu$, \ the angle to the line of sight. These parameters are also meant to absorb the residual contribution from higher-order derivatives and velocity biases, as well as the breakdown of the perturbative approach {at} ultraviolet modes. Leading-order stochastic contributions are included in the \eft model via an extra parameter, $\alpha_{\rm P}^i$, quantifying deviations from the Poisson limit. At next-to-leading order, the model requires the {inclusion} of higher-order parameters: an extra counter-term $c_{\rm nlo}^i$, and two scale-dependent shot-noise parameters $\alpha_{\rm P,2}^i$ and $\alpha_{\rm P,3}^i$.\footnote{In the main analysis of this paper, we reduce the dimensionality of the parameter space by fixing certain parameters based on physically-motivated relations and/or values. Specifically, we constrain the non-local bias parameters using relations derived from the excursion-set formalism and the assumption of conserved tracer evolution (coevolution) as outlined in previous works \citep[see e.g.][]{Eggemeier:2021cam, PhysRevD.104.043531}. Additionally, we do not vary the parameters $c_{\rm nlo}^i$, $\alpha_{\rm P,2}^i$, and $\alpha_{\rm P,3}^i$, keeping them fixed at the fiducial values of the data vectors. The validity of these assumptions will be tested in a dedicated future study (Euclid Preparation: Moretti et al., in prep.).} Finally, we include the impact of the purity of the redshift sample as an additional parameter, $f_{\rm out}^i$, which quantifies the fraction of outliers, e.g. interlopers, contaminating the main H$\alpha$ sample. We assume that the \Euclid science purity requirement will be fulfilled with an accuracy of at least 1\%.

\section{Synthetic data vectors}\label{sec:synthetic_data}
To simulate observations from $\Euclid$, we generate synthetic data in the form of angular power spectra for the photometric probes, and power spectrum Legendre multipoles for the spectroscopic probe. These data products aim to emulate the real data that will be provided by the Euclid Consortium Science Ground Segment \citep{Tessore24, EuclidSkyOverview}. We use the same theoretical specification{s} detailed in \cref{subsec:theoretical_specs} {and} the fiducial values in \cref{tab:fiducial_model} {to produce the synthetic data}. We {generate} noiseless data vectors, meaning no experimental errors were added to the theoretical predictions.

\subsection{Photometric data}

\begin{table} [hbp!]
\caption{Survey specifications to generate the synthetic photometric data.} 
  \centering
  \renewcommand{\arraystretch}{1.3}
    % \rule{0pt}{12pt}
    \begin{tabular}{p{3.75cm}>{\raggedright\arraybackslash}p{3.75cm}}
         %\hline\hline
         \noalign{\vskip 1pt}
         \textbf{Specification} & \textbf{Fiducial value} \\
         \hline\hline
         Survey area & 13\,245 deg$^2$ \\
         \hline
         $f_{\rm{sky}}$ & 0.321 \\
         \hline
         $\sigma_{\epsilon}$ & 0.368\\
         \hline
         Limiting magnitude & 24.5\\
         \hline
         $\bar{z}_i$& \{0.27575, 0.37635, 0.44634, 0.54284, 0.62145, 0.70957, 0.7986, 0.86687, 0.97753, 1.09136, 1.24264, 1.47918, 1.89264\}\\
         \hline
         %\hline
  \end{tabular}
  \label{tab:survey_specs}\\
  \justifying \hspace{-18pt} \footnotesize{Note: The survey area corresponds to the expected area for \drthree. $\sigma_{\epsilon}$ is the variance of the total intrinsic ellipticity dispersion of galaxy sources. The magnitude limit has been given in the optical $i$ band. $\bar{z}_i$ is the mean of the galaxy redshift bin distribution, and $\ell$ are the multipoles for the binned angular power spectra $\cl{\ell}[ij][AB]$.}
\end{table}

\begin{figure}[h!]
\centering
\includegraphics[width=0.49\textwidth]{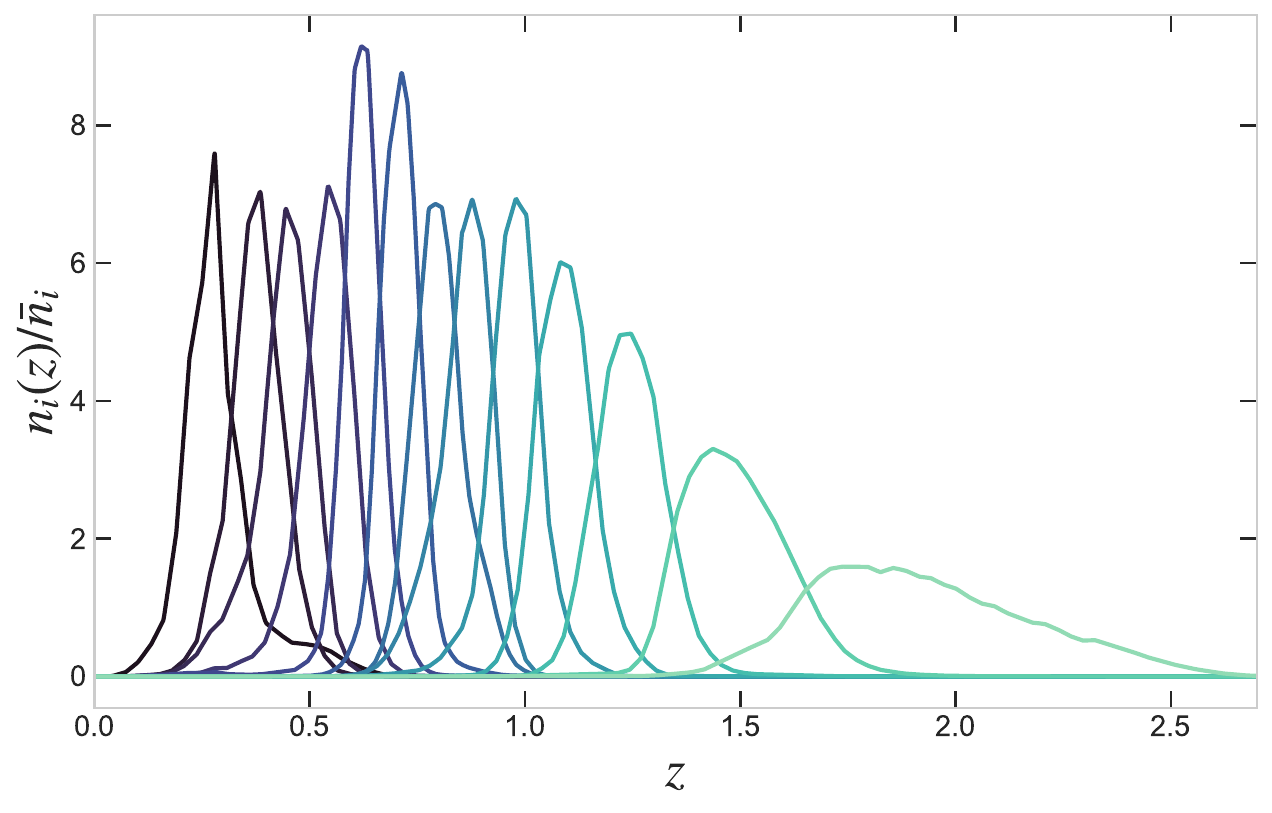}
\caption{Normalised equi-populated photometric galaxy redshift distributions, $n_i(z)/\bar{n}_i$, corresponding to \drthree.}
\label{fig:nz}
\end{figure}

To simulate $\Euclid$ \drthree survey specifications, which correspond to the full Euclid Wide Survey at the end of \Euclid's operations, we assume 13 equi-populated redshift bins in $z\in[0.2, 2.5]$, with the redshift distributions measured from the Flagship 2 simulation. The resulting redshift bin distributions $n_i(z)$ are shown in \cref{fig:nz}, where $\bar n_i=\int\de z\,n_i(z)$.  
Details of the mean redshift of each bin, $\bar{z}_i$, the survey area, shape noise $\sigma_\epsilon$, and limiting galaxy magnitude can be found in \cref{tab:survey_specs}. 

\begin{figure*}[htbp!]
\centering
\includegraphics[width=\textwidth]{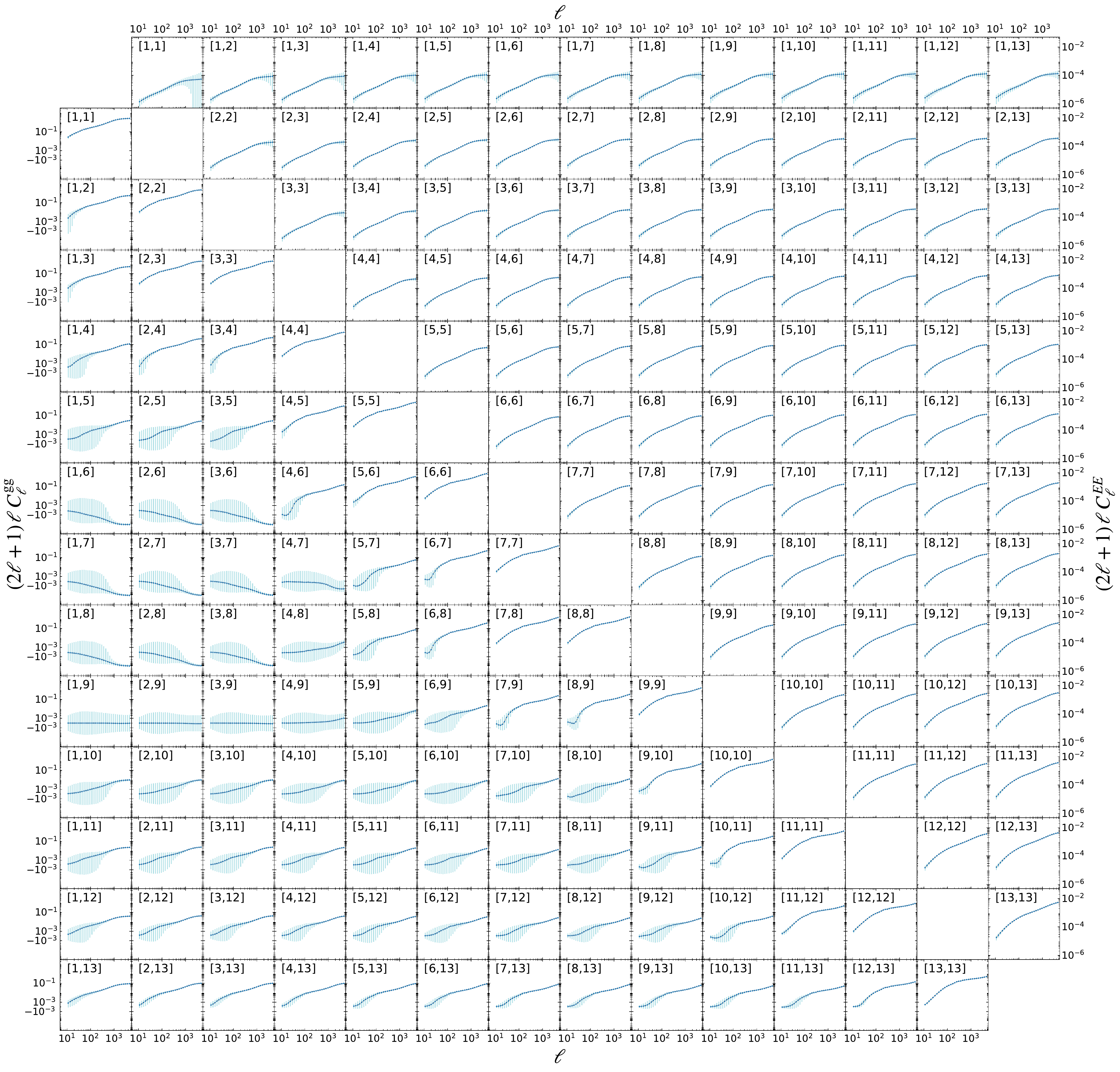}
\caption{Synthetic harmonic-space weak lensing, $\cl{\ell}[ij][EE]$ (upper triangle), and angular clustering power spectra, $\cl{\ell}[ij][gg]$ (lower triangle), for the auto and cross-correlations between the 13 photometric redshift bins. The shaded error bars show the corresponding uncertainty given by the corresponding analytical covariance matrix. We have used the $C_{\ell}$ notation in the axis labels, as here they refer to the discrete measured power spectra at binned ranges of $\ell$'s.} 
\label{fig:CLOE_euclid_probes_WL}
\end{figure*}

We generate synthetic angular power spectra $\cl{\ell}[ij][\rm AB]$ for the weak lensing, photometric galaxy clustering, and galaxy-galaxy lensing probes using \CLOE itself (Paper 2), where the combination of indices $AB$ {denote either} WL, GCph, or XC (corresponding to $EE$, $gg$ and $gE$ respectively in the axis labels of \cref{fig:CLOE_euclid_probes_WL,fig:CLOE_euclid_probes_GCsp,fig:CLOE_euclid_probes_XC}, for {compatibility} with \Euclid Science Ground Segment nomenclature). Here we have assumed the Limber approximation \citep{Limber1953, 1992ApJ...388..272K, LA2008}. In \cref{fig:CLOE_euclid_probes_WL}, we plot the angular weak lensing and galaxy-clustering auto power spectra, while the angular cross-correlation power spectrum XC is shown in \cref{fig:CLOE_euclid_probes_XC}. {In the latter, we present only the $\cl{\ell}[ij][gE]$ spectra for $i\leq j$, although we note that $\cl{\ell}[ij][gE] \neq \cl{\ell}[ji][gE]$ .}

The calculated power spectra are then binned in 32 logarithmically spaced multipole bins in $\ell\in[10,5000]$. The fiducial values for the galaxy- and magnification-bias polynomial coefficients, as well as the per-bin redshift shifts, are also obtained {from the} Flagship 2 simulation. \rsd are included in the production of the synthetic data set.\\

\subsection{Spectroscopic data}
\label{sec:spectro_data}

\begin{figure*}
\centering
\includegraphics[width=1.\textwidth]{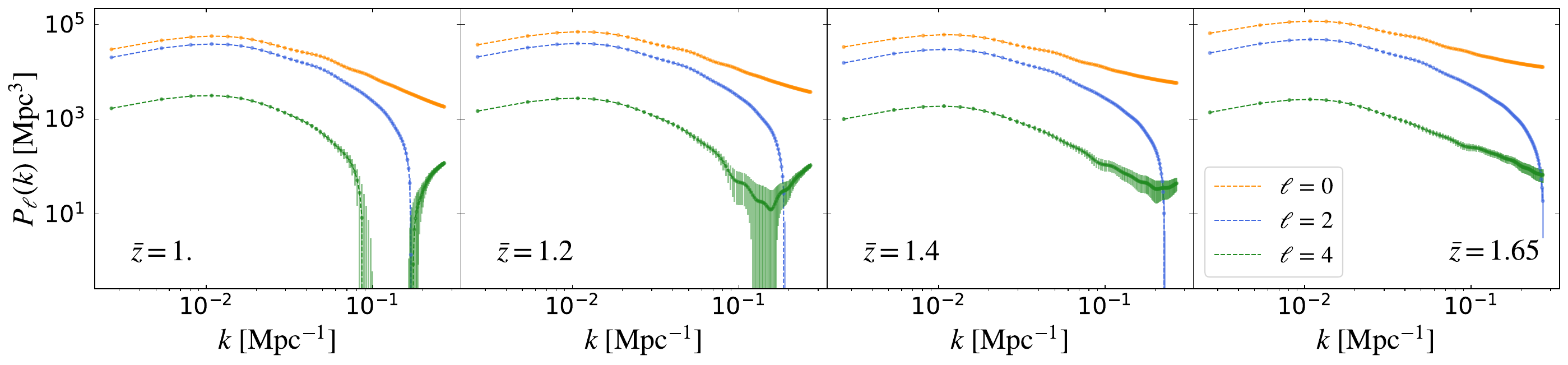}
\caption{Galaxy clustering power spectrum Legendre multipoles, $\pl{k}{\ell}$, as expected from the spectroscopic-survey data within four redshift bins (see values at \cref{tab:GCspectro_bins}). The plots show the monopole ($\ell = 0$, dark red dots), quadrupole ($\ell = 2$, blue dots), and hexadecapole ($\ell = 4$, green dots), together with their error bars as given by the corresponding Gaussian covariance matrix. The Poissonian shot noise has been subtracted from the monopole for clarity of plotting. }
\label{fig:CLOE_euclid_probes_GCsp}
\end{figure*}

As described in \cref{subsec:theoretical_specs}, we generate synthetic data vectors for the power spectrum Legendre multipoles $\pl{k}{\ell}$ adopting the \eft framework. Our reference setup consists of four spectroscopic bins across the redshift range $0.9\le z\le1.8$, and the $\pl{k}{\ell}$ are generated at the mean redshift of each bin, corresponding to the values $\bar z_i=\{1.0,1.2,1.4,1.65\}$. The multipoles are then sampled over 75 linearly spaced $k$ bins from $k_{\rm min}=0.004\,\kMpc$ to $k_{\rm max}=0.3\,\kMpc$. In all cases, we consider only the first three even terms of the multipole expansion, i.e.\ $\ell\in\{0,2,4\}$.\footnote{The presence of non-linear corrections leads to the appearance of higher order multipoles, starting with $\pl{k}{6}[z]$. Traditionally, these corrections are not considered in a likelihood analysis since they are mostly noise-dominated, and do not add any additional constraining power.} 

The reference values of the cosmological parameters and the ones of the \eft expansion in each spectroscopic bin are listed in \cref{tab:fiducial_model}. The latter has been calibrated from a synthetic model for the luminosity function of H$\alpha$ emitters \citep[Model 3 in][]{PozHirGea2016} as implemented in the Flagship Simulation 1, which is a previous en-suite simulation developed before the Flagship Simulation 2 \citep{EuclidSkyFlagship}. We do not include Alcock--Paczynski (AP) parameters in the computation of the spectroscopic synthetic data vectors, as we computed the synthetic data vectors on the same fiducial cosmology. See Paper 1 for the definition of the AP parameters.

We plot the resultant data vectors in \cref{fig:CLOE_euclid_probes_GCsp}. In all cases, the Poissonian shot noise, defined as the inverse of the target number density in the specific redshift bin, $P_{\rm SN}=1/\bar{N}$, has been subtracted from the monopole $\pl{k}{0}$. We do this to minimise the impact of the lower number density of detectable H$\alpha$ galaxies, as shown in the corresponding rows of \cref{tab:GCspectro_bins}. At the same time, the overall amplitude of the monopole is similarly influenced by the excess of non-Poissonian shot noise, which is determined by the parameters $\aP$. Also in this case, high-redshift snapshots exhibit a larger value of this parameter (see \cref{tab:fiducial_model}), hence the larger relative importance of shot noise at small scales.

In terms of observational systematics, we model the impact of catastrophic outliers in the observed spectroscopic H$\alpha$ sample with a scale-independent damping factor to the anisotropic galaxy power spectrum. This recipe assumes that line (S\,[{\sc iii}] and O\,[{\sc iii}]) and noise interlopers do not cluster among themselves and with the underlying H$\alpha$ population. For a more realistic analysis, such as the one that will be realised with \Euclid data, the presence of interlopers will be accounted for either at the level of the power spectrum estimator or in the modelling \citep{EP-Risso}. The reference values for the fraction of H$\alpha$ emitters, line, and noise interlopers, as estimated from dedicated end-to-end simulations, are listed in \cref{tab:GCspectro_bins}.

Finally, the effect of spectroscopic redshift uncertainties manifests itself in the damping of the small-scale galaxy power spectrum, which is caused by the smearing of the galaxy density field along the line of sight. This is included in the synthetic data vectors, assuming a Gaussian damping with $\sigma_z=0.002$.

\section{Covariance matrices}\label{sec:covmat}
In this section, we describe the modelling of the covariance matrices for both photometric and spectroscopic probes. Potential cross-covariances between these observables are not included, as they have been shown to be negligible \citep{2022PhRvD.106f3536T,EP-Paganin}.

\subsection{Covariance for photometric observables}\label{subsec:photo_covmat} 
In this work, we follow an analytical prescription to model the harmonic-space covariance matrix, which is composed of both Gaussian and non-Gaussian contributions. The former corresponds to the covariance of a Gaussian-distributed random field, while the latter arises due to the coupling of the Fourier modes of the density field caused by small scale non-linear evolution. Labelling the modes with wavelengths shorter (larger) than the linear survey size as sub- (super-) survey, we can divide the non-Gaussian covariance into super-sample covariance (SSC; \citealt{Takada2013}), arising from sub- to super-survey mode coupling, and connected non-Gaussian covariance, arising from sub- to sub-survey mode coupling \citep{Scoccimarro99}.

In the present work, we only account for the super-sample covariance term, since the connected non-Gaussian term has been suggested to be subdominant in \cite{Barreira2018}. We have also ignored the off-diagonal terms induced by survey shape and masking, and accounted for survey shape in both the Gaussian and SSC terms by rescaling the covariance matrices by the appropriate $f_{\rm{sky}}$ \citep{Knox:1996cd}. The analytical expression for the Gaussian covariance is given by \citep[see e.g.][]{Blanchard-EP7}
\begin{align}\label{eq: covgauss}
\vspace{-100pt}
{\rm Cov_{\rm G}} & \left[\cl{\ell}[ij][AB], \cl{\ell^\prime}[kl][CD]\right]=
\left[(2\,\ell + 1)\,f_{\rm sky} \, \Delta \ell \right]^{-1}
\delta_{\ell \ell^{\prime}}^{\rm K}  \nonumber \\
 & \times \Bigg\{ \left[\cl{\ell}[ik][AC] + {N}_{ik}^{AC}(\ell)\right]
\left[\cl{\ell^\prime}[jl][BD] + N_{jl}^{BD}(\ell^\prime) \right]  \nonumber \\
& + \left[\cl{\ell}[il][AD] + N_{il}^{AD}(\ell) \right]
\left[\cl{\ell^\prime}[jk][BC] + N_{jk}^{BC}(\ell^\prime) \right] 
\Bigg\}  \; ,
\end{align}
where the noise power spectra $ N_{ij}^{AB}(\ell)$ for the different probe combinations are
\begin{equation}
N_{ij}^{AB}(\ell) = \left \{
\begin{array}{ll}
\displaystyle{(\sigma_{\epsilon}^2/2\bar{n}_{i}^{\rm L}) \, \delta_{ij}^{\rm K}} & \displaystyle{A = B = {\rm L} \;\; ({\rm WL})} \\
 & \\
\displaystyle{0} & \displaystyle{A \neq B} \\
 & \\
\displaystyle{(1/\bar{n}_{i}^{\rm G}) \, \delta_{ij}^{\rm K}} & \displaystyle{A = B = {\rm G} \;\; ({\rm GCph}}) \\
\end{array}
\right . \;.
\label{eq: noiseps}
\end{equation}

In the above equations, the Kronecker delta symbols $\delta^{\rm K}$ enforce the absence of cross-multipole covariance and of cross-bin noise. The term $\sigma_{\epsilon}^2$ accounts for the total intrinsic ellipticity dispersion of the sources, with $\sigma_{\epsilon} = \sqrt{2}\,\sigma_\epsilon^{(i)}$, $\sigma_\epsilon^{(i)}$ being the ellipticity dispersion per component of the galaxy ellipticity. Finally, $\bar{n}_i^A$ is the average density of objects for weak lensing (L) and photometric galaxy clustering (G) in the $i$th redshift bin, while $\Delta \ell$ is the width of the multipole bin centred on a given $\ell$. This expression accounts for the finite survey volume via a rescaling by the fraction of the total sky area covered by the survey, $f_{\rm sky}$, which is expected to be sufficiently accurate for large survey areas.

As for SSC, we follow the modelling of \citet{Takada2013}, adapted to the multi-probe case in \citet{Krause2017} and used to forecast the SSC impact in \citet{EP-Sciotti},
\begin{multline}\label{eq: covSSCintermediate}
   {\rm Cov_{SSC}}\left[\cl{\ell}[ij][AB],\cl{\ell^\prime}[kl][CD]\right]\simeq \frac{1}{f_{\rm sky}}
   \int \diff V_1 \diff V_2 \; W^A_i(z_1)\,W^B_j(z_1)\, \\
   \times W^C_k(z_2)\, W^D_l(z_2) 
   \frac{\partial \pl{k_\ell}{AB}[z_1]}{\partial \delta_{\rm b}}\,
   \frac{\partial \pl{k_\ell^\prime}{CD}[z_2]}{\partial \delta_{\rm b}}\,
   \sigma^2(z_1, z_2) \; .
\end{multline}
In the expression above, the two main ingredients needed for SSC appear. The former, called probe response, is the derivative of the cross-spectrum $P_{AB}$ between probes $A$ and $B$ with respect to a change in the background density $\delta_{\rm b}$ induced by super-survey modes. The latter is the covariance of $\delta_{\rm b}$, given by the linear power spectrum as these modes are in the linear regime, i.e.\
\begin{equation}\label{eq: sigma}
   \sigma^2(z_{1}, z_{2}) = \frac{1}{2 \pi^{2}} \int \diff k \; k^{2} \,P_{\rm mm}^{\, \rm lin}\left(k , z_{12}\right)\, {\rm j}_{0}\left(k r_{1}\right)\, {\rm j}_{0}\left(k r_{2}\right) \; .
\end{equation}
For this term, we follow the modelling of \cite
{LacasaRosenfeld2016}, which does not assume $\sigma^2(z_1, z_2 \neq z_1)=0$ as commonly done in the literature to speed up the SSC computation; to this end, we employ a new code, \texttt{Spaceborne} (Euclid Preparation: Sciotti et al., in prep.). The above expression is valid for the full, curved-sky case, and the partial sky coverage is accounted for via a normalisation by $f_{\rm sky}$, which has been shown in \cite{beauchamps2021} to be accurate for large survey areas. Further details on the harmonic-space covariance matrix modelling and its numerical implementation using \texttt{Spaceborne} are given in Euclid Preparation: Sciotti et al. (in prep.).\\

\subsection{Covariance for spectroscopic observables}\label{subsec:spectro_covmat}

\begin{table}
\caption{Specifications for the analytical covariance matrices.}
\centering
\small
\renewcommand{\arraystretch}{1.15}
\setlength{\tabcolsep}{3pt}
\begin{tabular}{lcccc}
\textbf{Parameter} & \textbf{bin 1} & \textbf{bin 2} & \textbf{bin 3} & \textbf{bin 4} \\
\hline \hline
$\bar z$ & 1.0 & 1.2 & 1.4 & 1.65 \\
range & [0.9,1.1] & [1.1,1.3] & [1.3,1.5] & [1.5,1.8] \\
$V_{\rm shell}$ [Gpc$^3$] & 7.03 & 8.10 & 8.90 & 14.36 \\
$f^{\rm P}_{\rm H\alpha}$ & 0.805 & 0.796 & 0.694 & 0.879 \\
$f^{\rm P}_{\rm SIII}$ & 0.021 & 0.028 & 0.101 & 0.049 \\
$f^{\rm P}_{\rm OIII}$ & 0.029 & 0.109 & 0.118 & 0.020 \\
$f^{\rm P}_{\rm noise}$ & 0.145 & 0.067 & 0.087 & 0.052 \\
$f^{\rm C}_{\rm H\alpha}$ & 0.387 & 0.638 & 0.724 & 0.997 \\
$f^{\rm C}_{\rm SIII}$ & 0.010 & 0.022 & 0.105 & 0.056 \\
$f^{\rm C}_{\rm OIII}$ & 0.014 & 0.088 & 0.123 & 0.023 \\
$f^{\rm C}_{\rm noise}$ & 0.070 & 0.054 & 0.091 & 0.059 \\
$\bar n_{\rm true}$ [$h^3$Mpc$^{-3}$] & $14.28\!\times\!10^{-4}$ & $9.49\!\times\!10^{-4}$ & $6.44\!\times\!10^{-4}$ & $4.03\!\times\!10^{-4}$ \\
$\bar n_{\rm meas}$ [$h^3$Mpc$^{-3}$] & $6.87\!\times\!10^{-4}$ & $7.61\!\times\!10^{-4}$ & $6.71\!\times\!10^{-4}$ & $4.57\!\times\!10^{-4}$ \\
\hline
\end{tabular}
\label{tab:GCspectro_bins}
\end{table}

For spectroscopic galaxy clustering, as mentioned in \cref{sec:spectro_data}, we assume four redshift bins with the same geometric specifications adopted in \cite{Blanchard-EP7}. This corresponds to the nominal final-mission angular footprint of $13\,245\,\sqdeg$ and four spectroscopic bins spanning $z\in[0.9,1.8]$, as shown in \cref{tab:GCspectro_bins}. We determine the covariance matrix for the power spectrum multipoles within the Gaussian approximation, following the procedure detailed in \cite{GriSanSal2016}. The per-mode covariance of each multipole combination $\ell=\{0,2,4\}$ can be defined as
\be
    \begin{split}
        \sigma^2_{\ell_1 \ell_2}(k\,|\,z) = \;& \frac{(2\ell_1+1)\,(2\ell_2+1)}{V_{\rm shell}} \\
        & \times \int_{-1}^{\,1}\left[P_{\rm gg}(k,\mu\,|\,z)+\frac{1}{\bar{n}}\right]{\cal L}_{\ell_1}(\mu)\,{\cal L}_{\ell_2}(\mu)\,\diff\mu\;,
    \end{split}
    \label{eq:sigma2}
\ee
where $V_{\rm shell}$ and $\bar{n}$ are the volume and number density of the spectroscopic sample under consideration, and ${\cal L}_{\ell}$ is the $\ell$th-order Legendre polynomial. The bin-averaged multipole covariance can be derived from the previous expression as
\be
    C_{\ell_1\ell_2}(k_i,k_j)=\frac{2\,(2\pi)^4}{V_{k_i}^{\,2}}\,\delta^{\rm K}_{ij}\int_{k_i-\Delta k / 2}^{\,k_i+\Delta k/2}\sigma^2_{\ell_1\ell_2}(k)\,k^2\,\diff k\;,
\ee
where $V_{k_i}=4\pi/3\,\{[(k_i+\Delta k/2)]^3-[(k_i-\Delta k/2)]^3\}$ is the volume of a spherical shell centred at $k_i$ of width $\Delta k$.

The Kronecker symbol marks the diagonal nature of the covariance matrix, under the Gaussian approximation. While this assumption is bound to break on sufficiently small scales, recent analyses \citep[e.g.,][]{BloCroSef2019, WadIvaSco2020} have proven how the statistical constraint on cosmological parameters is only marginally affected ($\lesssim 10\%$) by the presence of non-linear corrections to the multipole covariance matrix \citep{Scoccimarro99, SefCroPue2006, BloCorAli2015, BloCorAme2016, BerSchSol2016, WadSco2020}. We adopt this approach for deriving these forecasts, whereas a more complete model shall be adopted for future analyses, such as the one of the DR1 data.

We account for the impact of observational systematic effects on the covariance matrix by adopting the same fractions of true H$\alpha$ (with a good quality redshift or not) line, and noise interlopers as calibrated with the FastSpec simulator \citep{Cagliari24} using a Euclid Wide Survey configuration.

The completeness of the sample is defined as the fraction of correctly targeted H$\alpha$ galaxies to the total H$\alpha$ population, while the purity of the sample is defined as the fraction of H$\alpha$ galaxies with correct redshift to the total observed sample.\footnote{All references to H$\alpha$ fluxes are flux measurements of the unresolved H$\alpha$+N\,[{\sc iii}] lines. In addition, the selection criteria always include a flux limit of $f_{{\rm H}\alpha}>2\times10^{-16}\,{\rm erg \,s^{-1}\, cm^{-2}}$, a sharp cut in redshift to select objects between $0.9 \le z \le 1.8$, and additional selections based on the quality of the observed spectra.}  In turn, the latter is expected to contain S\,[{\sc iii}] and O\,[{\sc iii}] contaminants due to line mis-identification, and noise interlopers due to the presence of catastrophic redshift errors. These two falsely detected populations modify the total clustering amplitude such that the total observed number density $n_{\rm obs}(z)$ and clustering amplitude $P_{\rm obs}(k,\mu)$ differs from the true quantities $n_{\rm true}(z)$ and $P_{\rm true}(k,\mu)$, as
\be
    n_{\rm obs}(z)=n_{\rm true}(z)\,\frac{1-f_{\rm inc}}{1-f_{\rm out}}\;,
\ee
and
\be
    P_{\rm obs}(k,\mu) = (1-f_{\rm out})^2\,P_{\rm true}(k,\mu)\;.
\ee
In the above equations, $f_{\rm inc}$ and $f_{\rm out}$ represent the fraction of undetected and falsely detected H$\alpha$ galaxies respectively. The observed number density and galaxy power spectrum are the quantities that are ultimately used in \cref{eq:sigma2} to obtain the per-mode covariance of the power spectrum multipoles.\\

\section{Forecast results} \label{sec:results} 

\begin{table}[h!]
\caption{\fom for the dark energy parameters \wo and \wa for different cosmological models and \Euclid probes.}
\centering
\small
\renewcommand{\arraystretch}{1.2}
\setlength{\tabcolsep}{4pt}
\begin{tabular}{lcccc}
\textbf{Cosmological model} & \textbf{WL} & \textbf{GCsp} & \textbf{3$\times$2pt} & \textbf{3$\times$2pt+GCsp} \\
\hline\hline
\wowaCDM (flat) & 21 & 35 & 380 & 500 \\
\wowaCDM (non-flat) & 11 & 13 & 186 & 331 \\
\wowaCDM + \gammag\ (flat) & 9 & 24 & 243 & 327 \\
\hline
\end{tabular}
\label{tab:FoM}
\footnotesize\\[2pt]
\justify
Note: The \wo--\wa posterior contours for WL, GCsp, 3$\times$2pt, and 3$\times$2pt+GCsp are shown in Figs.\ \ref{fig:2D_WL}, \ref{fig:2D_GCsp}, \ref{fig:2D_3x2pt}, and \ref{fig:2D_3x2pt_GCsp}.
\end{table}

\begin{table*}[h!]
    \centering
    \caption{68\% confidence limits for the cosmological nuisance parameters in the \wowaCDM model, assuming a flat geometry.}
    \renewcommand{\arraystretch}{1.2}
\begin{tabular} { l  c c c c}
%\noalign{\vskip 3pt}\hline\noalign{\vskip 1.5pt}\hline\noalign{\vskip 5pt}
 \multicolumn{1}{c}{\bf } &  \multicolumn{1}{c}{\bf WL} &  \multicolumn{1}{c}{\bf GCsp} &  \multicolumn{1}{c}{\bf 3\texttimes2pt} &  \multicolumn{1}{c}{\bf 3\texttimes2pt + GCsp}\\
%\noalign{\vskip 3pt}\cline{1-5}\noalign{\vskip 3pt}
\hline\hline
% Parameter &  68\% limits &  68\% limits &  68\% limits &  68\% limits\\
%\hline
{$\ln{(10^{10}\,A_{\rm s})}$} & $3.02^{+0.11}_{-0.13}      $ & $3.007\pm 0.061            $ & $3.039\pm 0.020            $ & $3.041\pm 0.013            $\\

{$n_{\rm s}      $} & $0.957^{+0.044}_{-0.059}   $ & $0.960^{+0.018}_{-0.016}   $ & $0.9656^{+0.0069}_{-0.0061}$ & $0.9662\pm 0.0037          $\\

{$\Omega_{{\rm b}}\,h^2$} & $0.02269\pm 0.00036        $ & $0.02269\pm 0.00032        $ & $0.02271\pm 0.00027        $ & $0.02268\pm 0.00019        $\\

{$\Omega_{{\rm c}}\,h^2$} & $0.125^{+0.014}_{-0.016}   $ & $0.1230^{+0.0028}_{-0.0032}$ & $0.1221^{+0.0019}_{-0.0022}$ & $0.12181\pm 0.00097        $\\

{$w_0            $} & $-0.98\pm 0.20             $ & $-0.92^{+0.19}_{-0.25}     $ & $-1.003\pm 0.051           $ & $-1.004\pm 0.046           $\\

{$w_a            $} & $-0.10^{+0.68}_{-0.47}     $ & $-0.28^{+0.74}_{-0.50}     $ & $0.01\pm 0.16              $ & $0.01\pm 0.13              $\\

{$S_8            $} & $0.842\pm 0.010            $ & $0.838\pm 0.019            $ & $0.8421\pm 0.0026          $ & $0.8419\pm 0.0024          $\\

{$\Omega_{\rm b} $} & $0.0495^{+0.0050}_{-0.0070}$ & $0.0507^{+0.0033}_{-0.0039}$ & $0.0499\pm 0.0011          $ & $0.04993\pm 0.00080        $\\

{$\Omega_{\rm m} $} & $0.321^{+0.018}_{-0.021}   $ & $0.327^{+0.021}_{-0.028}   $ & $0.3196\pm 0.0045          $ & $0.3195\pm 0.0043          $\\

{$\sigma_8       $} & $0.815^{+0.019}_{-0.017}   $ & $0.804^{+0.028}_{-0.025}   $ & $0.8158\pm 0.0039          $ & $0.8159\pm 0.0036          $\\

{$h$} & $0.681\pm 0.040            $ & $0.671\pm 0.024            $ & $0.6746\pm 0.0073          $ & $0.6740\pm 0.0050          $\\
\hline
\end{tabular}
    \label{tab:CL_cosmo}\\
\justifying \hspace{-14pt}\footnotesize{Note: The table shows both constraints on sampled and derived parameters.}
\end{table*}

\corr{
\begin{figure}[h!]
    \centering
    \includegraphics[width=1.\linewidth]{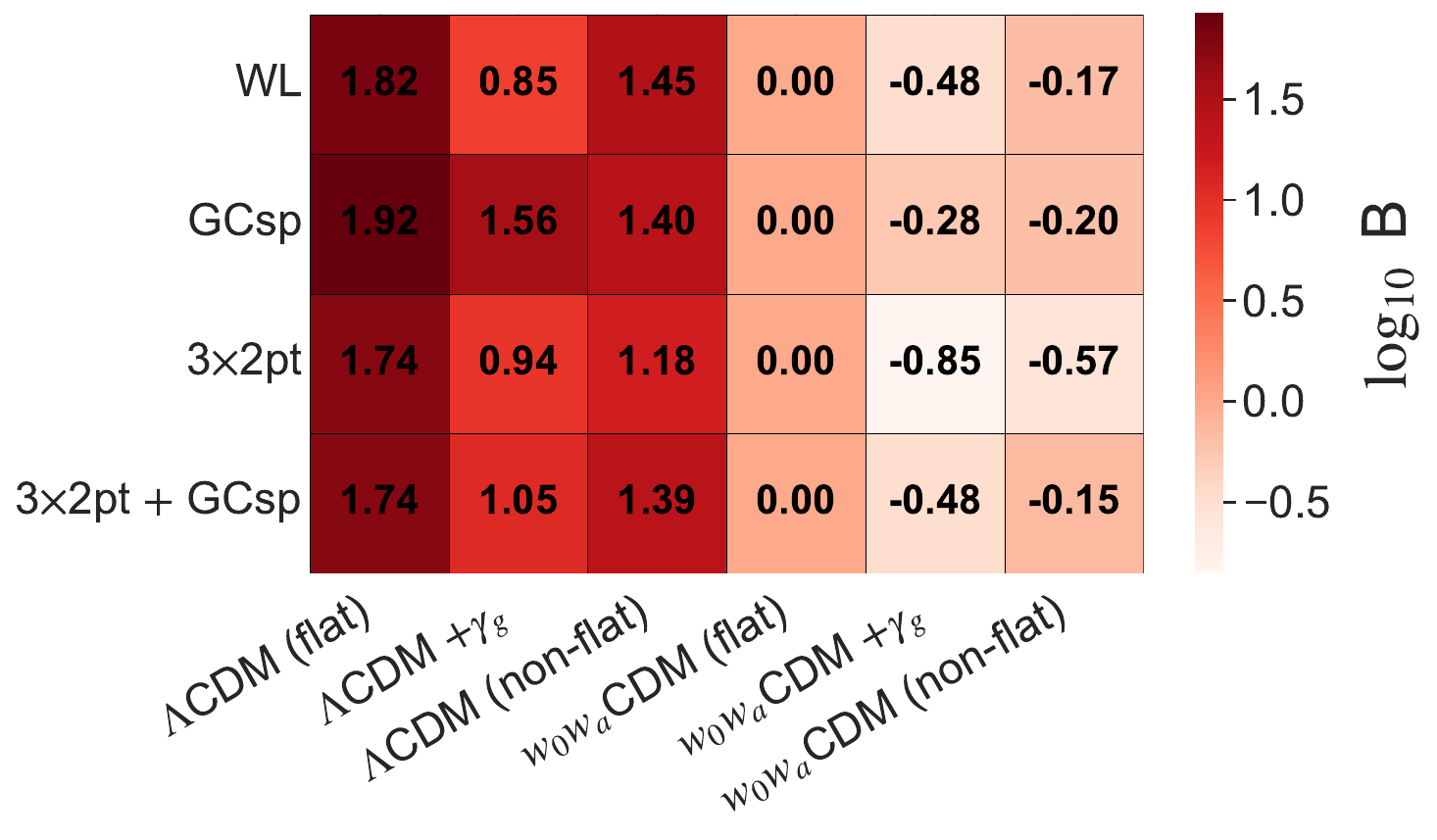}
    \caption{\color{black}{Heat map of the logarithm of the Bayes factors ($\log_{10} B$) for different cosmological models relative to \wowaCDM (flat), across various probes: weak lensing (WL), galaxy clustering (GCsp), 3$\times$2pt, and their combination. Darker red shades indicate models strongly disfavoured by the data relative to the reference, while lighter shades indicate models more compatible with the data. The Bayes factor quantifies the relative evidence for each model, automatically penalising unnecessary complexity. Overall, the data generally favour simpler \LCDM models over extensions such as \wowaCDM or models including $\gamma_g$ or curvature. $\ln \evid$ values are found in \cref{tab:computational_resources}}.
}
    \label{fig:bayesfactor}
\end{figure}}

We present now the main forecast results for all of \Euclid's primary probes (WL, 3\texttimes2pt, GCsp, and 3\texttimes2pt + GCsp) across six different cosmological models, assuming \Euclid specifications for \drthree \citep{EuclidSkyOverview}, which corresponds to the nominal mission duration. The parameter space explored in this analysis, accounting for 13 tomographic bins for the photometric probes and 4 spectroscopic bins for GCsp, along with the inclusion of additional systematic nuisance parameters for each bin, ranges from 21 to 61 sampled parameters. The prior used for each cosmological and nuisance parameter is listed in \cref{tab:fiducial_model}. In total, we analysed 24 different cases. For each case, we also compute derived parameters from sampled parameters such as $\sigma_8$ and $S_8$ using \texttt{CAMB} or \cref{eq:S8}, respectively, as well as the nuisance parameters specified in \cref{tab:fiducial_model}. 
Details of each run, including the size of the parameter space, allocated computational resources, and computational time, are listed in \cref{app:computational_resources}, specifically in \cref{tab:computational_resources}.

Recent studies in the LSS community have emphasised the impact of prior volume effects on the statistical inference of cosmological parameters \citep[see, e.g.][for applications to spectroscopic and photometric probes, respectively]{CarMorPor2023, HadWolAlo2023}. Specifically, the priors imposed on certain nuisance parameters -- typically marginalised over -- can influence the marginal posterior distributions of cosmological parameters, making these priors effectively informative. This is particularly relevant for the \rsd counter-terms within the \eft framework.

To address this, we apply external information from \bbn\ in the form of a Gaussian prior on the baryon density parameter $\Ob\,h^2$ \citep{Cooke18,Schoneberg:2022ggi,Schoneberg:2024ifp} in all configurations involving GCsp. For consistency, this prior is used across all probes in the reference setup for the exploration of all the models in \cref{sec:results}. Furthermore, we marginalise only over a limited set of nuisance parameters: $b_{\rm G}^1$, $b_{\rm G}^2$, $\aP$, and $f_{\rm out}$. We fixed the two non-local bias parameters, $b_{{\rm G}_2}$ and $\bGthree$, according to coevolution relations \citep{EP-Pezzotta}. For the remaining nuisance parameters, we adopt an optimistic setup by fixing them to the fiducial values used to generate the data vectors (see third column from left of \cref{tab:fiducial_model}). While this approach may seem overly optimistic, we argue that with \drthree, some \eft counter-terms could be constrained robustly using realistic cosmological simulations exploring diverse galaxy population models. A more comprehensive exploration of prior volume effects, incorporating all \eft nuisance parameters, will be detailed in future work (Euclid Preparation: Moretti et al., in prep.)).

Given the staggering quantity of results generated, we focus on presenting only the key highlights in this section to ensure clarity and conciseness. This approach allows us to streamline the discussion while still providing access to the full breadth of information for in-depth analysis. In all posterior distribution contour plots, the fiducial values are indicated by dashed grey lines. 
As part of our validation process, we conducted a series of sanity checks to ensure the robustness of our results. The samplers were carefully tested to confirm that they correctly recover the maximum of the likelihood at the fiducial model values, which were used for the generation of the synthetic data vectors. Additionally, we verified that the posterior distribution is properly sampled by employing a sufficiently large number of live points (see \cref{sec:methodology} for details), and ensured that the parameter priors are sufficiently broad to avoid hitting the boundaries during the sampling process. To reach convergence, a total number from $\mathcal{O}(10^5)$ up to $\mathcal{O}(10^7)$ likelihood evaluations is requested, for single probes (WL, GCsp, 3\texttimes2pt) and joint ones (3\texttimes2pt + GCsp), respectively.

For all the cases in which we explore an evolving dark energy EoS (flat and non-flat \wowaCDM, and flat \wowaCDM+ \gammag), we apply a logical prior on the \wo$-$\wa parameter space to avoid exploring a dark energy fluid that would not lead to an accelerated expansion in the post-inflationary epoch ($a\ll 1$), also known as the strong energy condition \citep{Chevallier:2000qy, PhysRevLett.90.091301},
\begin{equation}
\label{eq:strong_condition} 
w_0 + w_a \leq -1/3\;. 
\end{equation}
This choice also enables more efficient sampling of the parameter space, as it is imposed in the analysis through an external likelihood. The final goal of the analysis is then to calculate the dark energy \fom for the various configurations that we explore. The obtained values are presented in \cref{tab:FoM}. When comparing our results to those of \citet{Blanchard-EP7}, we note that (1) the theoretical predictions in this paper are based on a more complex implementation than those used in \citet[see Paper 1 for details]{Blanchard-EP7}, and (2) we sample the full posterior distribution of the parameters of interest (Eq.~\ref{eq:fisher-general}), rather than using the Fisher formalism for forecasting as outlined in \cref{eq:likelihood_gaussian}. The statistical best-fits and 68\% confidence intervals for both sampled and derived cosmological parameters for the \wowaCDM model, following a flat geometry, are found in \cref{tab:CL_cosmo}. 

\corr{Finally, benefiting from the nested sampling algorithm, we quote the Bayes factor values using $w_0 w_a$CDM (flat) as the reference model. The Bayes factor, defined as
\begin{equation}
B_{i,\rm ref} = \frac{\mathcal{Z}_i}{\mathcal{Z}_{\rm ref}} = \frac{P(D \mid M_i)}{P(D \mid M_{\rm ref})} = \exp[\Delta\!\ln \evid],
\end{equation}
quantifies the relative evidence for model $M_i$ compared to the reference model $M_{\rm ref}$, taking into account both the fit to the data $d$ and the model complexity (i.e. number of parameters). In terms of Bayes factors, a difference of $\Delta\!\ln \evid \sim 2.0$ corresponds to a Bayes factor of $B \sim \mathrm{e}^2 \approx 7$, which is generally considered worth mentioning. A stronger difference of $\Delta\!\ln \evid \sim 5.0$ translates into $B \sim \mathrm{e}^5 \approx 150$, which falls into the decisive category according to Jeffreys' scale. Across all probes -- WL, GCsp, 3$\times$2pt, and their combination -- the results consistently favour simpler \LCDM models. Models including additional parameters, such as curvature or the growth modification $\gamma_{\rm g}$, are generally penalised, with Bayes factors below unity, indicating that the data do not require these extensions. This hierarchical comparison highlights the constraining power of each probe individually and in combination, emphasising that while some extended models remain marginally allowed, the evidence strongly prefers the minimal \LCDM framework (see \cref{fig:bayesfactor}).}

\subsection{Weak lensing}\label{sec:results_WL}
\begin{figure*}[h!]
    \centering
    \begin{minipage}{0.5\textwidth}
        \centering
        \includegraphics[width=1.0\textwidth]{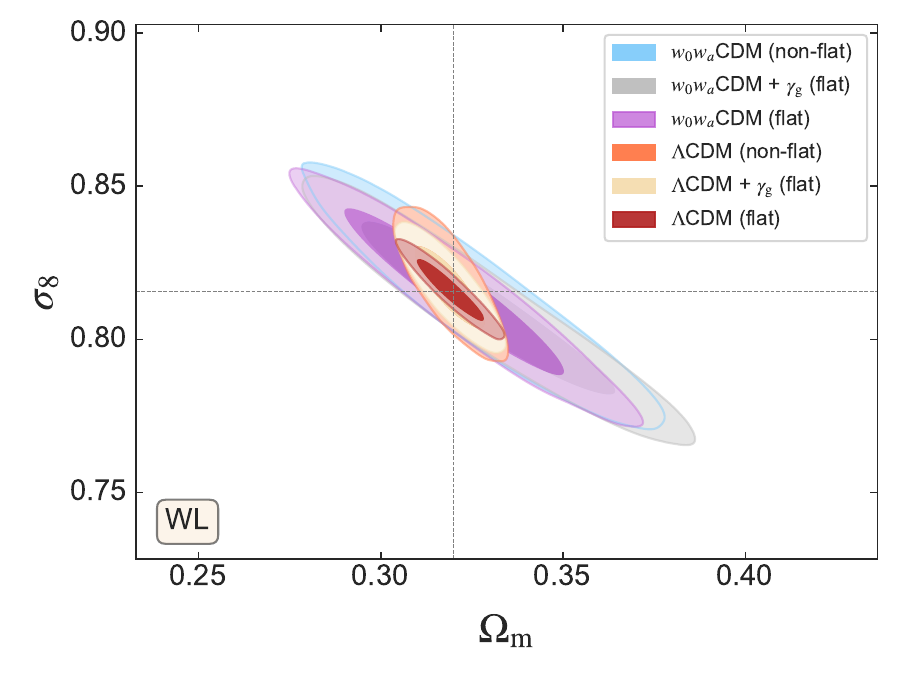} % first figure itself
        
    \end{minipage}\hfill
    \begin{minipage}{0.5\textwidth}
        \centering
        \includegraphics[width=1.0\textwidth]{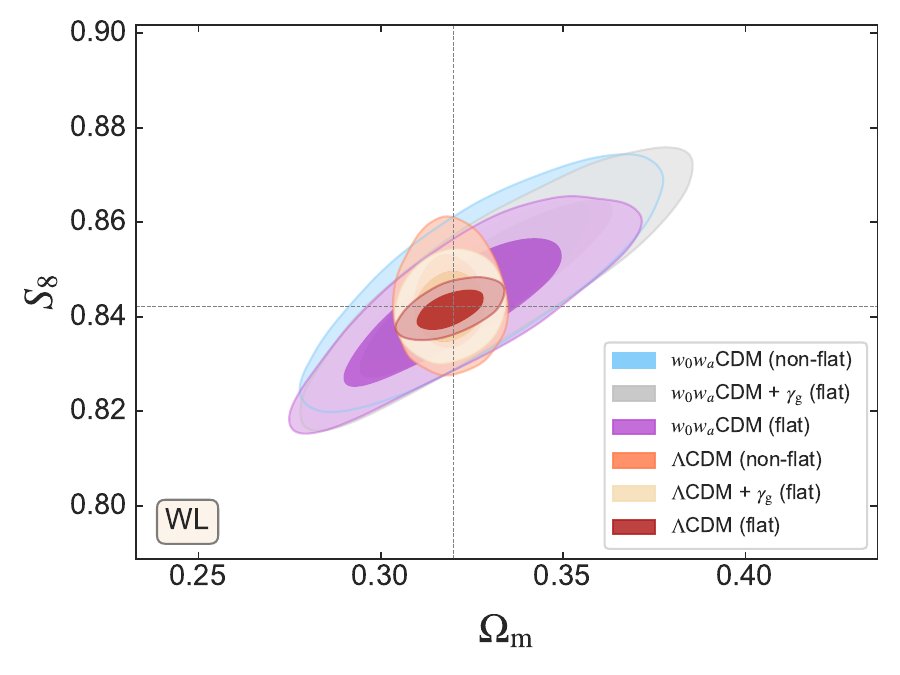} 
 
    \end{minipage}
        \begin{minipage}{0.5\textwidth}
        \centering
        \includegraphics[width=1.0\textwidth]{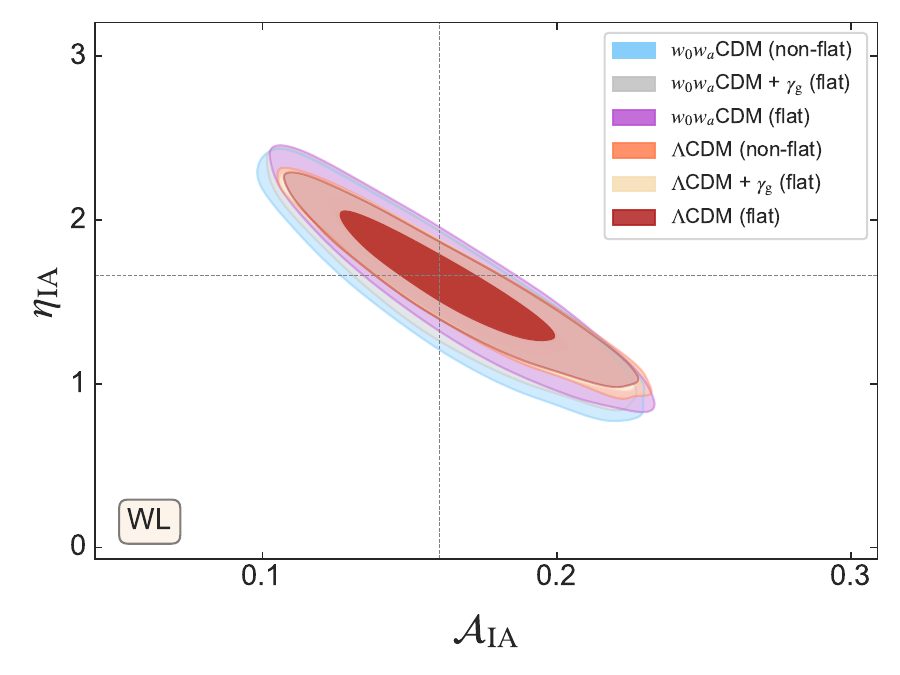} % first figure itself
        
    \end{minipage}\hfill
    \begin{minipage}{0.5\textwidth}
        \centering
        \includegraphics[width=1.0\textwidth]{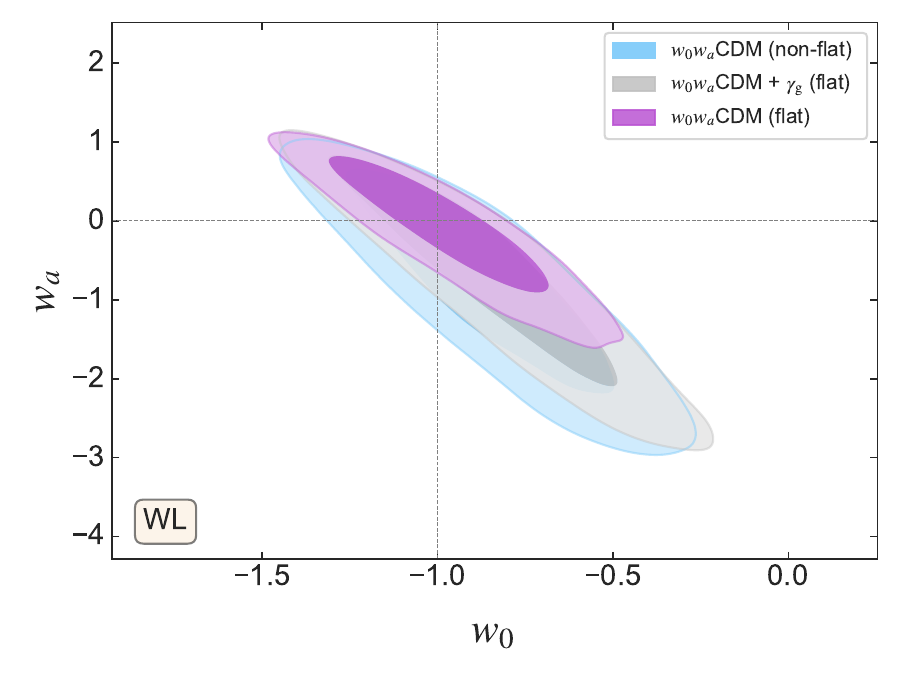} 
        
    \end{minipage}
    \caption{Two-dimensional posterior distributions for the cosmological parameters \Om, \sotto and $S_8$, the eNLA Intrinsic Alignment parameters \aIA and \etaIA, and the dark energy parameters \wo and \wa, using only the weak lensing probe (WL) for the six cosmological models. The values for the \fom of \wo and \wa can be found in \autoref{tab:FoM}.}
    \label{fig:2D_WL}
\end{figure*}
\begin{figure}[h!]
\centering
\includegraphics[width=0.48\textwidth]{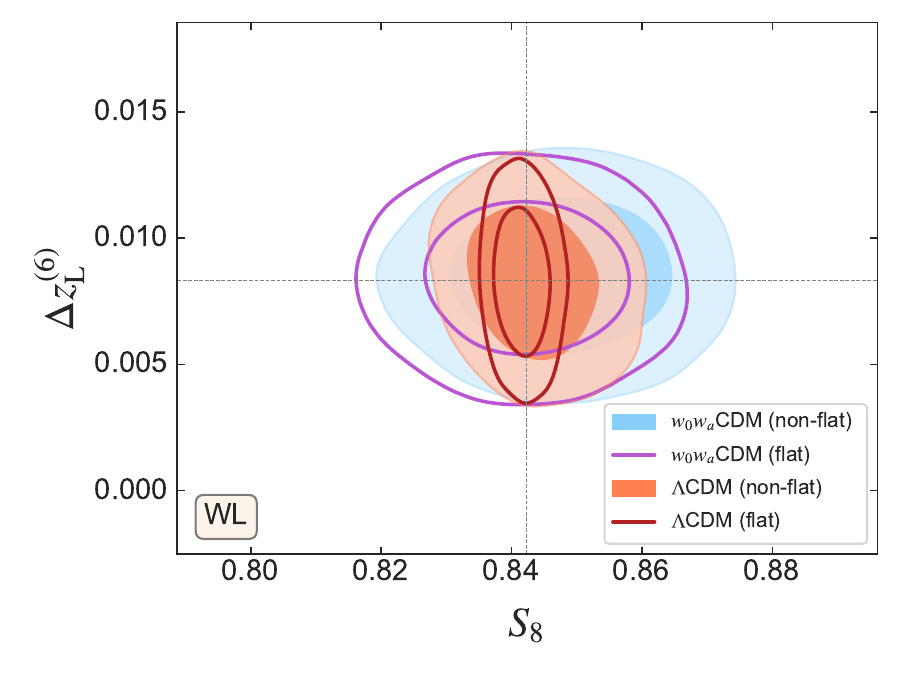}
\caption{Two-dimensional posterior distributions for the cosmological parameter $S_8$ and the per-bin redshift shift nuisance parameter $\Delta z_{\rm L}^{(6)}$, using only the Weak Lensing probe for four different cosmological models. The correlation between \( S_8 \) and \( \Delta z_{\rm L}^{(6)} \) is about \(-0.1\) in \LCDM, while it is close to zero in the \wowaCDM model. Similar behaviour is found for $\Delta z_{\rm L}^{(i)}$ vs. $S_8$ regardless of the redshift bin $i$.}
\label{fig:WL_S8_deltaZ_deg}
\end{figure}
Cosmic shear is a powerful probe for constraining the combination of the total matter density, \Om, and the amplitude of matter fluctuations, $\sigma_8$. However, due to the strong degeneracy between these two parameters in cosmic shear measurements, weak lensing is particularly sensitive to their combination, $S_8$ (as defined in Eq.~\ref{eq:S8}), which is better constrained and more robustly measured in this context. In \cref{fig:2D_WL}, we present the two-dimensional sampled posterior distributions for these parameters, as well as for the dark energy parameters, \wo and \wa, and the intrinsic alignment parameters, \aIA and \etaIA.\footnote{See Eq. (116) of Paper 1 for the definition of the intrinsic alignment model used in this work.} Beyond \LCDM models, these posteriors exhibit non-Gaussian behaviour and mild prior volume effects (see \cref{subsubsec:3x2pt_projection_effects} for detailed examples of projection effects in photometric probes). In these two-dimensional posteriors, we observe a subtle rotation in the correlation between $\sigma_8$ and \Om, as well as between $S_8$ and 
\Om, for the non-flat \LCDM and \LCDM + $\gamma_{\rm g}$ models compared to the others. This rotation arises from the additional redshift freedom introduced by the parameters \wo and \wa, which induce parameter correlations between $S_8$ and the per-bin redshift shifts, $\Delta z_{\rm L}^i$ (see \cref{fig:WL_S8_deltaZ_deg}). As expected, the spectral index \ns cannot be constrained using the cosmic shear probe alone, since the recovered posterior distribution remains uninformative and essentially mirrors the uniform prior assumed for this parameter. Additionally, this probe fails to constrain non-flat cosmologies, since the marginalised posterior distribution of \OK is mostly prior-dependent. As expected, we do not observe the usual degeneracies between the Hubble parameter $H_0$ and the baryon density \Ob \citep[see][for instance]{DESY3shear} due to the inclusion of the \bbn prior in the analysis.
\begin{figure*}[htp!]
    \centering
    \begin{minipage}{0.5\textwidth}
        \centering
        \includegraphics[width=1.0\textwidth]{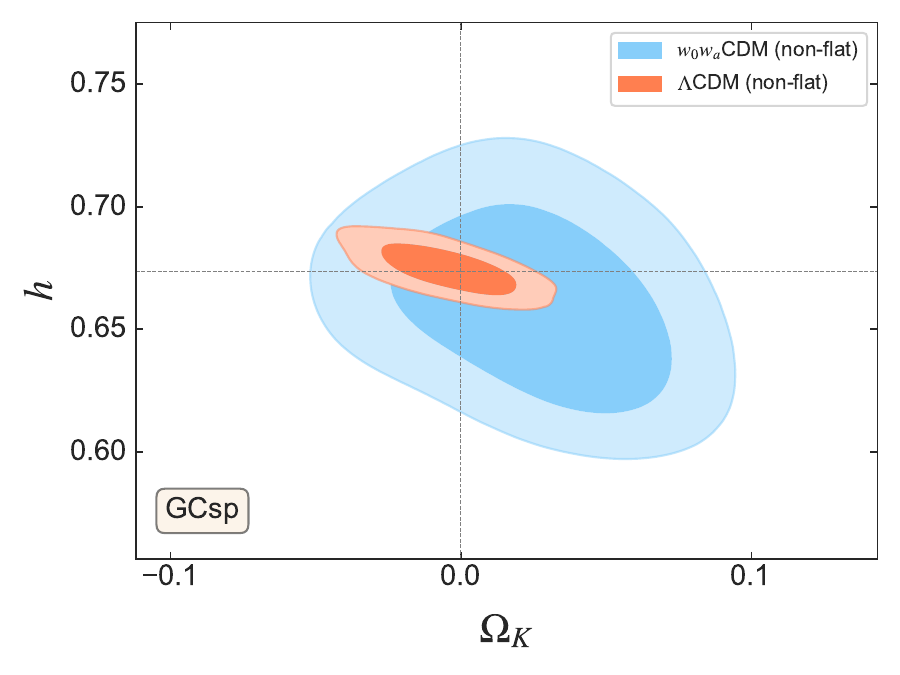} % first figure itself
        
    \end{minipage}\hfill
    \begin{minipage}{0.5\textwidth}
        \centering
        \includegraphics[width=1.0\textwidth]{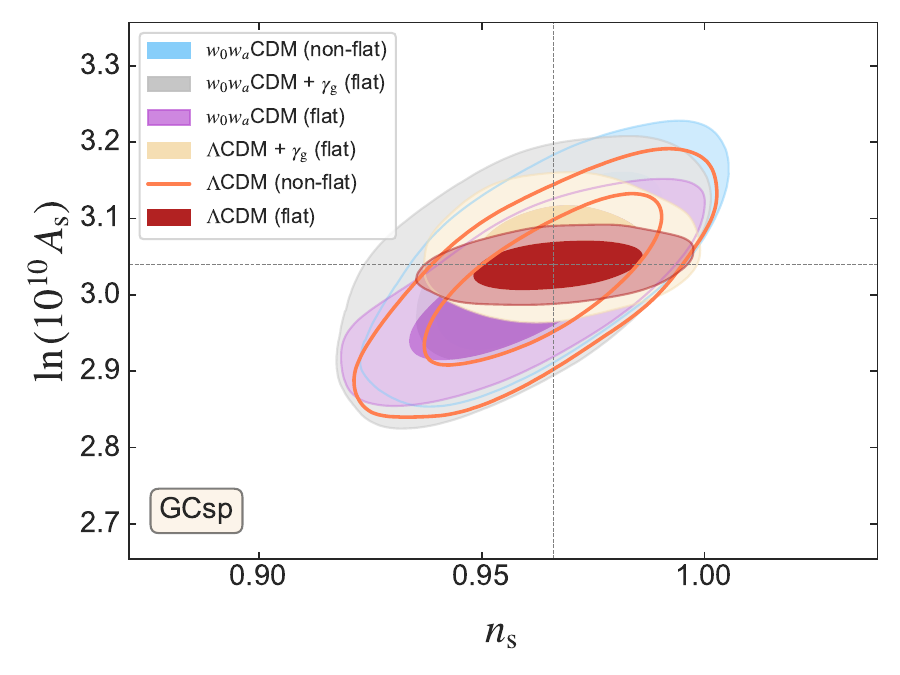} 
    \end{minipage}\hfill
    
        \begin{minipage}{0.5\textwidth}
        \centering
        \includegraphics[width=1.0\textwidth]{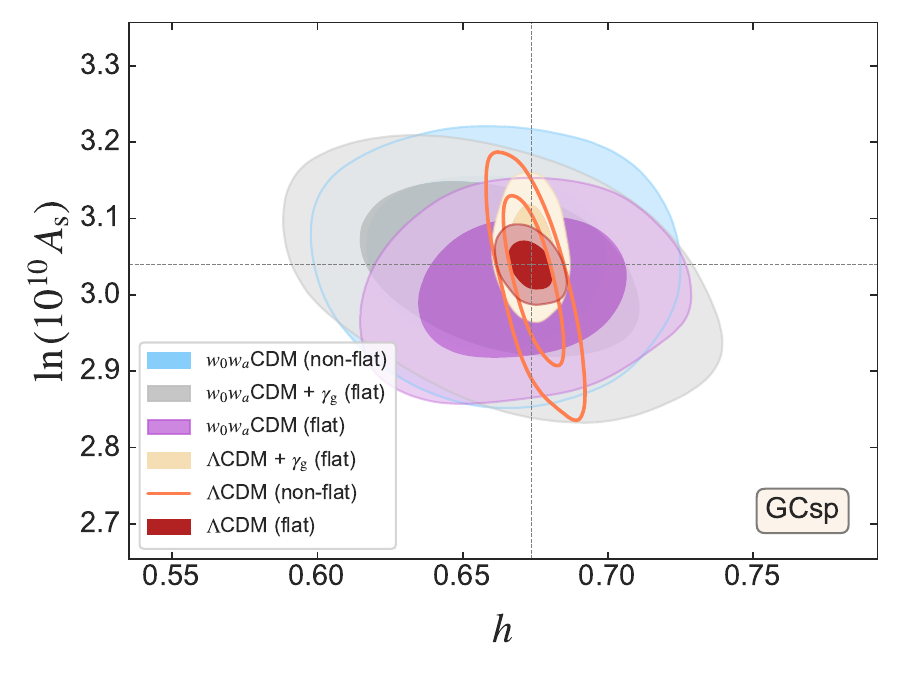} % first figure itself
        
    \end{minipage}\hfill
    \begin{minipage}{0.5\textwidth}
        \centering
        \includegraphics[width=1.0\textwidth]{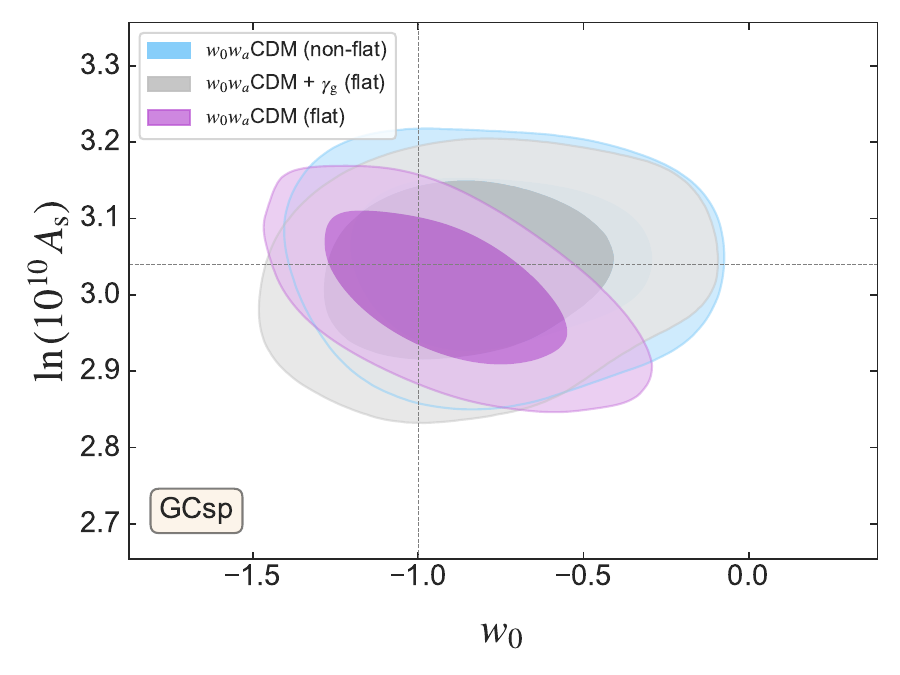} 
        
    \end{minipage}
            \begin{minipage}{0.5\textwidth}
        \centering
        \includegraphics[width=1.0\textwidth]{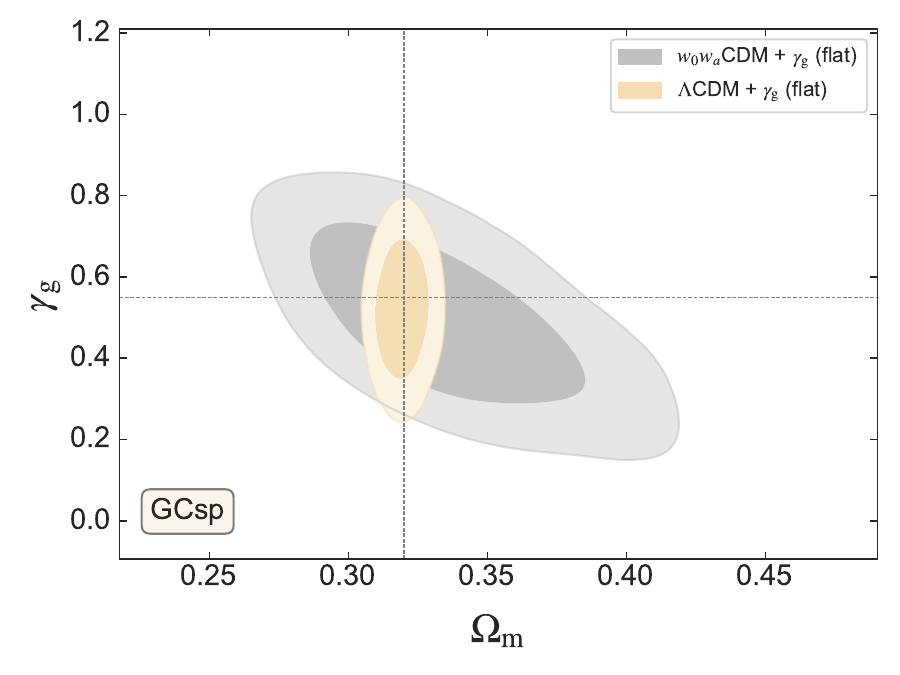} % first figure itself
        
    \end{minipage}\hfill
    \begin{minipage}{0.5\textwidth}
        \centering
        \includegraphics[width=1.0\textwidth]{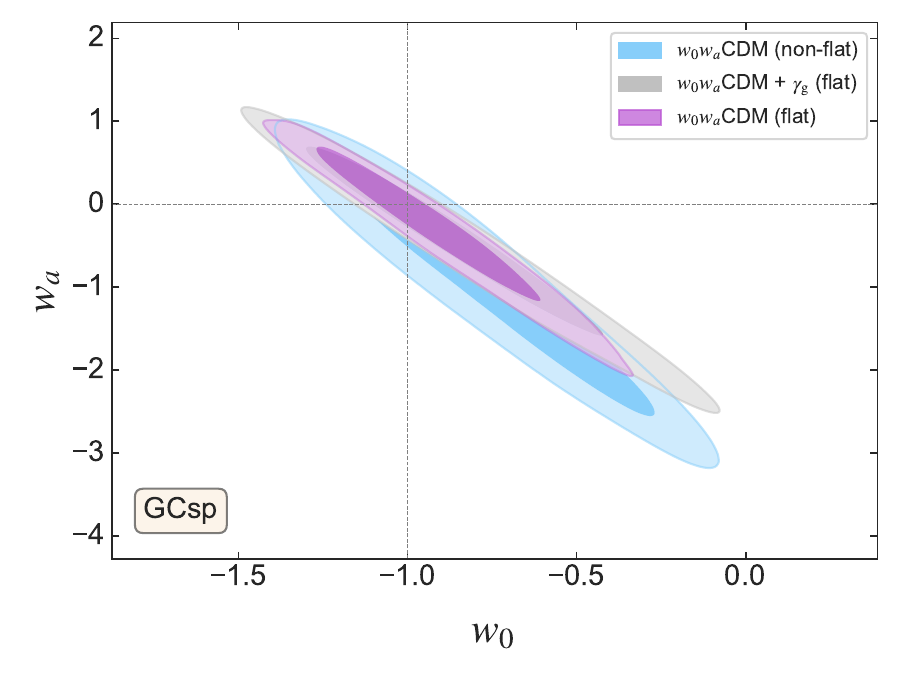} 
        
    \end{minipage}
    \caption{Two-dimensional posterior distributions for the cosmological parameters \OK and $h$, the primordial parameters \ns and \lnAs, $h$ and \lnAs, the dark energy parameter \wo and \lnAs, the modified-gravity Linder parameter \gammag, and the dark energy parameters \wo and \wa, using only the spectroscopic galaxy clustering probe (GCsp), for the six cosmological models. The values for the \fom of \wo and \wa can be found in \autoref{tab:FoM}.}
    \label{fig:2D_GCsp}
\end{figure*}

The constraining power on the dark energy parameters \wo and \wa is limited, with the corresponding posterior distributions displaying strongly non-Gaussian behaviour (see \cref{fig:2D_WL}, lower right panel). The \fom values obtained for \wo and \wa using this probe (\cref{tab:FoM}, first column) are consistent with those reported in \citet{Blanchard-EP7}. Specifically, our forecasts align with the predictions for the pessimistic scenario presented in their paper, despite using a higher value for the maximum multipole $\ell_{\rm max}$ in our analysis. This likely occurs because we employ a more sophisticated modelling approach for the cosmic shear probe in this work, including the sampling of multiplicative bias and redshift-bin shifts as systematic nuisance parameters. This increases the size of the parameter space, thereby widening our constraints.

The posterior distributions of the systematic nuisance parameters, such as the redshift bin shifts and multiplicative bias parameters, are well constrained. Most of the posteriors are either informative or prior dominated, and they remain consistent with the fiducial values (see detailed discussion in \cref{subsubsec:3x2pt_nuisance}). In particular, by examining the full posterior distributions shown in \cref{app:nuisance}, Figs. \ref{fig:triangle_w0waCDM_nuisance2} and \ref{fig:triangle_w0waCDM_nuisance3} for the redshift shifts and multiplicative biases, we conclude that these parameters are prior dominated. This is expected, as the priors adopted reflect the \Euclid\ science requirements for the primary photometric probe -- $3\times2$pt -- which are more stringent than for shear-only analyses.

Specifically, the results in \cref{fig:2D_WL} (lower left panel) demonstrate that the constraints on the intrinsic alignment parameters in the eNLA model are cosmological model-independent, with no significant differences in the constraints across the six models studied. This indicates that \aIA and \etaIA are not degenerate with the dark energy parameters \wo and \wa, or with the modified gravity parameter \gammag, and that they are independent of cosmological geometry. This is due to the definition of the IA model in the theoretical prescription of the cosmic shear probe (see Paper 1, Eq. 116), where the IA parameters contribute both linearly and quadratically to the cosmic-shear power spectrum. We {also show} the correlation matrices between the intrinsic alignment parameters, and \Om and $S_8$ to investigate further degeneracies. The correlations among the four key parameters \aIA, \etaIA, \Om, and \(S_8\) exhibit similar qualitative behaviour in both \LCDM and \wowaCDM models. In particular, the matter density \Om and the clustering amplitude parameter \(S_8\) continue to show a positive correlation, reflecting their joint influence on structure growth. However, the introduction of the dark energy parameters in the \wowaCDM introduces additional degeneracies that mildly affect the strength of these correlations. This results in a slight broadening of parameter degeneracies.

Finally, we note that our results show a rotation in the direction of degeneracy between \sotto and \Om as compared to $S_{8}$ and \Om in the two-dimensional posteriors, relative to previous Stage-III results \citep[e.g.][which shows the same direction of degeneracy in both the \sotto-\Om and the $S_{8}$-\Om plane]{DESY3}. We investigated this rotation and found that it results from using a \bbn prior and the increased constraining power of this probe for \Euclid-like surveys.

For \Euclid \drthree, we anticipate an improvement by one order of magnitude over previous surveys\footnote{We note that the DES Y3 analysis quoted here employs a slightly different cosmological model, including a more advanced intrinsic alignment treatment, while the KiDS-1000 legacy constraints are derived using power spectrum bandpowers and COSEBIs rather than angular power spectra.} in constraining \sotto and $S_8$ for \LCDM KiDS-1000 legacy survey \citep{kids-legacy} and DES Y3 \citep{DESY3shear}: 
\begin{align*}
\textbf{\Euclid} \hspace{20pt} & \sigma_8 = 0.816^{+0.013}_{-0.013}, \quad S_8 = 0.8416^{+0.0056}_{-0.0052}. \\
\textbf{DES Y3} \hspace{24pt} & \sigma_8 = 0.863 \pm 0.096, \quad S_8 = 0.793^{+0.038}_{-0.025}. \\
\textbf{KiDS legacy} \hspace{22pt} & S_8 = 0.815^{+0.016}_{-0.020}. \\
\end{align*}
\Euclid achieves an improvement of one order of magnitude in \( S_8 \), with a precision of \( \sigma(S_8) = 0.0054 \), compared to \( \sigma(S_8) \approx 0.0315 \) from DES Y3 and \( \sigma(S_8) \approx 0.018 \) from KiDS-1000. This corresponds to an improvement of approximately \( 83\% \) relative to DES Y3 and \( 70\% \) relative to KiDS-1000.

\subsection{Spectroscopic galaxy clustering}\label{sec:results_GCsp}

The large-scale distribution of galaxies can be used to effectively put constraints on cosmological parameters via the growth of cosmic structures (through \rsd) and the geometrical information (through the AP cosmological test). In \cref{fig:2D_GCsp}, we show the two-dimensional sampled posterior distributions for $H_0$, \Om, \OK, the primordial parameters \ns and \lnAs, the modified gravity parameter \gammag, as well as the dark energy parameters \wo and \wa for the spectroscopic galaxy clustering (GCsp) probe. Overall, in beyond-\LCDM (flat) model extensions, the posterior distributions are strongly non-Gaussian and show projection effects.

Contrary to cosmic shear-only, we observe that spectroscopic galaxy clustering is able to constrain the primordial parameter \lnAs significantly. The constraint on \ns arises primarily because the majority of the nuisance parameters are fixed, and high GCph scale cuts are assumed. In the absence of these conditions, \ns exhibits significant degeneracy with the nuisance parameters. For more details, we refer the reader to the future work of Euclid Preparation: Moretti et al. (in prep.).

Interestingly, in the two-dimensional posteriors, we observe a change in the degeneracy direction between \ns and \lnAs for the \wowaCDM models, as compared to flat and non-flat \LCDM (see \cref{fig:2D_GCsp}, upper right panel). This degeneracy can be explained when the two-dimensional posterior of $H_0$ and \lnAs is illustrated (see \cref{fig:2D_GCsp}, middle left panel), where the presence of the dark energy parameters \wo and \wa introduces extra degrees of freedom and broader contours for \lnAs (see \cref{fig:2D_GCsp}, middle right panel).
\begin{figure*}[ht!]
    \centering
    \begin{minipage}{0.5\textwidth}
        \centering
        \includegraphics[width=1.\textwidth]{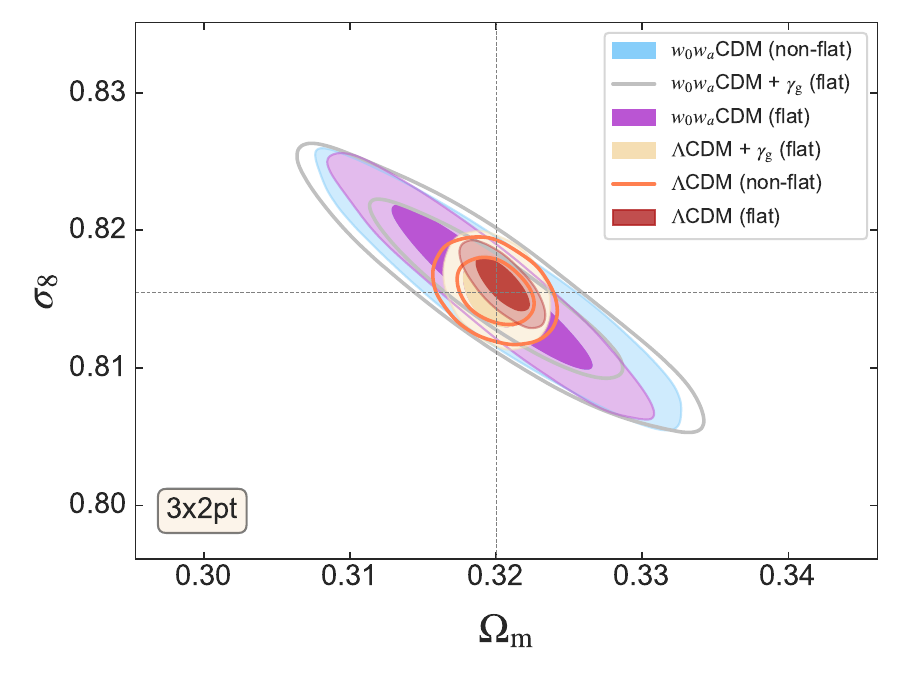} % first figure itself
        
    \end{minipage}\hfill
    \begin{minipage}{0.5\textwidth}
        \centering
        \includegraphics[width=1.\textwidth]{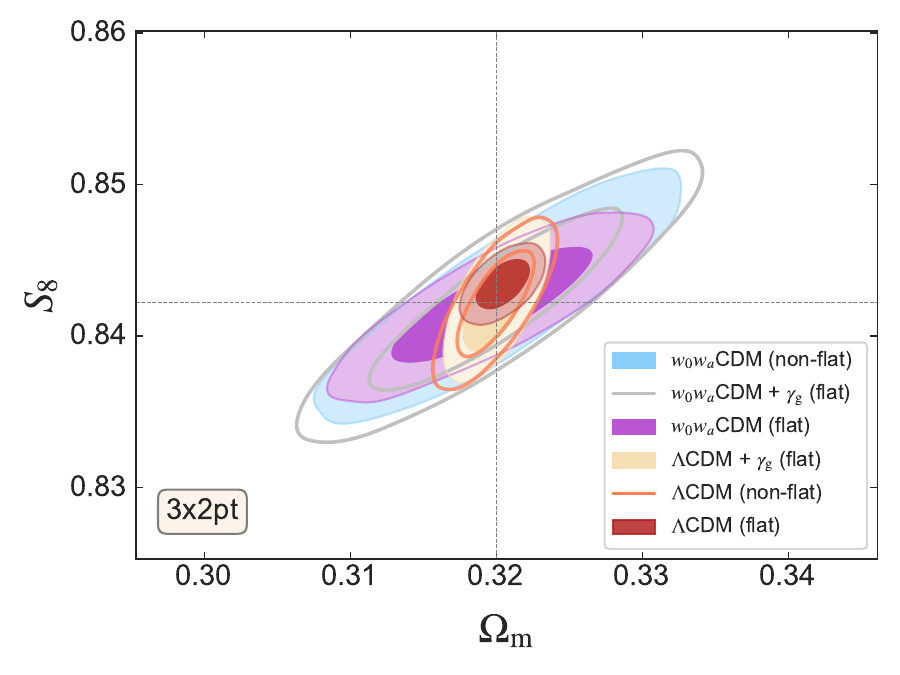} 
    \end{minipage}
        \begin{minipage}{0.5\textwidth}
        \centering
        \includegraphics[width=1.\textwidth]{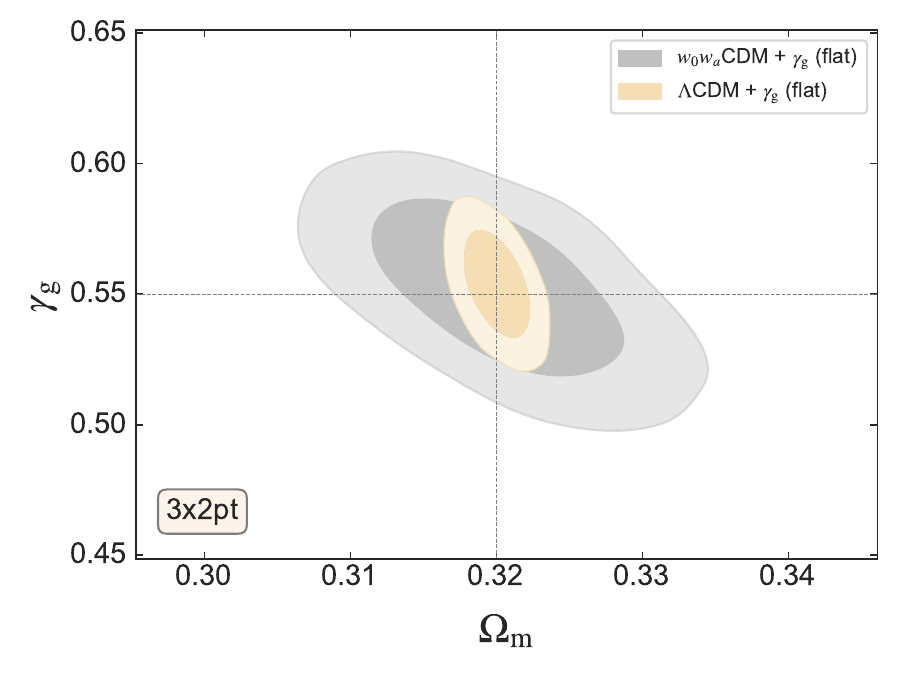} % first figure itself
        
    \end{minipage}\hfill
    \begin{minipage}{0.5\textwidth}
        \centering
        \includegraphics[width=1.\textwidth]{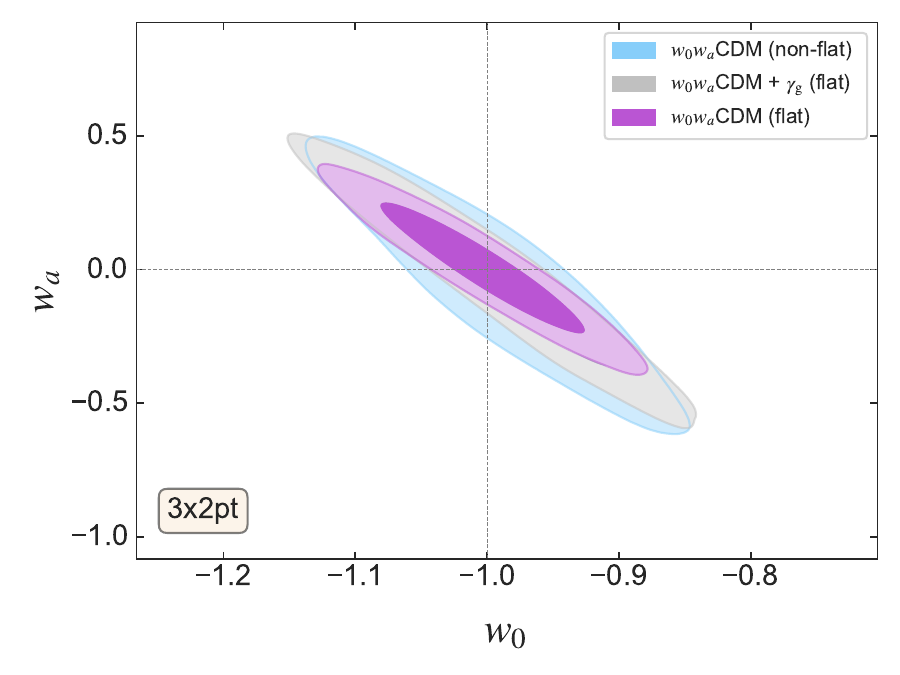} 
        
    \end{minipage}

    \caption{Two-dimensional posterior distributions for the cosmological parameters \Om and \sotto, as well as $S_8$, \wo and \wa, and \Om and \gammag, using the full photometric probe combination 3\texttimes2pt. The values for the \fom of \wo and \wa can be found in \cref{tab:FoM}.}
    \label{fig:2D_3x2pt}
\end{figure*}
Spectroscopic galaxy clustering alone exhibits substantially weaker constraining power on \( S_8 \) compared to cosmic shear, as expected given the sensitivity of each probe. In the context of the \LCDM model, specifically, the relative uncertainty on \( S_8 \) from GCsp (\( S_8 =0.841^{+0.030}_{-0.029} \)) is approximately 446\% larger than that from WL alone (\( S_8 = 0.8416^{+0.0056}_{-0.0052} \)), the latter being more directly sensitive to the amplitude of matter fluctuations. Similarly to the cosmic-shear only case, the BBN priors break the degeneracy between the Hubble parameter $H_0$ and the baryon density \Ob.

Notably, spectroscopic galaxy clustering on its own provides significant constraints on non-flat cosmologies (\cref{fig:2D_GCsp}, upper left panel). When comparing the BAO-only results from \citet{DESIDR1} to our predictions for the full-shape analysis of spectroscopic galaxy clustering, both results are consistent, and we forecast a 50\% improvement in the measurement of \OK with \Euclid only. Furthermore, we also predict that the modified gravity \gammag parameter (\cref{fig:2D_GCsp}, lower left panel) can be constrained with full shape galaxy clustering only. In both the \LCDM and \wowaCDM extensions, \gammag is measured with an uncertainty of approximately 0.2 and 0.3, respectively, indicating that it is better constrained in \LCDM. In relative terms, this corresponds to a 42\% uncertainty in \LCDM versus 57\% in \wowaCDM, showing that more information is gained on \gammag in the \LCDM scenario. For \Euclid \drthree, we anticipate an improvement by one order of magnitude in general over former surveys like eBOSS \citep{eBOSS} and first preliminary DESI results \citep{DESIDR1, DESIDR1_full_shape, DESIDR2}. 

The obtained \fom values for \wo and \wa using spectroscopic angular clustering (\cref{tab:FoM}, second column) fall within the pessimistic-optimistic constraint range presented in \citet{Blanchard-EP7}, aligning more closely with the predictions for the optimistic scenario. This is most likely due to the reduced parameter space sampled here, as many nuisance parameters of the theoretical modelling have been fixed to mitigate the impact of projection effects.

The posterior distributions corresponding to the systematic nuisance parameters that are freely sampled (e.g.\ bias, purity, and Poissonian shot noise parameters) are well-constrained and recover the fiducial values. Examples of these posterior distributions can be seen in Figs. \ref{fig:triangle_w0waCDM_nuisance5} and \ref{fig:triangle_w0waCDM_nuisance6}. In particular, the purity parameters, although sampled from a Gaussian prior distribution, are more tightly constrained than the prior in the first redshift bin. They also exhibit correlations across the remaining redshift bins, where the posterior distributions are broader. These correlations impact the cosmological parameters as well, with non-negligible degeneracies observed with \As, \ns, and $H_0$, especially for the purity parameters in the higher redshift bins. The strongest correlation is found between \As and the purity parameter in the third redshift bin, with a correlation coefficient of 0.2. \\

\subsection{3\texttimes2pt}\label{sec:results_3x2pt}

\begin{figure*}[h!]
    \centering
    \begin{minipage}{0.5\textwidth}
        \centering
        \includegraphics[width=1.\textwidth]{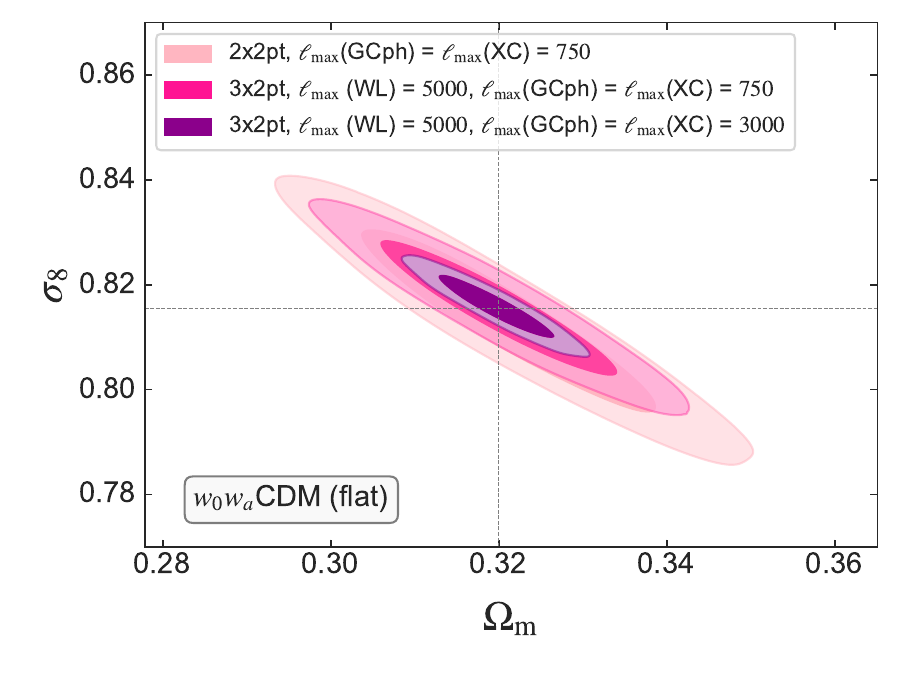}
        
    \end{minipage}\hfill
    \begin{minipage}{0.5\textwidth}
        \centering
        \includegraphics[width=1.\textwidth]{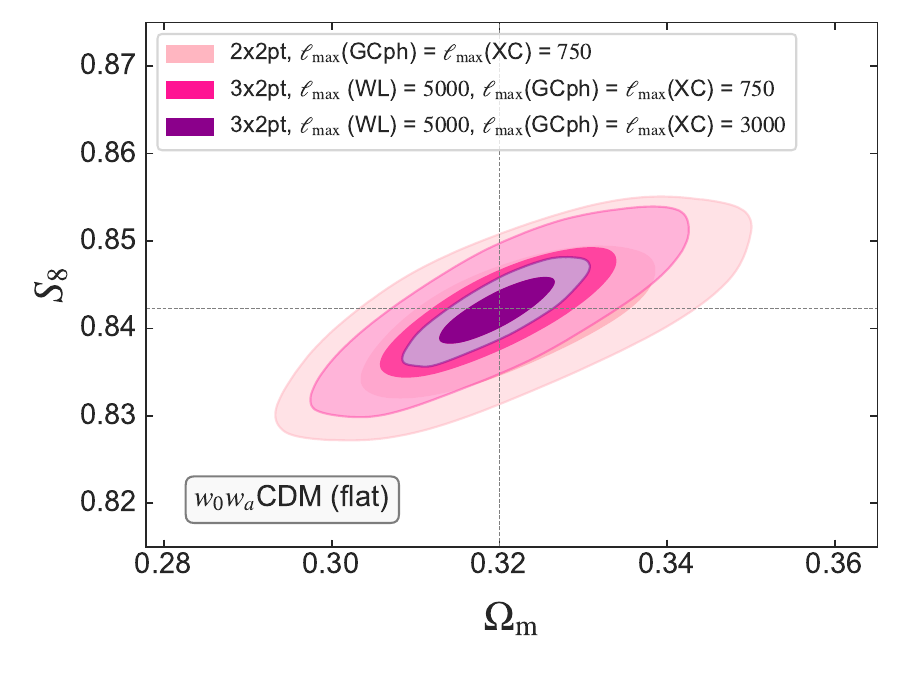} 
 
    \end{minipage}
        \begin{minipage}{0.5\textwidth}
        \centering
        \includegraphics[width=1.\textwidth]{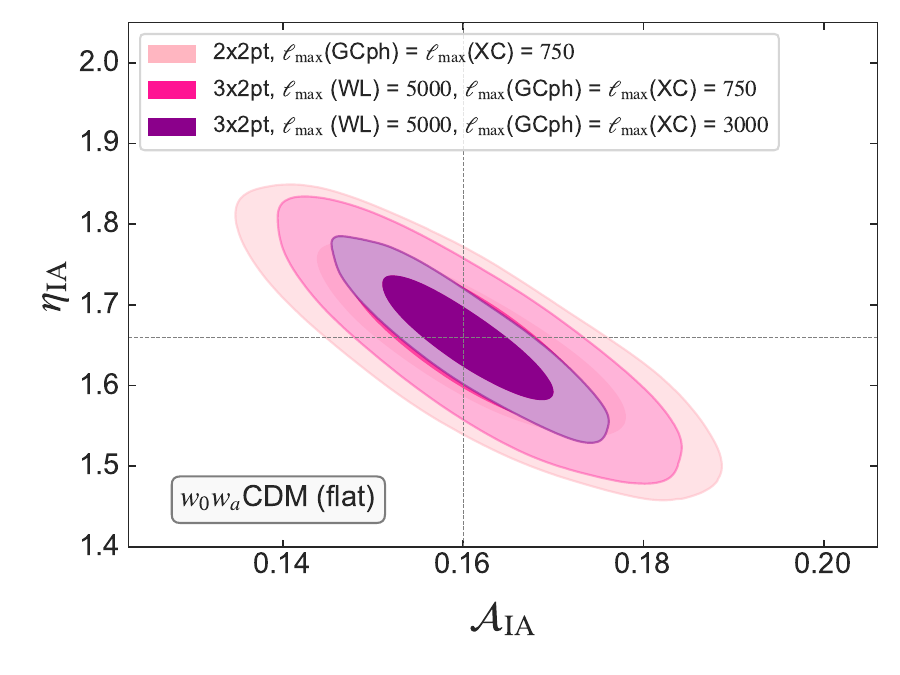} % first figure itself
        
    \end{minipage}\hfill
    \begin{minipage}{0.5\textwidth}
        \centering
        \includegraphics[width=1.\textwidth]{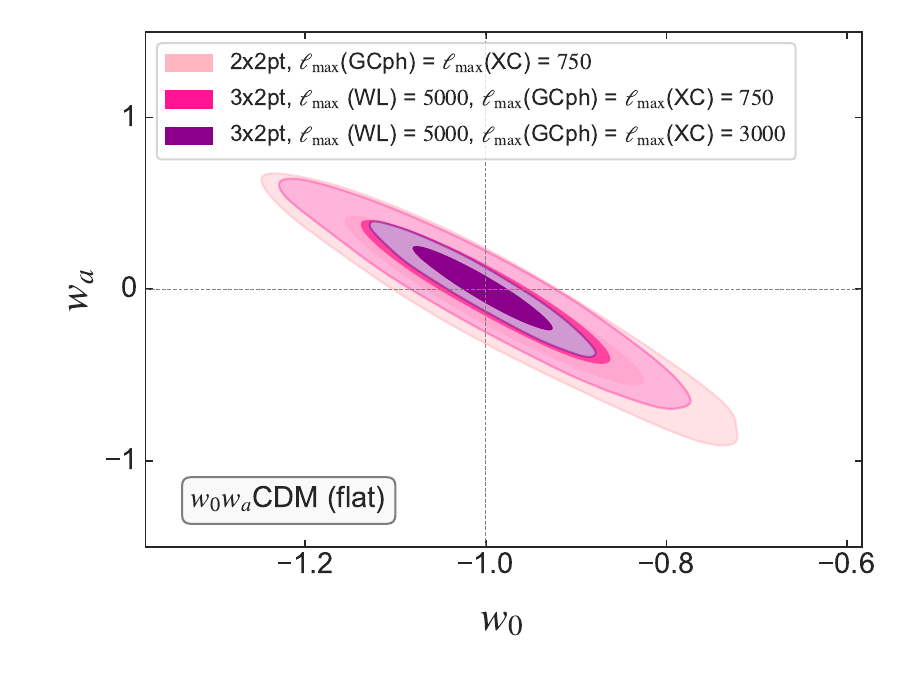} 
        
    \end{minipage}

    \caption{Two-dimensional posterior distributions for the cosmological parameters \Om, \sotto and $S_8$, \wo and \wa, and \etaIA and \aIA, only for the \wowaCDM model adopting a flat geometry, using the full photometric probe combination of 3\texttimes2pt, and the combination of only angular clustering and galaxy-galaxy lensing (2\texttimes2pt). The colour pattern indicates different scale cuts for angular clustering and galaxy-galaxy lensing (either $\ell_{\rm max} =$ 3000 or 750). The values for the \fom of \wo and \wa are 380 for 3\texttimes2pt ( $\ell_{\rm max} =$ 3000), 110 for 3\texttimes2pt ( $\ell_{\rm max} =$ 750) and 81 for 2\texttimes2pt. }
    \label{fig:2D_3x2pt_scale_cuts}
\end{figure*}

\begin{figure*}[h!]
\centering
\includegraphics[width=1.\textwidth]{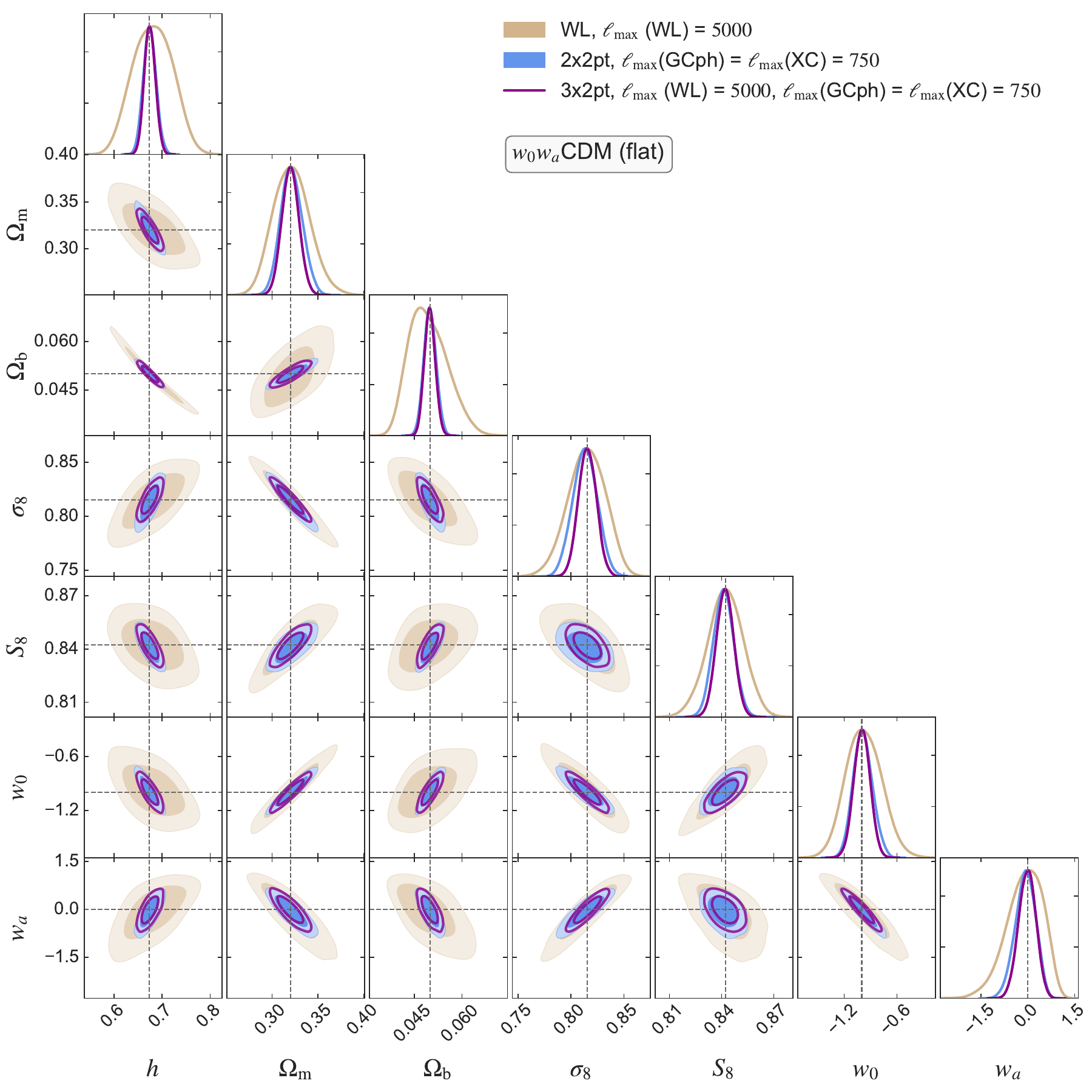}
\caption{Forecast of the constraints for the \wowaCDM cosmological model (adopting a flat geometry) using only \Euclid photometric probes: WL, galaxy-galaxy lensing and angular clustering, 2\texttimes2pt, and cosmic shear, galaxy-galaxy lensing and angular clustering, 3\texttimes2pt, as described in \cref{subsubsec:3x2pt_scale_cuts}. For the photometric probes, we used $\ell_{\rm max} = 5000$ for cosmic shear and $\ell_{\rm max} = 750$ for photometric angular clustering, and galaxy-galaxy lensing. The corresponding \fom obtained for each sample can be found either in \cref{tab:FoM} (WL) and in the text of the corresponding section (2\texttimes2pt and 3\texttimes2pt). The values for the \fom of \wo and \wa are 21 for WL, 110 for 3\texttimes2pt ( $\ell_{\rm max} =$ 750) and 81 for 2\texttimes2pt. }
\label{fig:triangle_w0waCDM_WL_2x2pt_3x2pt_750}
\end{figure*}

The 3\texttimes2pt probe, which combines angular galaxy clustering (GCph), galaxy-galaxy lensing (XC) and cosmic shear (WL), is an effective tool for constraining \Om and \sotto, similarly to WL, and \wo, \wa, and \gammag, as well as several nuisance parameters associated to the modelling of effects such as intrinsic alignment, galaxy bias, and magnification. In \cref{fig:2D_3x2pt}, we present the main highlights from the full posterior distributions by plotting the marginalised two-dimensional posteriors for the main parameter combination, as we did in Figs. \ref{fig:2D_WL} and \ref{fig:2D_GCsp}. In subsequent subsections, we analyse in detail the particular behaviour of 3\texttimes2pt for different scale cuts in the \wowaCDM model adopting a flat geometry (\cref{subsubsec:3x2pt_scale_cuts}). We study the impact of sampling experimental nuisance systematic parameters on the posterior distributions of the cosmological parameters (\cref{subsubsec:3x2pt_nuisance}), and evaluate the presence of possible projection effects in the \LCDM and \wowaCDM models also adopting a flat geometry (\cref{subsubsec:3x2pt_projection_effects}).

The expected constraints in the \sotto$--$\Om and $S_{8}$--\Om planes show an improvement by one order of magnitude with respect to WL only (see \cref{fig:2D_3x2pt}, upper row). As in the case of both WL and GCsp, we see a difference in the direction of degeneracy between \sotto and \Om compared to $S_{8}$ and \Om in the two-dimensional posteriors. Although not explicitly shown, 3\texttimes2pt also constrains the primordial parameters \ns and \lnAs. This joint probe is also able to constrain the Hubble parameter $H_0$, partially due to the extra prior information from the \bbn prior, and to the incorporation of angular clustering, which breaks parameter degeneracies. Similarly to WL only, for \Euclid \drthree, we anticipate an improvement by one order of magnitude with respect to completed surveys in terms of constraints on \sotto and $S_8$ \citep{Heymans:2020gsg, DESY3} within the \LCDM model:

\begin{align*}
\textbf{\Euclid} \hspace{24pt} & \sigma_8 = 0.8161^{+0.0026}_{-0.0025}, \quad S_8 = 0.8434^{+0.0022}_{-0.0021}. \\
\textbf{DES Y3} \hspace{24pt} & \sigma_8 = 0.733^{+0.039}_{-0.049}, \quad S_8 = 0.776 \pm 0.017. \\
\textbf{KiDS-1000} \hspace{24pt} & S_8 = 0.766^{+0.020}_{-0.01}.\\
\end{align*}
\Euclid achieves an improvement of one order of magnitude in \( S_8 \), with a precision of \( \sigma(S_8) \approx 0.0022 \), compared to \( \sigma(S_8) \approx 0.017 \) from DES Y3 and \( \sigma(S_8) \approx 0.02 \) from KiDS-1000. The improvement with respect to shear-only is driven by angular photometric galaxy clustering.

The increase in the constraining power on the dark energy parameters \wo and \wa  (see \cref{fig:2D_3x2pt}, lower right panel) for 3\texttimes2pt with respect to single probes is remarkable. The obtained \fom values for \wo and \wa using this probe (\cref{tab:FoM}, third column) are consistent with those reported in \citet{Blanchard-EP7}. Specifically, our forecasts align with the predictions for the pessimistic scenario, despite using high GCph scale cuts. Similarly to cosmic shear, the reason behind this is the increase in the number of sampled nuisance parameters in the analysis, which broadens the overall posterior distribution. Yet, we predict a \fom value of approximately 380 for the \wowaCDM (flat) model, which is one order of magnitude larger than the \fom calculated for single probes WL and GCsp.

The systematic nuisance parameters (e.g.\ shifts in the redshift bins and multiplicative bias parameters) and other physical effect parameters (intrinsic alignments, galaxy bias, magnification bias) are well-constrained given the informative priors provided by \Euclid science requirements, accurately recovering the fiducial values (Figs.\ \ref{fig:triangle_w0waCDM_nuisance2}, \ref{fig:triangle_w0waCDM_nuisance3}, and \ref{fig:triangle_w0waCDM_nuisance4}). The magnification biases and redshift bin shifts remain prior-dominated, with no significant improvement from the inclusion of angular photometric galaxy clustering; see \cref{fig:triangle_w0waCDM_nuisance4}. These parameters show minimal correlations or degeneracies with the cosmological parameters. However, moderate degeneracies, particularly at the level of approximately 0.2, are observed between the photometric galaxy and magnification biases and certain cosmological parameters, most notably \Om. Although not explicitly shown, similarly to the WL case, the resulting 2D-marginalised posterior distributions corresponding to the intrinsic alignment parameters in the eNLA model, \aIA and \etaIA, are cosmological-model independent. The photometric galaxy bias nuisance parameters are strongly correlated, showing correlation coefficients of at least 0.9.

\subsubsection{Impact of scale cuts for angular clustering and cosmic shear on 3\texttimes2pt probe}\label{subsubsec:3x2pt_scale_cuts}
\begin{figure*}
    \centering
    \begin{minipage}{0.5\textwidth}
        \centering
        \includegraphics[width=1.\textwidth]{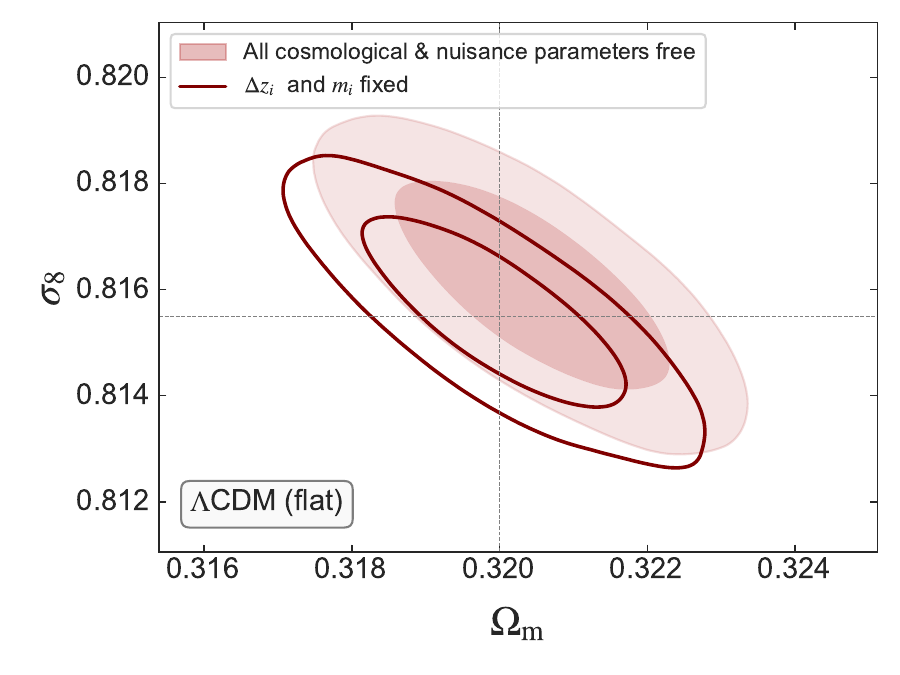}
        
    \end{minipage}\hfill
    \begin{minipage}{0.5\textwidth}
        \centering
        \includegraphics[width=1.\textwidth]{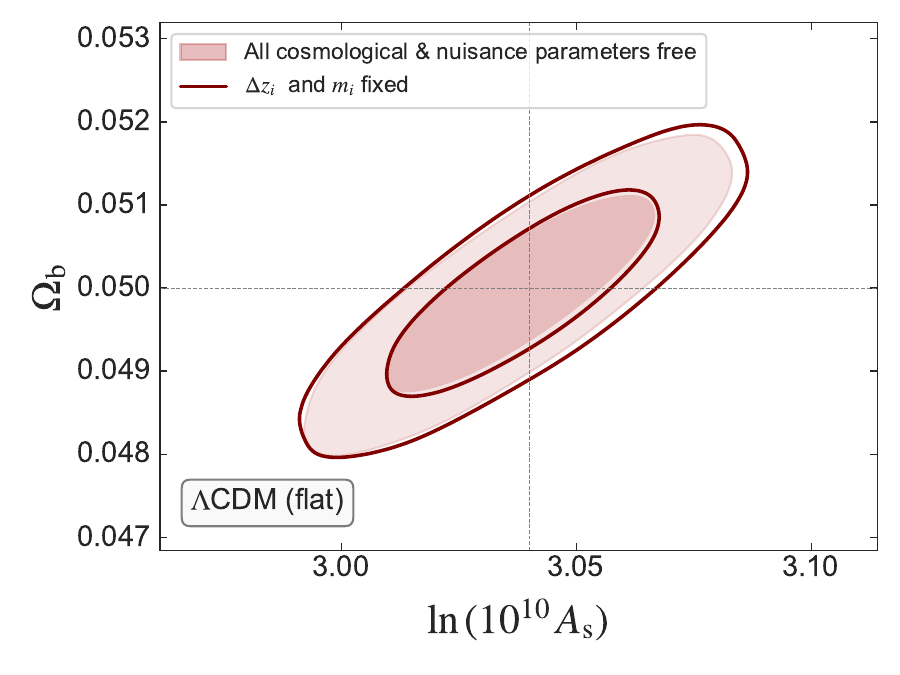} 
 
    \end{minipage}
        \begin{minipage}{0.5\textwidth}
        \centering
        \includegraphics[width=1.\textwidth]{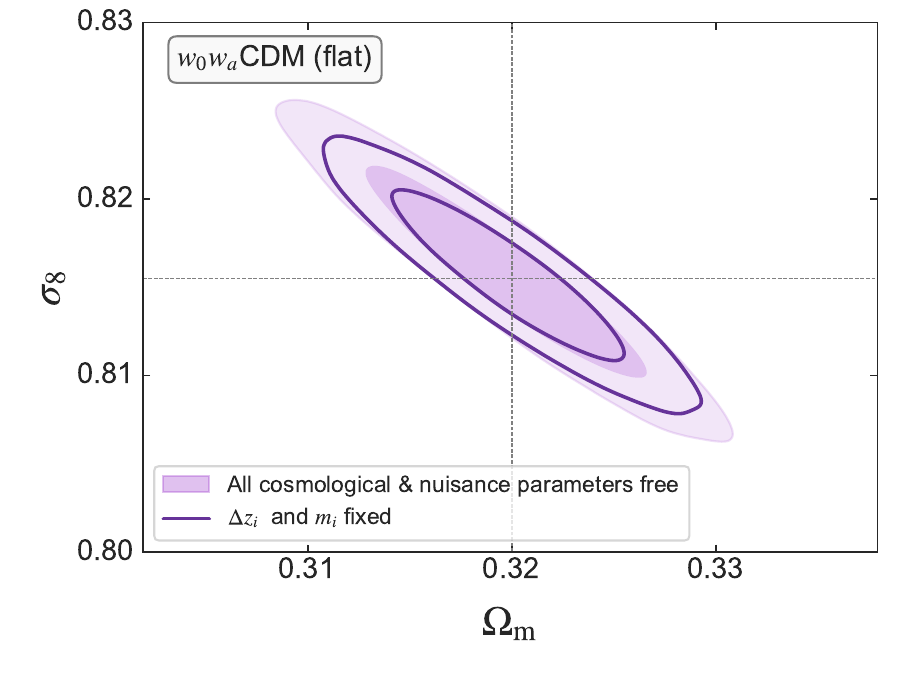} % first figure itself
        
    \end{minipage}\hfill
    \begin{minipage}{0.5\textwidth}
        \centering
        \includegraphics[width=1.\textwidth]{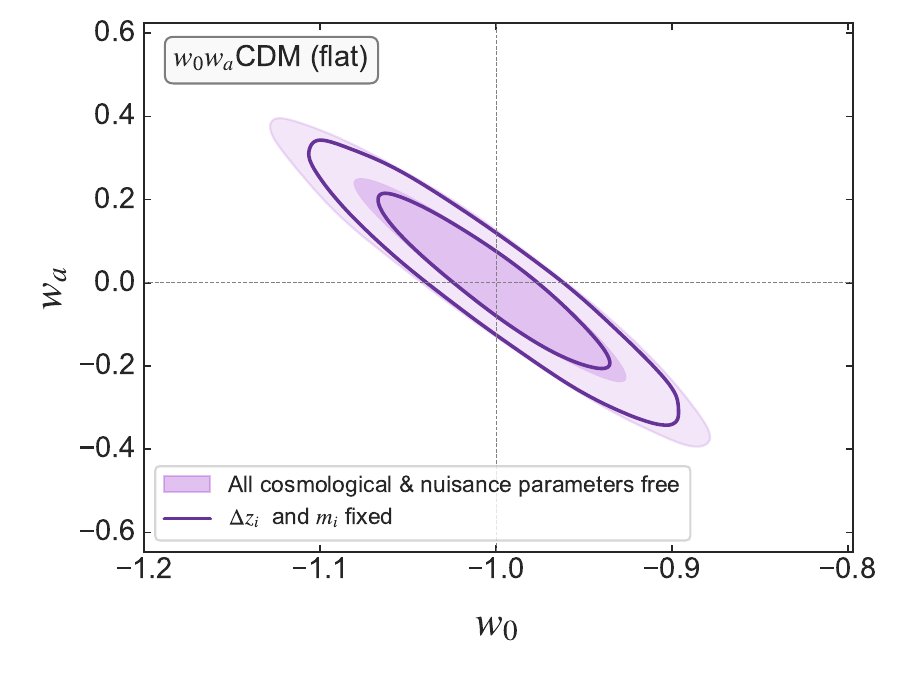} 
        
    \end{minipage}

    \caption{Two-dimensional posterior distributions for the cosmological parameters \Om, \sotto, \lnAs, \Ob, \wo, and \wa, for the models \LCDM and \wowaCDM adopting a flat geometry, when the systematic mean redshift-bin shifts $\Delta z^{\rm L}_i$ and shear multiplicative bias $m_i$ nuisance parameters were sampled (filled contours) or kept fixed (unfilled contours). The variation in the values of the obtained \fom for the dark energy parameters can be found in the text in \cref{subsubsec:3x2pt_nuisance}.}
    \label{fig:2D_3x2pt_impact_nuisance}
\end{figure*}
To understand the significant improvement in the dark energy \fom\ -- from approximately 20 for WL alone to around 380 for the full 3\texttimes2pt analysis -- we examine the role of photometric angular galaxy clustering modelling at different scales. Specifically, we investigate how conservative scale cuts in angular clustering and galaxy-galaxy lensing affect constraints on key cosmological parameters, including $\sigma_8$, $S_8$, and the dark energy parameters \wo and \wa, within the flat \wowaCDM model. The conservative cuts, referred to as `low GCph scale cuts' in \cref{tab:theoretical_probes}, are motivated by the need for realistic forecasts that reflect current theoretical limitations -- most notably, the simplified treatment of GCph (e.g. absence of non-linear galaxy bias modelling, proper modelling of the visibility mask and mixing matrices). To reflect this limited theoretical modelling or treatment of systematics, we introduce a more aggressive (smaller) scale cut for both angular clustering and galaxy-galaxy lensing, as described in \cref{tab:theoretical_probes}.

As expected, the constraining power decreases significantly when adopting a pessimistic setup, due to the reduced number of available data points and consequently, the diminished statistical strength of the analysis. This is evident in the two-dimensional posterior distributions shown in \cref{fig:2D_3x2pt_scale_cuts}, where the results with conservative scale cuts are compared to those when employing a higher GCph scale cuts. The results show a notable decrease in the dark energy \fom for the flat \wowaCDM model, dropping from 380 in the optimistic scenario to 110 in the more conservative analysis (see \cref{fig:2D_3x2pt_scale_cuts}, lower right panel, where the dark purple contours are compared to the pink ones). This highlights the importance of accurately modelling GCph within the 3\texttimes2pt probe in order to retain its full constraining power on dark energy parameters.

To further explore this assumption, we assess the influence of WL alone within the broader 3\texttimes2pt framework by conducting an additional run using the 2\texttimes2pt probe (combining angular clustering and galaxy-galaxy lensing) with the same conservative scale cuts. In this scenario, the overall cosmological constraints degrade, the uncertainties on key cosmological parameters increase, and the \fom drops further to a value of 81 (see \cref{fig:2D_3x2pt_scale_cuts}, lower right panel, soft pink contours) -- still well above the dark energy \fom for weak lensing alone, which is 21. This confirms that most of the constraining power in the 3\texttimes2pt analysis arises from the 2\texttimes2pt combination. These findings underscore the importance of accurately modelling angular clustering to achieve robust cosmological constraints in \Euclid-like surveys. The full one- and two-dimensional posterior distributions of the cosmological parameters for the three cases (WL, 2\texttimes2pt, and 3\texttimes2pt), using $\ell_{\rm max} = 750$ for angular clustering and galaxy-galaxy lensing, are shown in \cref{fig:triangle_w0waCDM_WL_2x2pt_3x2pt_750}.

\begin{figure*}
    \centering
    \begin{minipage}{0.5\textwidth}
        \centering
        \includegraphics[width=1.\textwidth]{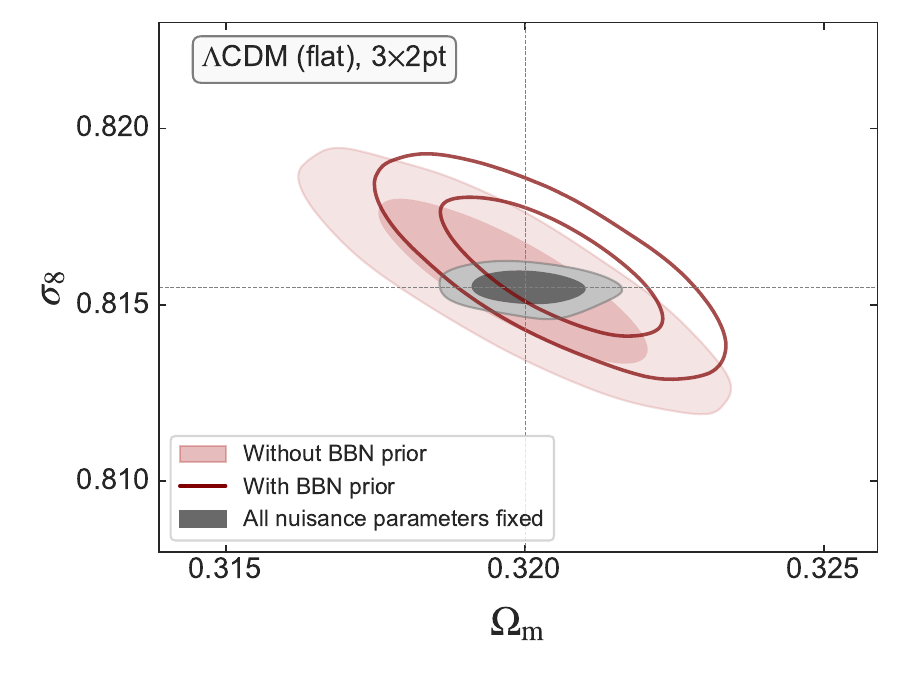}
        
    \end{minipage}\hfill
    \begin{minipage}{0.5\textwidth}
        \centering
        \includegraphics[width=1.\textwidth]{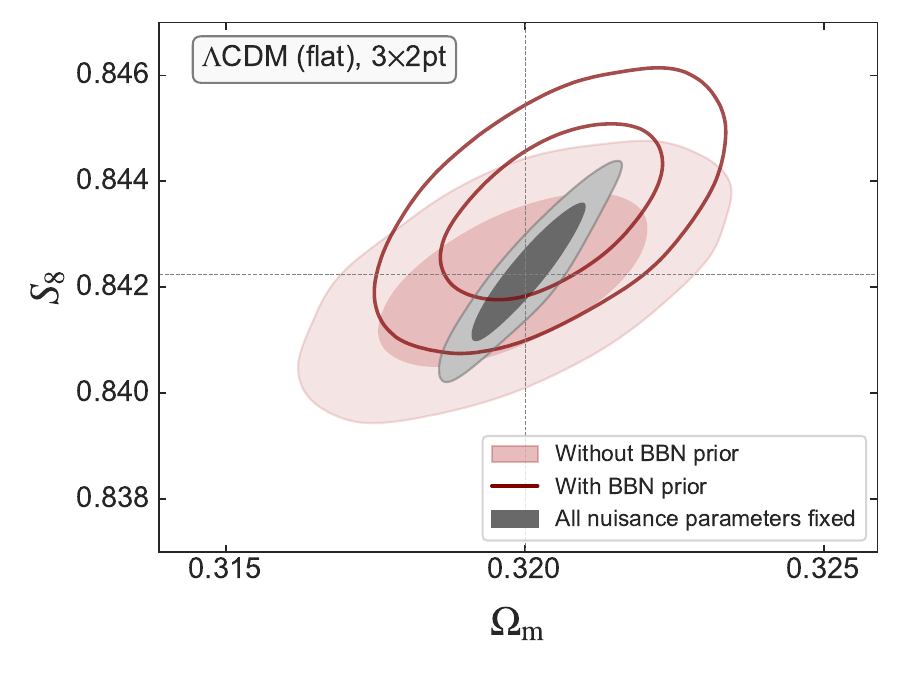} 
 
    \end{minipage}
        \begin{minipage}{0.5\textwidth}
        \centering
        \includegraphics[width=1.\textwidth]{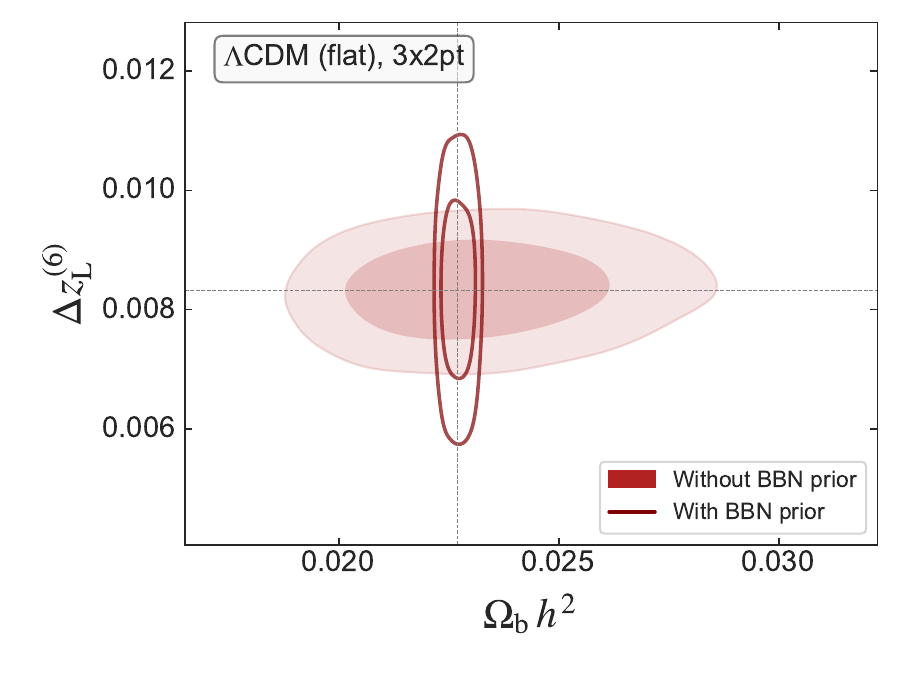} % first figure itself
        
    \end{minipage}\hfill
    \begin{minipage}{0.5\textwidth}
        \centering
        \includegraphics[width=1.\textwidth]{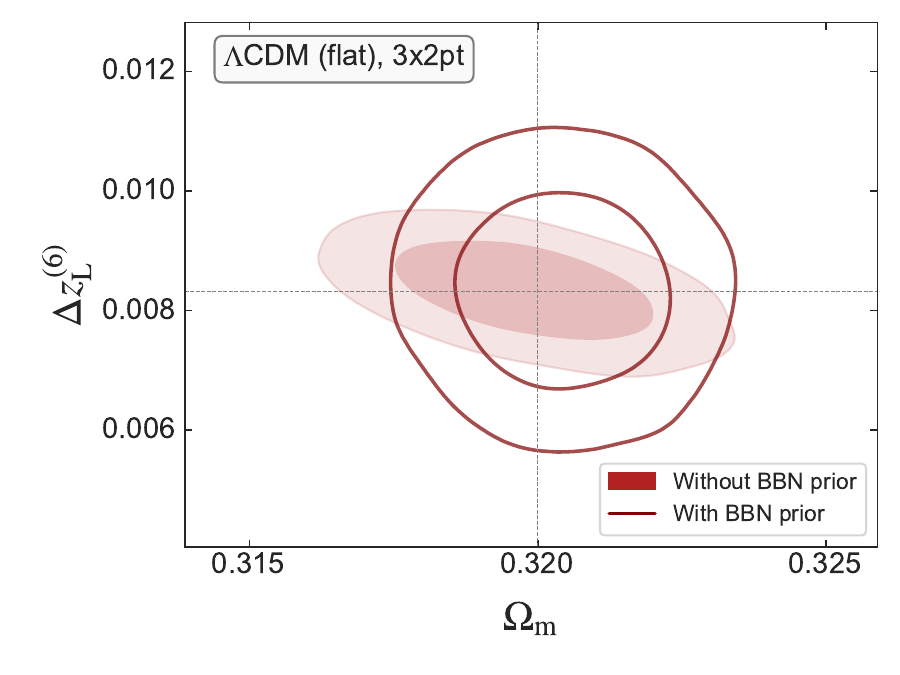} 
        
    \end{minipage}

    \caption{Two-dimensional posterior distributions for the cosmological parameters \Om and \sotto, as well as $S_8$, $\Ob\,h^2$, and six redshift bin shift mean $\Delta z^{(6)}_{\rm L}$ for the lensing probe, for the model \LCDM. We show cases with and without \bbn prior (dark red unfilled and light red filled contours, respectively), keeping all nuisance parameters free, and with the \bbn prior but keeping all nuisance parameters fixed (light grey filled contours). The case with \bbn prior and free nuisance parameters shows significant projection effects in \sotto and $S_8$.}
    \label{fig:2D_3x2pt_projection_effects}
\end{figure*}

\subsubsection{Impact of the experimental nuisance parameters on the cosmological parameters}\label{subsubsec:3x2pt_nuisance}

In this section, we evaluate the impact of fixing or sampling systematic nuisance parameters on the constraints on cosmological parameters within the context of a 3\texttimes2pt analysis, using high GCph scale cuts and a \bbn prior. The focus is on the cosmological parameters for both the \LCDM and \wowaCDM models, under the assumption of a flat geometric configuration. Specifically, we assess how the treatment of two key systematic nuisance parameters, the per-bin shear multiplicative bias $m_i$ and the mean redshift shifts $\Delta z^{\rm L}_i$, influences the resulting cosmological constraints. These forecasts give insightful information about the current scientific requirements for \Euclid, which were previously developed in \citet{Amara:2007as} and \citet{2013MNRAS.429..661M} and provide a historical perspective of the mission design.

While some systematic nuisance parameters, such as intrinsic alignment, magnification, and galaxy biases are inherently dependent on the underlying cosmological or astrophysical models, $m_{\rm L}^i$ and $\Delta z_{\rm L}^i$ are primarily determined by the experimental setup and the data processing pipeline. The \Euclid data analysis pipeline and calibration efforts impose stringent requirements on these parameters, and this information is used to define informative priors during the cosmological inference exercise (see \cref{tab:fiducial_model} for the imposed Gaussian priors on $m_{\rm L}^i$ and $\Delta z_{\rm L}^i$). Our objective is to investigate whether these nuisance parameters can be ignored if \Euclid's requirements are met in \drthree, and therefore, if they can be fixed during the inference analysis.

To address this question, we performed analyses with the same configuration as presented in \cref{fig:2D_3x2pt}, examining scenarios where $m_{\rm L}^i$ and $\Delta z_{\rm L}^i$ are kept fixed for both the flat \LCDM and \wowaCDM models (see \cref{fig:2D_3x2pt_impact_nuisance}). The resulting posterior distributions reveal that the results are model-dependent. In the \wowaCDM model, fixing these nuisance parameters leads to a significant impact on the marginalised uncertainties of key cosmological parameters such as \sotto and $S_8$, and the dark energy EoS parameters \wo and \wa. This in turn results in a measurable change in the dark energy \fom, with a higher value of 454, thus highlighting the importance of accounting for these systematics in the analysis.

In contrast, in the \LCDM model, the effect of fixing these nuisance parameters during the inference analysis is minimal. The marginalised uncertainties for the cosmological parameters remain virtually unchanged, both at the 68 and 95 percent confidence levels. This result suggests that, for \LCDM, the influence of these particular nuisance parameters is less critical, allowing us to reduce the sampled parameter space significantly without inducing extra errors on the cosmological uncertainties.\footnote{We observe in \cref{fig:2D_3x2pt_impact_nuisance}, a shift in the \Om$--$\sotto contour when all parameters are free. These findings are consistent with so-called projection effects, which are further studied in \cref{subsubsec:3x2pt_projection_effects}. To draw the conclusions presented in this section, we have focused on the lower and upper marginalised confidence limits obtained using \getdist, \corr{where the fractional difference is $\lesssim 10^{-4}$}.} We have left for future work the analysis of the impact of experimental nuisance parameters in other cosmological models beyond the \LCDM and \wowaCDM models. 

\subsubsection{Projection effects in \LCDM}\label{subsubsec:3x2pt_projection_effects}

\begin{figure*}[htp!]
    \centering
    \begin{minipage}{0.5\textwidth}
        \centering
        \includegraphics[width=1.\textwidth]{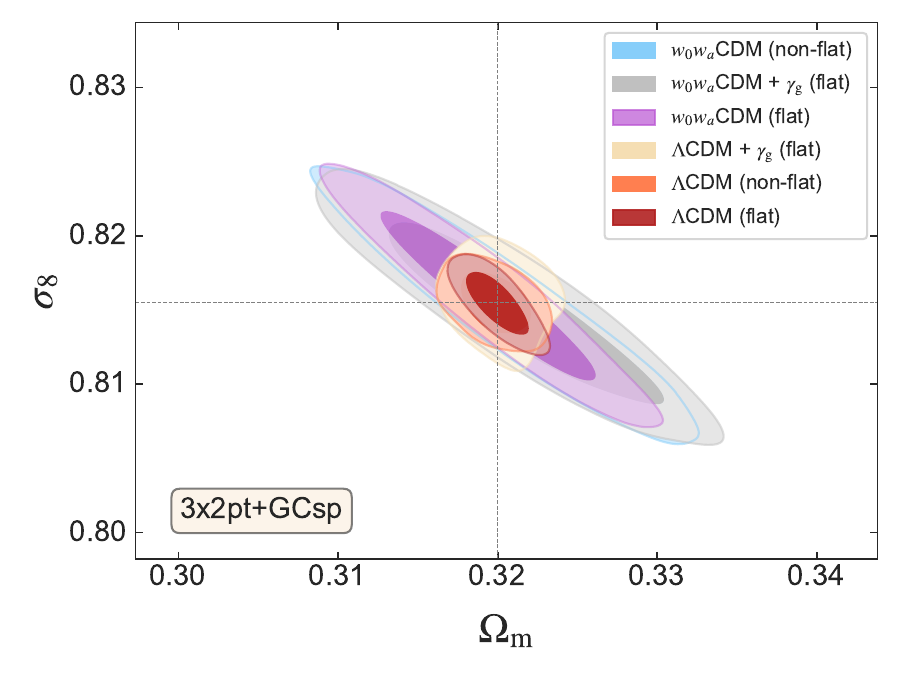} % first figure itself
        
    \end{minipage}\hfill
    \begin{minipage}{0.5\textwidth}
        \centering
        \includegraphics[width=1.\textwidth]{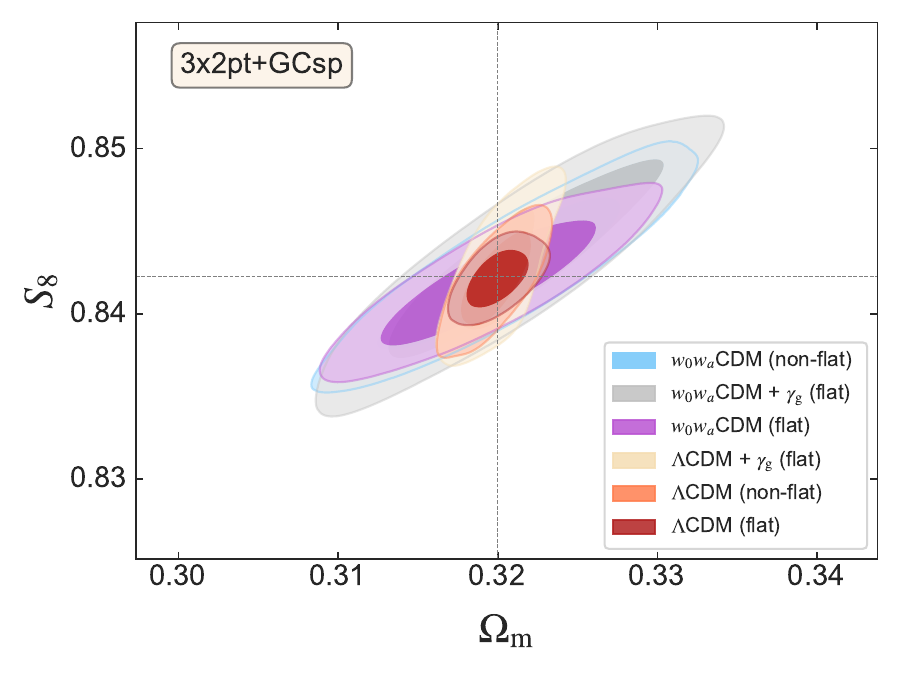} 
 
    \end{minipage}
        \begin{minipage}{0.5\textwidth}
        \centering
        \includegraphics[width=1.\textwidth]{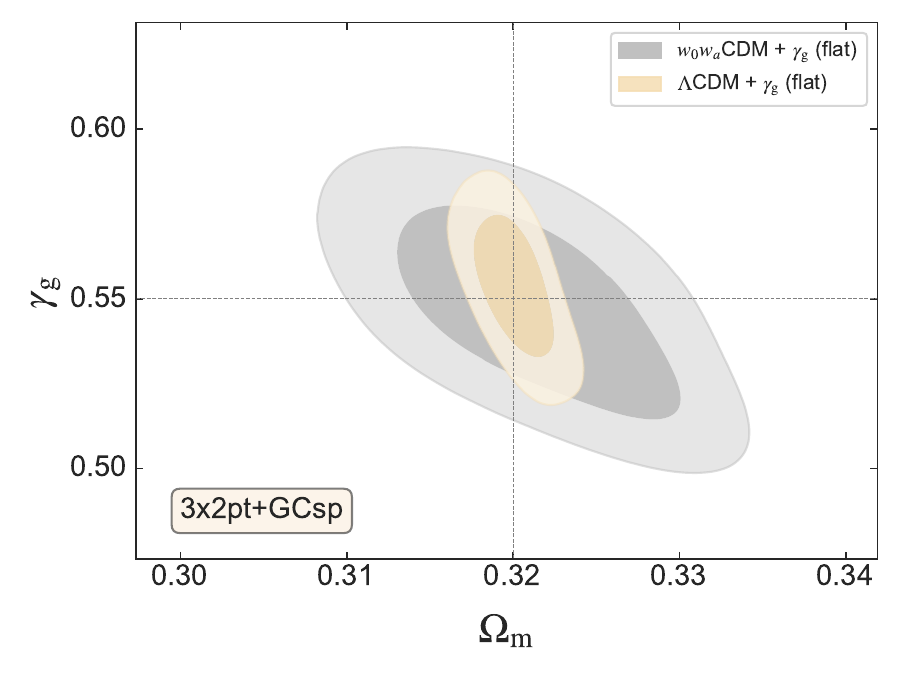} 
        
    \end{minipage}\hfill
    \begin{minipage}{0.5\textwidth}
        \centering
        \includegraphics[width=1.\textwidth]{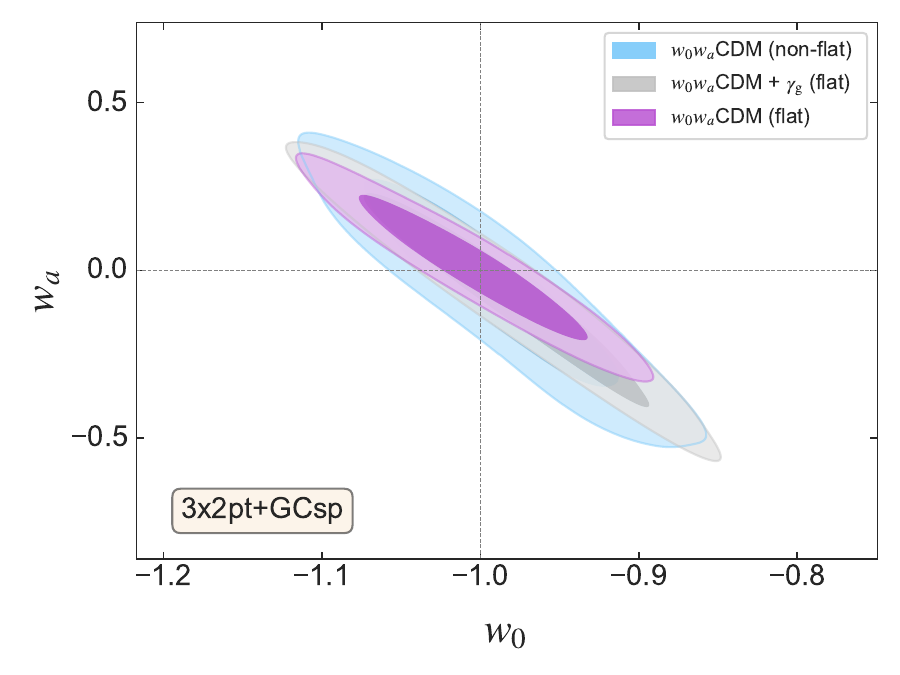} 
        
    \end{minipage}
    \caption{Two-dimensional posterior distributions for the cosmological parameters \Om and \sotto, as well as $S_8$, \wo and \wa, and \Om and \gammag, using the full photometric probe combination (3\texttimes2pt) in combination with spectroscopic galaxy clustering full shape (GCsp). The values for the \fom of \wo and \wa can be found in \cref{tab:FoM}.}
    \label{fig:2D_3x2pt_GCsp}
\end{figure*}
When analysing the 3\texttimes2pt results in \cref{fig:2D_3x2pt}, we observe an evident shift in the one-dimensional posterior distributions of \sotto and $S_8$ for the flat \LCDM model. These shifts are due to the presence of a shift in the one-dimensional marginalised posterior distribution in the sampled parameters $\Oc\,h^2$ and \lnAs. We do not observe this behaviour in the results of the other cosmological models, most likely given that the uncertainties associated with the cosmological parameters in those cases are larger.

To verify if this is indeed a projection effect, we checked that the best-fit (that is, the minimum of the $\chi^2$) is recovered when the sampled parameters adopt the fiducial set of values. Moreover, we have checked if this effect is due to possible under-sampling of the parameter space, by increasing the number of live points and the number of trained neural networks of \nautilus, finding the same resulting posterior distributions. Furthermore, we have validated \nautilus for the \Euclid study case against \polychord, and we did not find significant differences in the sampled posterior distributions.

While the presence of projection effects has been reported in the usual sampled parameter space of LSS when conservative scale cuts and/or non-linear galaxy bias are used to model the 3\texttimes2pt probe and the nuisance parameters are left free, it is still novel to find projection effects within a \LCDM framework. To assess the origin of these projection effects, we investigated two possible sources:
\begin{enumerate}
    \item The large parameter space to sample: To check this hypothesis, we run several cases where we progressively fix all nuisance parameters and evaluate the marginalised posterior distributions for cosmological parameters.\\
    \item Possible degeneracies between the experimental systematic nuisance parameters and the cosmological parameters in the presence of a \bbn prior: To verify this hypothesis, we re-run the same 3\texttimes2pt \LCDM case with all parameters free, but imposing a flat prior on $\Ob\,h^2$.
\end{enumerate}

\begin{figure*}[ht!]
\centering
\includegraphics[width=1.\textwidth]{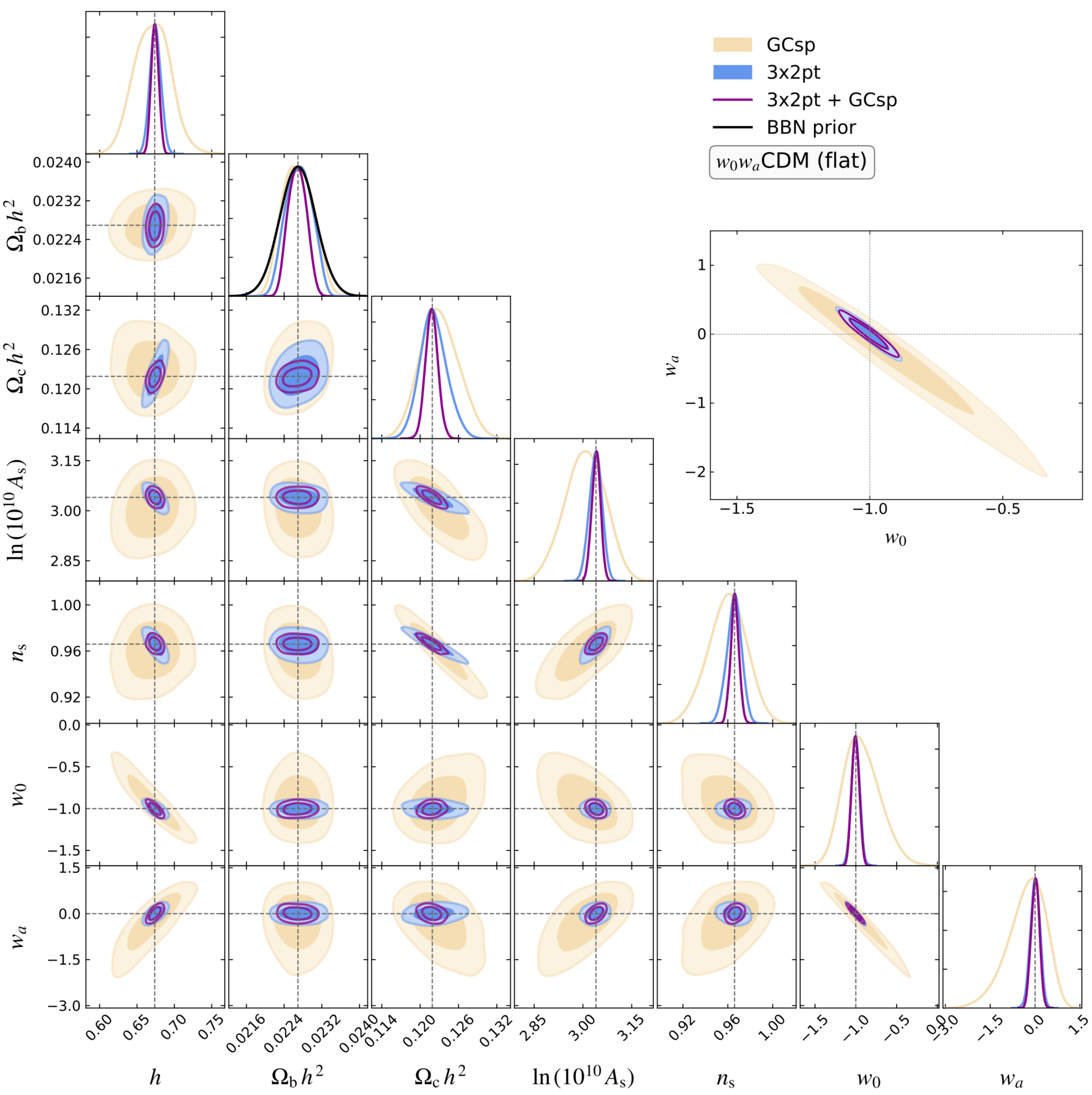}
\caption{Forecast of the constraints for the \wowaCDM cosmological model (adopting a flat geometry) using only \Euclid photometric and spectroscopic probes: spectroscopic galaxy clustering (GCsp), cosmic shear, galaxy-galaxy lensing and angular clustering, 3\texttimes2pt, and the combination of both of them as described in \cref{sec:results_full}. For the photometric probes, we used $\ell_{\rm max} = 5000$ for cosmic shear and $\ell_{\rm max} = 3000$ for photometric angular clustering, and galaxy-galaxy lensing. The corresponding \fom obtained for each sample can be found at \cref{tab:FoM}. We show, as a zoom-in, the 2-dimensional posterior distribution corresponding to the dark energy parameters \wo and \wa.}
\label{fig:triangle_WL_GCsp_3x2pt_GCsp_w0waCDM}
\end{figure*}
The results of our hypothesis testing cases can be found in \cref{fig:2D_3x2pt_projection_effects}, where we show the corresponding two-dimensional posterior distributions for key cosmological parameters, as well as the degeneracy between some cosmological parameters and the per-bin redshift bin shift parameters.\footnote{We illustrate the results with the redshift bin shift for tomographic bin 6, $\Delta z_{\rm L}^{(6)}$ as in \cref{fig:nz}. However, the same behaviour is found in all 13 redshift bin shift means.} We conclude that the projection effects disappear when all nuisance parameters are kept fixed and the \bbn prior is still imposed (see grey filled contours shown in the upper row of \cref{fig:2D_3x2pt_projection_effects}). Although not explicitly shown, the disappearance of projection effects is also found when all nuisance parameters are kept free except for the per-bin redshift bin shift mean parameters and when the \bbn prior is still imposed within the analysis. On the other hand, when all nuisance parameters are kept open, but a flat prior is imposed on $\Ob\,h^2$, there are no projection effects (see red unfilled contours shown in the upper row of \cref{fig:2D_3x2pt_projection_effects}). 
We conclude that projection effects are therefore a consequence of the degeneracy between $\Ob\,h^2$ and all the 26 per-bin redshift bin shift mean parameters (for both cosmic shear and angular clustering), as observed in the lower left panel of \cref{fig:2D_3x2pt_projection_effects}. The \bbn prior imposes tight one-dimensional constraints on $\Ob\,h^2$, allowing more freedom for $\Delta z_{\rm L}^i$ to explore a broader parameter-value range within its Gaussian prior, despite the two-dimensional volume (light red area within the contour shown in the lower left panel of \cref{fig:2D_3x2pt_projection_effects}) being larger. \\

\subsection{Full analysis: 3\texttimes2pt + GCsp}\label{sec:results_full}

Combining photometric and spectroscopic probes unlocks the full potential of \Euclid's primary probes (see \cref{fig:2D_3x2pt_GCsp}, and confidence intervals for all the cosmological and nuisance parameters for the \wowaCDM model in \cref{app:CL}). The photometric primary probes alone (3\texttimes2pt) tightly constrain \Om and \sotto, while spectroscopic galaxy clustering primarily constrains \lnAs, $H_0$ and \OK. For this analysis, we assume no correlation between spectroscopic and photometric probes \citep{EP-Paganin}. The inclusion of GCsp data alongside the 3\texttimes2pt analysis significantly enhances constraints on several cosmological parameters and degeneracies are significantly reduced, particularly those related to the shape and amplitude and shape of the matter power spectrum. 

The full one and two-dimensional posterior distributions for the cosmological parameters given the \wowaCDM flat model for spectroscopic galaxy clustering, 3\texttimes2pt and the full combination are shown in \cref{fig:triangle_WL_GCsp_3x2pt_GCsp_w0waCDM}. Including spectroscopic galaxy clustering leads to a marked improvement in the confidence intervals for all cosmological parameters compared to the 3\texttimes2pt case. This includes tighter constraints on \Ob (where the posterior becomes more informative than the BBN prior), \Om, $H_0$ and \ns. Notably, the uncertainty on $\Oc\,h^2$ improves by approximately 53\%, highlighting the strong sensitivity of galaxy clustering to the matter density. Similarly, the scalar spectral index $n_{\rm s}$ and the amplitude parameter $\ln(10^{10} A_s)$ see relative improvements of 43\% and 35\%, respectively. The Hubble parameter $h$ and the baryon density $\Oc\,h^2$ also benefit from tighter constraints, with reductions in uncertainty of about 32\% and 30\%. These gains demonstrate the complementarity of GCsp with weak lensing and galaxy-galaxy lensing, especially in breaking degeneracies and enhancing sensitivity to early-universe parameters (see \cref{tab:CL_cosmo}). The latter improvement on \ns arises from fixing several EFT parameters, as detailed in \cref{sec:results_GCsp}. 

The full combination further constrains the sampled EFTofLSS spectroscopic nuisance parameters (see \cref{app:nuisance}, in particular Figs.~\ref{fig:triangle_w0waCDM_nuisance5} and \ref{fig:triangle_w0waCDM_nuisance6}) by breaking degeneracies with the cosmological parameters. In contrast, the spectroscopic galaxy clustering probe does not significantly improve constraints on the photometric nuisance parameters (see Figs.~\ref{fig:triangle_w0waCDM_nuisance2} and \ref{fig:triangle_w0waCDM_nuisance5}), as these are already dominated by their priors -- specifically the photometric multiplicative bias and the per-bin redshift parameters given by \Euclid science requirements. Best-fits of nuisance parameters for the \wowaCDM flat model are found in \cref{tab:CL_nuis_photo} and \cref{tab:CL_nuis_spectro}. 

In our analysis, we find that certain cosmological parameters exhibit significant correlations with nuisance parameters, which can impact the robustness of cosmological inferences. Notably, $\Oc\,h^2$ shows strong negative correlations with several nuisance parameters associated with spectroscopic calibration, in particular the scale-independent shot noise terms, {with values ranging} around $-0.8$. Additionally, the spectral index $n_{\rm s}$ itself is positively correlated with these parameters with values close to $-0.7$. These correlations, with absolute values often exceeding 0.7, indicate that marginalising over nuisance parameters can substantially affect constraints on key cosmological quantities. Such findings underscore the importance of careful modelling and mitigation of systematics in cosmological analyses to avoid biased parameter estimation, as well as the importance of priors in these nuisance parameters.

We assess model selection for \Euclid joint analyses of cosmological models using the Bayesian evidence $\ln \evid$, provided by \nautilus, where higher values indicate preferred models (see \cref{tab:computational_resources}). Across all probes, the flat \LCDM model consistently yields the highest evidence values, serving as the reference for model comparison. Extensions such as the inclusion of a time-varying dark energy EoS (\wowaCDM), non-flat geometries, or modifications to the growth index (\gammag) lead to lower $\ln \evid$ values, reflecting a statistical penalty for the increased model complexity. Nevertheless, within the \wowaCDM family, the differences in $\ln \evid$ between the flat and non-flat models, or between those with and without \gammag, remain relatively small -- typically $\Delta\ln \evid \lesssim 2$ -- indicating that these extensions are not strongly disfavoured. This is consistent with \cref{fig:2D_3x2pt_GCsp}, which shows that the inclusion of non-flatness or modified gravity does not significantly affect the constraints on $(w_0, w_a)$. This result is encouraging in the context of \Euclid, as it demonstrates the capability of the survey to constrain multiple \LCDM extensions simultaneously, rather than needing to isolate them due to strong degeneracies.

The \fom values obtained for \wo and \wa using the full combination of probes (see \cref{tab:FoM}, last column) are consistent with the results presented in \citet{Blanchard-EP7}. As in previous cases, our findings align with the predictions for the pessimistic scenario discussed in their study, despite adopting a higher maximum photometric multipole for 3\texttimes2pt in our analysis. This result can primarily be attributed to the extensive parameter space sampled in the current analysis, which spans 58 to 61 dimensions for the full statistical sampling exercise. Yet, we recover a \fom value for \wo and \wa above 400 for the flat \wowaCDM model, in agreement with the scientific performance specifications presented in the \Euclid definition study report \citep{Laureijs11}.

Altogether, the combination of all \Euclid primary probes underscores that the mission will set the future benchmark for multi-probe large-scale structure cosmological analyses. Its constraining power on cosmological parameters is comparable to, or potentially surpasses, that of \Planck. This conclusion holds robustly across all studied cosmological models, extending beyond the flat \wowaCDM scenarios.

\section{Conclusions}\label{sec:conclusions}
In this study, we have presented a comprehensive analysis of the forecast results for \Euclid's primary probes across various cosmological models using \CLOE. We have demonstrated that \CLOE, when used with \cobaya and \nautilus, is able to produce robust forecasts that are easily reproducible by using Open Science tools such as \texttt{ESA datalabs}.

Our findings demonstrate the complementarity of weak lensing and spectroscopic galaxy clustering in constraining key cosmological parameters. We observed that, while WL provides strong constraints on \Om and \sotto, GCsp offers enhanced precision on primordial parameters such as \lnAs and on the geometrical curvature of the Universe \OK. The 3\texttimes2pt combination of photometric probes further enhances the constraining power of \Euclid, increasing the figure of merit for dark energy parameters by an order of magnitude compared to the single probes WL and GCsp. It also achieves at least an order of magnitude improvement in the uncertainty constraints for \sotto and \Om, and an order of magnitude improvement with respect to current galaxy surveys such as DES and KiDS.

We have demonstrated that the combination of angular clustering and galaxy-galaxy lensing provides most of the cosmological constraining power in the photometric 3\texttimes2pt analysis when compared with shear only. This conclusion is reached after studying the behaviour of GCph with low scale cuts. This highlights that further refinements in the theoretical modelling of galaxy clustering, including non-linear scales and systematic effects (visibility), are essential to reduce systematic uncertainties and improve the reliability of \Euclid's primary probes, thereby unlocking the full potential of the mission.

We have shown that model selection critically influences the interpretation of \Euclid data, as different cosmological models impact posterior distributions and parameter degeneracies across all probes. Using Bayesian evidence $\ln \evid$ computed with \nautilus (see \cref{tab:computational_resources}), the flat \LCDM model consistently emerges as the preferred baseline. Although extensions such as \wowaCDM, non-flat geometries, and modified growth index (\gammag) face a statistical penalty for complexity, their evidence differences remain small, namely $\Delta \ln \evid \lesssim 2$, indicating that they are not strongly disfavoured. This, along with minimal impact on $(w_0, w_a)$ constraints, underscores \Euclid's ability to simultaneously constrain multiple \LCDM extensions without significant degeneracies. \corr{In any case, we note that our model selection results are derived under the assumption that the likelihood peaks exactly at \LCDM, the model for which the synthetic data vectors were generated. In practice, real survey and noise realisations may yield fluctuations that occasionally favour extended models such as \wowaCDM, even if the ensemble average still prefers \LCDM. This effect, which is not captured in our current forecasts, is expected to moderate the decisive evidence differences reported here and will need to be carefully reassessed with real \Euclid data}.

Additionally, the sheer scale of the parameter space explored in this work has highlighted the importance of developing a robust strategy to secure and manage the computational resources needed to carry out a study of such a scale. In fact, the complexity of the models and the large number of parameters require significant computational power, which must be efficiently managed to ensure timely and accurate results. Our analysis also points to the need for exploring alternative sampling methods. The traditional approaches may struggle with the high dimensionality and non-Gaussian posteriors encountered in some of these analyses, which will, most likely, become more important when non-standard cosmological models are explored. Employing more sophisticated sampling techniques, such as Simulation-Based Inference \citep{vonWietersheim-Kramsta:2024cks, DES:2024xij, Franco-Abellán_2024} and Hamiltonian Monte Carlo \citep{2024OJAp....7E..10B,Ruiz-Zapatero:2023hdf, 2023OJAp....6E..20P}, could offer improved convergence and accuracy in parameter estimation, particularly for scenarios with a large number of free parameters.

Finally, our work suggests that achieving a realistic \fom greater than 400 requires the full combination of \Euclid's primary probes: 3\texttimes2pt and GCsp. Yet, while 2-point summary statistics of \Euclid's primary probes provide valuable insights on the dark energy parameters \wo and \wa, incorporating information from \Euclid higher-order statistical probes and external cosmological data sets could substantially enhance the overall statistical constraining power on a given cosmological model and reduce degeneracies among parameters. This will be crucial in pushing the boundaries of precision cosmology and extracting the maximum amount of information from future data releases. 

\section{Data availability}
The analysis scripts and notebooks can be found at the corresponding GitHub repository\footnote{\url{https://github.com/gcanasherrera/Forecasting-Euclid}}, which also provides a direct link to the associated Zenodo archive\footnote{\url{https://doi.org/10.5281/zenodo.18610864}} where all the results produced in this work are publicly available.

\begin{acknowledgements}
\AckEC
GCH thanks J. U. Lange for useful exchanges about the performance of the \nautilus sampler. GCH acknowledges support through the ESA research fellowship programme. During part of this work, AMCLB was supported by a Paris Observatory-PSL University Fellowship, hosted at the Paris Observatory. SC acknowledges support from the Italian Ministry of University and Research (\textsc{mur}), PRIN 2022 `EXSKALIBUR – Euclid-Cross-SKA: Likelihood Inference Building for Universe'DESY3s Research', Grant No.\ 20222BBYB9, CUP D53D2300252 0006, and from the European Union -- Next Generation EU.
SD acknowledges support from the Italian Ministry of University and Research (\textsc{mur}), PRIN 2022 `LaScaLa - Large Scale Lab', Grant No.\ 20222JBEKN, CUP I53D23000800006 founded by the European Union -- Next Generation EU.
SP acknowledges support through the \textit{Conception Arenal Programme} of the Universidad de Cantabria and funding from the proiect UC-LIME
(PID2022-140670NA-I00), financed by MCIN AEI/ 10.13039/501100011033/FEDER, UE. DNG acknowledges support from the European Research Council (ERC) under the European Union’s Horizon 2020 research and innovation program with Grant agreement No. 101053992.
%%%%%%%%%%
% COMPUTATIONAL ACKNOWLEDGEMENTS
%%%%%%%%%%
The computations were performed at the University of Geneva using Baobab and Yggdrasil HPC service. This work was performed using the compute resources from the Academic Leiden Interdisciplinary Cluster Environment (ALICE) provided by Leiden University and the Xmaris cluster provided by the Lorentz Institute of Theoretical Physics. We acknowledge EuroHPC Joint Undertaking for awarding us access to MeluXina at LuxProvide, Luxembourg. The runs were performed on the Luxembourg national supercomputer MeluXina. The authors gratefully acknowledge the LuxProvide teams for their expert support. \texttt{ketelmeer} is funded thanks to the European Research Council (ERC) under the European Union’s Horizon 2020 research and innovation program (Grant Agreement No. 101053992). This work has made use of the Infinity Cluster hosted by Institut d'Astrophysique de Paris. This project was provided with computing HPC and storage resources by GENCI at CINES thanks to the grant 2024-AD010415184 on the supercomputer Adastra's GENOA partition. This work was granted access to the HPC resources of MesoPSL financed by the Region Ile-de-France and the project Equip@Meso (reference ANR-10-EQPX-29-01) of the programme Investissements d’Avenir supervised by the Agence Nationale pour la Recherche. Part of this work was carried out using the Feynman cluster of the Institut de recherche sur les lois fondamentales de l'Univers (Irfu) at CEA Paris-Saclay. Computations were performed with computing resources granted by RWTH Aachen University under project rwth1304.
%%%%%%%%%%
% WORKSHOP
%%%%%%%%%%
The authors acknowledge the contribution of the Lorentz Center (Leiden), and of the European Space Agency (ESA), where the workshop "Making \CLOE shine" and the "\CLOE workshop 2023" were held.

\end{acknowledgements}

\bibliography{Euclid}

\begin{appendix}
  \onecolumn %If you don't want single column for the Appendix, please
             %comment this out

\section{Fiducial cosmology}\label{app:fiducial_cosmo}
In this appendix, we describe the reference parameters adopted for producing the synthetic data vectors and illustrate the resulting galaxy–galaxy lensing (XC) probe.

\begin{table*}[h!]
\renewcommand{\arraystretch}{1.18}
\caption{Reference values and prior distributions for the cosmological and nuisance parameters for the models listed in \cref{sec:models}.}
\centering
\begin{tabularx}{\textwidth}{>{\raggedright\arraybackslash}m{7cm}m{2.5cm}>{\raggedright\arraybackslash}m{4.cm}m{3.cm}}%{lXXX}
\multicolumn{2}{l}{\textbf{Parameters}} & \textbf{Fiducial value} & \textbf{Prior} \\
\hline
\hline
\multicolumn{4}{c}{Cosmology} \\
\hline
Dimensionless Hubble constant & $h$ & $0.6737$ & $\mathcal U(0.55, 0.91)$ \\
Present-day physical baryon density & $\Ob\,h^2$ &  $0.0227$ & $\mathcal N(0.0227, 0.00038)$\\
Present-day physical cold dark matter density & $\Oc\,h^2$ & $0.1219$ & $\mathcal U(0.01, 0.37)$\\
Dark energy equation-of-state parameters & $\{\wo,\wa\}$ & $\{-1,0\}$ & $\{\mathcal U(-3.0, -0.5),$ $\mathcal U(-3.0, 3.0)\}$\\
Slope of primordial curvature power spectrum  & \ns & $0.966$ & $\mathcal U(0.87, 1.07)$ \\
Amplitude of the primordial curvature power spectrum  & $\ln(10^{10}\,\As)$ & 3.04 & $\mathcal U(1.6, 3.9)$ \\
Growth index & \gammag & $0.545$ & $\mathcal U(0.01, 1.1)$ \\
Present-day curvature density & \OK & $0.0$ & $\mathcal U(-0.1, 0.1)$ \\
Baryonic feedback efficiency factor of the HMCode emulator & $\log_{10}(T_{\rm AGN}/\mathrm{K})$ & $7.75$ & $\mathcal N(7.75, 0.17825)$ \\
\hline
\hline
\multicolumn{4}{c}{Photometric sample} \\
\hline
Amplitude of intrinsic alignments & \aIA &  $0.16$ & $\mathcal U(-2, 2)$\\
Power-law slope of intrinsic alignment redshift evolution & \etaIA &  $1.66$ & $\mathcal U(0.0, 3.0)$\\
Coefficients of cubic polynomial for clustering bias & $b_{\mathrm{G}, \,i\in\{0, 3\}}$ & $\{1.33291,$ $-0.72414,$ $1.01830,$ $-0.14913\}$ & $\mathcal U(-3, 3)$\\
Coefficients of cubic polynomial for magnification bias & $b_{\mathrm{mag}, \,i\in\{0, 3\}}$ & $\{-1.50685,$ $1.35034,$ $0.08321,$ $0.04279\}$ & $\mathcal U(-3, 3)$\\
Per-bin shear multiplicative bias & $m_{\rm L}^{i\in\{1,13\}}$ & $0.0$ & $\mathcal N(0.0, 0.0005)$\\
Per-bin mean redshift shift  & $\Delta z_{\rm L}^{i\in\{1,13\}}$ & $\{-0.025749,$
 $0.022716,$
 $-0.026032,$
 $0.012594,$
 $0.019285,$
 $0.008326,$
 $0.038207,$
 $0.002732,$
 $0.034066,$
 $0.049479,$
 $0.066490,$
 $0.000815,$
 $0.049070\}$ & $\mathcal N\left[z_i^{\rm fid}, 0.002\,(1+z_i^{\rm fid})\right]$\\
\hline
\hline

\multicolumn{4}{c}{Spectroscopic sample} \\
\hline
Per-bin linear bias & $b^1_{\rm G,\,i\in\{1,4\}}$& $\{1.412,$ $1.769,$ $2.039,$ $2.496\}$ & $\mathcal U(1.0, 3.0)$\\
Per-bin quadratic bias & $b^2_{\rm G,\,i\in\{1,4\}}$& $\{0.695,$ $0.870,$ $1.162,$ $2.010\}$ & $\mathcal U(-5.0, 5.0)$\\
Per-bin non-local quadratic bias & $b_{{\rm G}_2,\,i\in\{1,4\}}$ & $\{-0.156,$ $-0.299,$ $-0.400,$ $-0.555\}$ & Derived using ex-set relation\\
Per-bin non-local cubic bias & $b_{\Gamma_3,\,i\in\{1,4\}}$ & $\{0.323,$ $0.621,$ $0.827,$ $1.137\}$ & Derived using coevolution\\
Scale-independent shot noise & $\alpha_{\rm P}^{i\in\{1,4\}}$ & $\{0.056,$ $0.152,$ $0.144,$ $0.309\}$ & $\mathcal U(-1.0, 2.0)$\\
Scale-dependent shot noise $[k^2\,\mu^0]$ $[(\mathrm{Mpc} / h)^2]$ & $\alpha_{\rm P,2}^{i\in\{1,4\}}$ & $\{0.0,$ $0.0,$ $0.0,$ $0.0\}$ & Derived\\
Scale-dependent shot noise $[k^2\,\mu^2]$ $[(\mathrm{Mpc} / h)^2]$ & $\alpha_{\rm P,3}^{i\in\{1,4\}}$ & $\{0.0,$ $0.0,$ $0.0,$ $0.0\}$ & Derived\\
Per-bin leading-order counter-term $[k^2\,\mu^0]$ $[(\mathrm{Mpc} / h)^2]$ & $c_{0}^{i\in\{1,4\}}$ & $\{11.603,$ $14.475,$ $15.667,$ $26.413\}$ & Derived\\
Per-bin leading-order counter-term $[k^2\,\mu^2]$ $[(\mathrm{Mpc} / h)^2]$ & $c_{2}^{i\in\{1,4\}}$ & $\{35.986,$ $44.914,$ $43.819,$ $62.353\}$ & Derived \\
Per-bin leading-order counter-term $[k^2\,\mu^4]$ $[(\mathrm{Mpc} / h)^2]$ & $c_{4}^{i\in\{1,4\}}$ & $\{56.943,$ $55.443,$ $44.214,$ $42.89\}$ & Derived \\
Per-bin next-to-leading-order counter-term $[k^4]$ $[(\mathrm{Mpc} / h)^4]$ & $c_{{\rm nlo}}^{i\in\{1,4\}}$ & $\{0.0,$ $0.0,$ $0.0,$ $0.0\}$ & Derived \\
Per-bin purity factor (assuming Poisson distributed interlopers) & $f_{{\rm out}}^{i\in\{1,4\}}$ & $\{0.195,$ $0.204,$ $0.306,$ $0.121\}$ & $\mathcal N\left(f_{{\rm out},\,i}^{\rm fid}, 0.01\right)$ \\
\hline
\end{tabularx}
\justifying \footnotesize{Note: These fiducial values are used to generate the synthetic data in \cref{sec:synthetic_data}. For photometric nuisance parameters, we apply a polynomial fit for both galaxy bias $b_{\rm {G},i}$ and magnification bias $b_{\mathrm{mag},i}$ with coefficients for $i = 0$ to 3. We use a constant multiplicative bias $m_{\rm L}^i$ and redshift bin shifts $\Delta z_{\rm L}^i$ per bin, with fiducials from the \Euclid Flagship Simulations 2. Spectroscopic nuisance parameters have one per bin; some are fixed to fiducials or physically motivated relations as in Sect. \ref{subsec:theoretical_specs}. Sampled parameters have priors that are either \textit{uniform} $\mathcal{U}(\text{min},\text{max})$ or \textit{Gaussian} $\mathcal{N}(\mu,\sigma)$.}
\label{tab:fiducial_model}
\end{table*}

\begin{figure*}[htbp!]
\centering
\includegraphics[width=\textwidth]{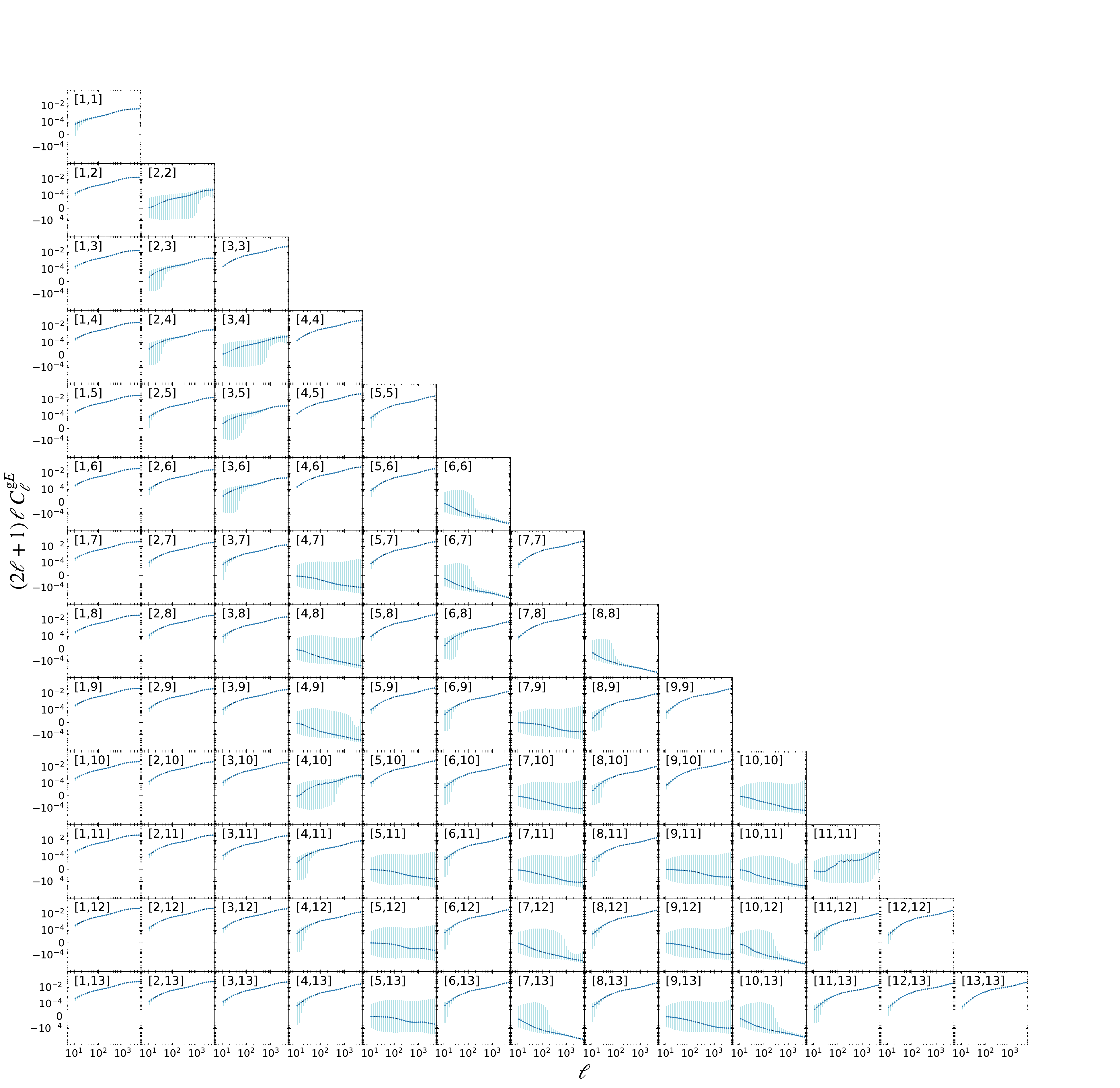}
\caption{Similar to \cref{fig:CLOE_euclid_probes_WL}  but showing the cross-correlation {galaxy-galaxy lensing XC probe} $\cl{\ell}[ij][gE]$. }
\label{fig:CLOE_euclid_probes_XC}
\end{figure*}

\section{Computational resources}\label{app:computational_resources}

\begin{table*}[h!]
\caption{Summary of the computational resources used during the forecast exercise using the sampler \nautilus, for all the observational probes and cosmological models.}
\centering
\begin{tabularx}{\textwidth}{l S S l l l S}

\multicolumn{2}{l}{\textbf{Cosmology}} & \textbf{Size} & \textbf{$\ln\evid$} & \textbf{Cluster} & \textbf{Wall-Time} & \textbf{CPUs}\\
\hline
\hline
%%%%%%%%%%%%%%%%%%%%%%%%%%%%%%%%%%%
\multicolumn{7}{c}{Weak Lensing (WL)} \\
\midrule
\LCDM (flat) &  & 34  & $-29.52$ & \texttt{yggdrasil} & 2 days, 1 hour, 42 minutes  & 24 \\
\LCDM + \gammag (flat) & & 35 & $-31.74$ & \texttt{yggdrasil} & 2 days, 11 hours, 14 minutes  & 24 \\
\LCDM (non-flat) & & 35 & $-30.47$ & \texttt{yggdrasil} & 2 days, 10 hours, 39 minutes  & 24 \\
\wowaCDM (flat) & & 36 & $-33.70$ & \texttt{yggdrasil} & 7 days, 2 hours, 32 minutes & 24 \\
\wowaCDM + \gammag (flat) & & 37 & $-34.81$ & \texttt{yggdrasil} & 3 days, 8 hours, 21 minutes  & 24\\
\wowaCDM (non-flat) & & 37 & $-34.07$ & \texttt{yggdrasil} & 7 days, 16 hours, 15 minutes  & 24 \\
\midrule
%%%%%%%%%%%%%%%%%%%%%%%%%%%%%%%%%%%
\multicolumn{7}{c}{Galaxy Clustering (GCsp)} \\
\midrule
\LCDM (flat) &  & 21 & $-80.67$ & \texttt{yggdrasil} & 5 days, 18 hours, 24 minutes  & 24 \\
\LCDM + \gammag (flat) &  & 22 & $-81.88$ & \texttt{yggdrasil} & 6 days, 23 hours, 8 minutes  & 24 \\
\LCDM (non-flat) &  & 22 & $-82.20$ & \texttt{yggdrasil} & 7 days, 12 hours, 41 minutes  & 24 \\
\wowaCDM (flat) &  & 23 & $-85.43$ & \texttt{yggdrasil} & 9 days, 14 hours, 37 minutes  & 24 \\
\wowaCDM + \gammag (flat) &  & 23 & $-86.29$ & \texttt{yggdrasil} & 15 days, 8 hours, 31 minutes  & 24 \\
\wowaCDM(non-flat) &  & 24 & $-86.07$ & \texttt{yggdrasil} & 12 days, 6 hours, 58 minutes  & 24 \\
\midrule
%%%%%%%%%%%%%%%%%%%%%%%%%%%%%%%%%%%
\multicolumn{7}{c}{3\texttimes2pt} \\
\midrule
\LCDM (flat) &  & 42 & $-124.69$ & \texttt{gouwezee} & 7 days, 7 hours, 45 minutes  & 50 \\
\LCDM (flat, no \bbn prior) &  & 42 & $-127.32$ & \texttt{ketelmeer} & 4 days, 21 hours, 0 minutes  & 50 \\
\LCDM (flat, fixed nuisance) &  & 6 & $-20.19$ & \texttt{gouwezee} & 8 hours, 21 minutes  & 50 \\
\LCDM (flat, fixed systematic nuisance) &  & 16 & $-82.83$ & \texttt{ketelmeer} & 3 days, 2 hours, 35 minutes  & 24 \\
\LCDM + \gammag (flat) &  & 43 & $-128.06$ & \texttt{eemmeer} & 9 days, 4 hours, 17 minutes   & 50 \\
\LCDM (non-flat) &  & 43 & $-127.09$ & \texttt{xmaris} & 10 days, 12 hours, 52 minutes  & 50 \\
%%%%%%%%%%%%
\wowaCDM (flat) &  & 44 & $-131.59$ & \texttt{amsteldiep} & 11 days, 16 hours, 29 minutes  & 50 \\
%\wowaCDM (flat, \polychord) &  & 44 & $??$ & \texttt{infinity} & ???  & ?? \\

\wowaCDM (flat, low GCph scale cuts) &  & 44 & $-107.44$ & \texttt{baobab} & 5 days, 6 hours, 16 minutes  & 50 \\
\wowaCDM (flat, fixed nuisance) &  & 8 & $-35.77$ & \texttt{gouwezee} & 1 day, 16 hours, 50 minutes  & 50 \\
\wowaCDM (flat, fixed systematic nuisance) &  & 18 & $-90.32$ & \texttt{ketelmeer} & 2 days, 15 hours, 43 minutes  & 50 \\
%%%%%%%%%%%%
\wowaCDM + \gammag (flat) &  & 45 & $-134.82$ & \texttt{markermeer} & 15 days, 5 hours, 19 minutes  & 50 \\
\wowaCDM (non-flat) &  & 45 & $-133.69$ & \texttt{eemmeer} & 21 days, 1 hours, 6 minutes  & 50 \\
\midrule
%%%%%%%%%%%%%%%%%%%%%%%%%%%%%%%%%%%
\multicolumn{7}{c}{2\texttimes2pt} \\
\midrule
%%%%%%%%%%%%
\wowaCDM (flat, low GCph scale cuts) &  & 
44 & $-104.48$ & \texttt{baobab} & 5 days, 2 hours, 27 minutes  & 50 \\
%%%%%%%%%%%%
\midrule
%%%%%%%%%%%%%%%%%%%%%%%%%%%%%%%%%%%
\multicolumn{7}{c}{3\texttimes2pt + GCsp} \\
\midrule
\LCDM (flat) &  & 58 & $-192.02$ & \texttt{meluxina} & 19 days, 2 hours, 32 minutes  & 100 \\
\LCDM + \gammag (flat) &  & 59 & $-196.26$ & \texttt{meluxina} & 9 days, 3 hours, 26 minutes  & 100 \\
\LCDM (non-flat) &  & 59 & $-195.46$ & \texttt{baobab} & 11 days, 19 hours, 58 minutes  & 100 \\
\wowaCDM (flat) &  & 60 & $-200.38$ & \texttt{ketelmeer} & 9 days, 11 hours, 39 minutes  & 100 \\
\wowaCDM + \gammag (flat) &  & 61 & $-202.58$ & \texttt{baobab} & 24 days, 14 hours, 53 minutes  & 100 \\
\wowaCDM (non-flat) &  & 61 & $-201.08$ & \texttt{baobab} & 28 days, 13 hours, 9 minutes  & 100 \\
\bottomrule
\end{tabularx}
\justifying \footnotesize{Note: We specify the size of the parameter space, the value of the log-evidence $\ln \evid$ when the run converges, the name of the cluster, the total computational wall running time and the number of CPUs used to parallelise the likelihood calls (pool multithreading). On top of these resources, if available (i.e: machines used to run 3\texttimes2pt and 3\texttimes2pt + GCsp), another 16 or 24 CPUs in total were employed to speed up the \camb transfer function calculation when necessary. If not specified, the scale-cuts used for GCph and XC were $\ell = 3000$. If specified as \textit{low GCph scale cuts}, they correspond to $\ell = 750$ for GCph and XC as listed in \cref{tab:theoretical_probes}.}
\label{tab:computational_resources}
\end{table*}

The computational effort required for this research work is considerable, reflecting the complexity and scale of cosmological modelling and data analysis involved. \corr{The main bottleneck in the calculation arises from the theory calculations (approximately 2 seconds for both 3\texttimes2pt and spectroscopic galaxy clustering full-shape analyses, including the Boltzmann solver call to obtain the matter power spectrum) and from sampling across the large parameter space}. In total, this paper utilises over half of a million CPU hours for a variety of cosmological models and cosmological probes, such as weak lensing (WL), spectroscopic galaxy clustering (GCsp), and the combination of different data sets (3\texttimes2pt analysis, and 3\texttimes2pt + GCsp). Details on the computational resources used for each exercise using the \nautilus sampler can be found at \cref{tab:computational_resources}. The clusters \texttt{gouwezee}, \texttt{ketelmeer}, \texttt{eemmeer}, \texttt{amsteldiep}, and \texttt{markermeer} are located at Leiden Observatory (The Netherlands). The clusters \texttt{yggdrasil} and \texttt{baobab} are hosted by the University of Geneva (Switzerland). The cluster \texttt{xmaris} (now part of the cluster \texttt{Alice}) is located at Leiden University (The Netherlands). The cluster \texttt{meluxina} was accessed through a EuroHPC Joint Undertaking grant at LuxProvide (Luxembourg).

For weak lensing alone, the computational cost reaches nearly 22\,000 CPU hours. Galaxy clustering analyses needed about 50\,000 CPU hours. The 3\texttimes2pt analysis, combining weak lensing, galaxy clustering, and galaxy-galaxy lensing for all the different tests carried out in this research represents the second-to-first most computationally demanding part of the research, summing up over 176\,000 CPU hours. Finally, the most computationally expensive case is the full \Euclid analysis combining 3\texttimes2pt with galaxy clustering, reaching up to 247\,000 CPU hours. Overall, the results indicate that the more complex the parameter space -- whether due to higher dimensionality or stronger degeneracies -- the longer the computational time required for sampling, although specific scalability tests might be indicated for future studies. If this forecast exercise had been done with \polychord, the total computational cost would have been roughly 2 million CPU hours. We found that the MH MCMC algorithm in \cobaya is not suitable for sampling such a large parameter space.

\section{Nuisance-parameter posterior distributions}\label{app:nuisance}
\begin{figure*}[h!]
\centering
\includegraphics[width=0.942\textwidth]{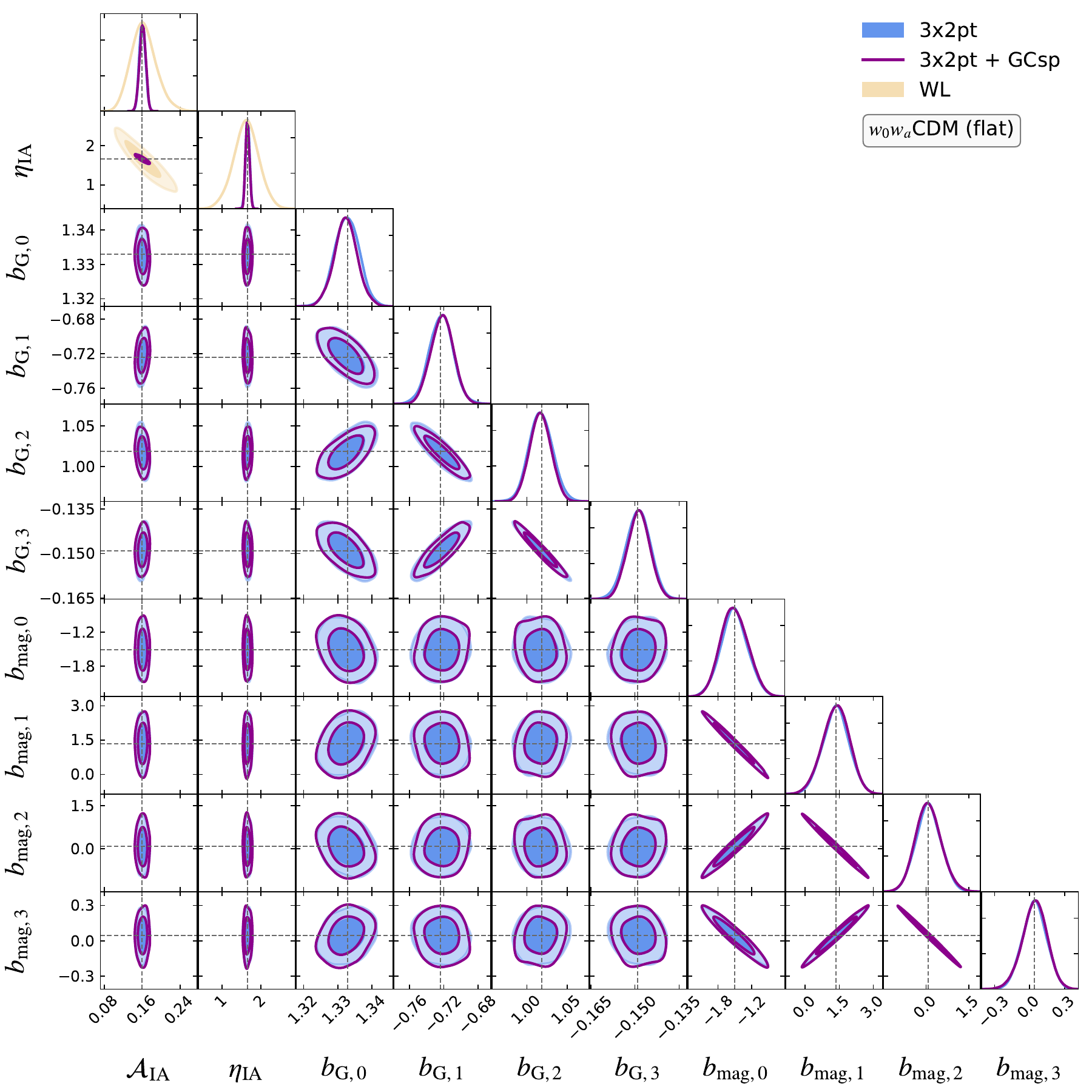}
\caption{Forecast of the constraints for the photometric nuisance parameters of the \wowaCDM cosmological model (adopting a flat geometry) using \Euclid photometric and spectroscopic probes: cosmic shear (WL), cosmic shear, galaxy-galaxy lensing, and angular clustering (3\texttimes2pt), and the combination of both of them as described in \cref{sec:results_full} (3\texttimes2pt + GCsp). For the photometric probes, we used $\ell_{\rm max} = 5000$ for cosmic shear and $\ell_{\rm max} = 3000$ for photometric angular clustering, and galaxy-galaxy lensing. We show, as zoom-in, the 2-dimensional posterior distribution corresponding to the intrinsic alignment parameters \aIA and \etaIA.}
\label{fig:triangle_w0waCDM_nuisance1}
\end{figure*}
\newpage

In this section, we present the posterior distributions of all the sampled and some derived nuisance parameters for WL, 3\texttimes2pt, and GCsp within the \wowaCDM cosmological model, including single probes, the combined 3\texttimes2pt analysis, and the joint 3\texttimes2pt + GCsp analysis. By providing the full set of posteriors, we aim to ensure transparency and enable reproducibility of our results in future work. Additionally, this comprehensive presentation highlights the complexity and vast dimensionality of the parameter space sampled in our analyses. Definitions of each parameter are detailed in \cref{tab:fiducial_model}. If a Gaussian prior was used, this is showcased as example.

\begin{figure*}
\centering
\includegraphics[width=1.\textwidth]{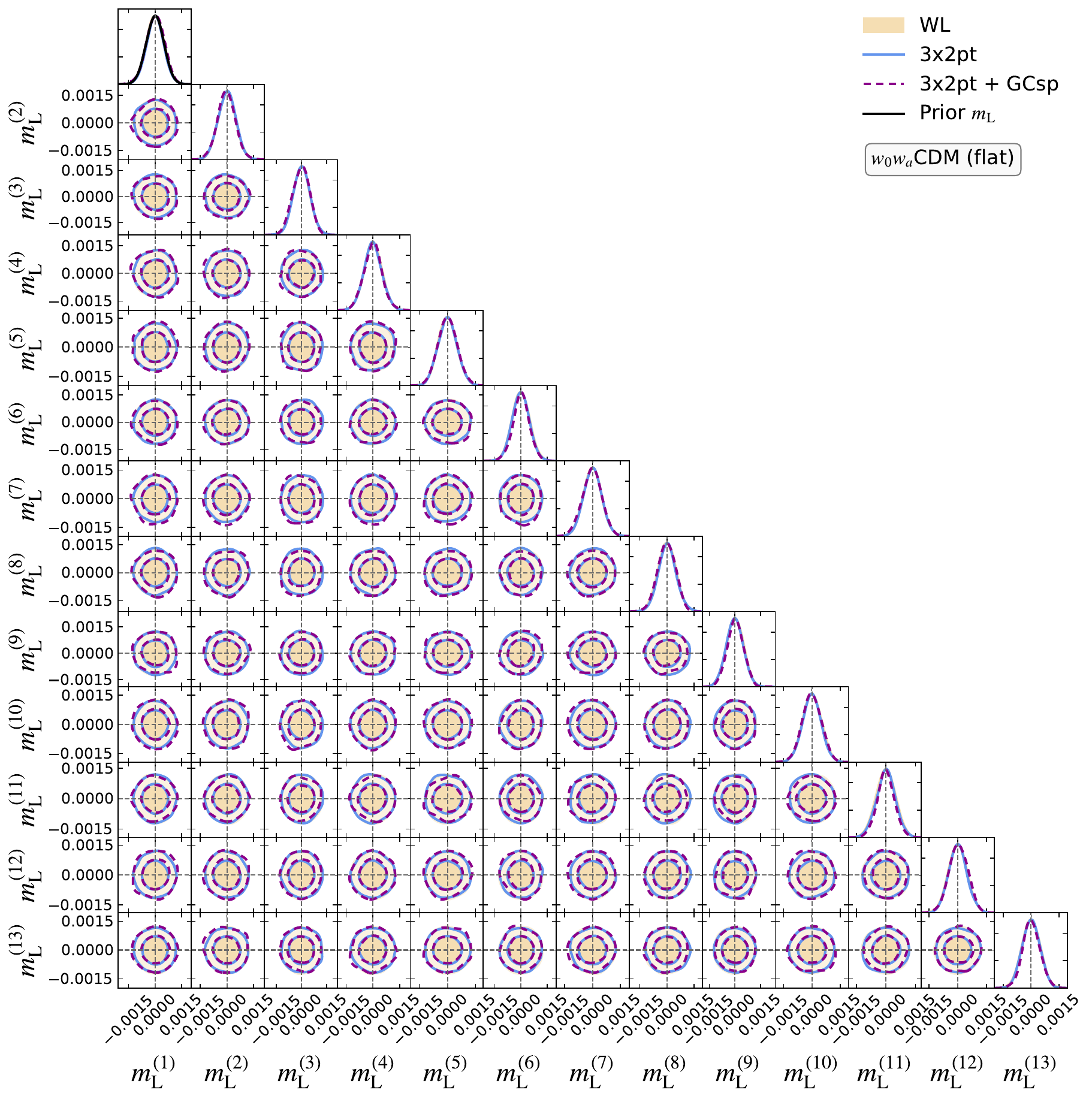}
\caption{Forecast of the constraints for the photometric nuisance parameters of the \wowaCDM cosmological model (adopting a flat geometry) using \Euclid photometric and spectroscopic probes: cosmic shear (WL), cosmic shear, galaxy-galaxy lensing, and angular clustering (3\texttimes2pt), and the combination of both of them as described in \cref{sec:results_full} (3\texttimes2pt + GCsp). For the photometric probes, we used $\ell_{\rm max} = 5000$ for cosmic shear and $\ell_{\rm max} = 3000$ for photometric angular clustering, and galaxy-galaxy lensing.}
\label{fig:triangle_w0waCDM_nuisance2}
\end{figure*}
\newpage

\begin{figure*}
\centering
\includegraphics[width=1.\textwidth]{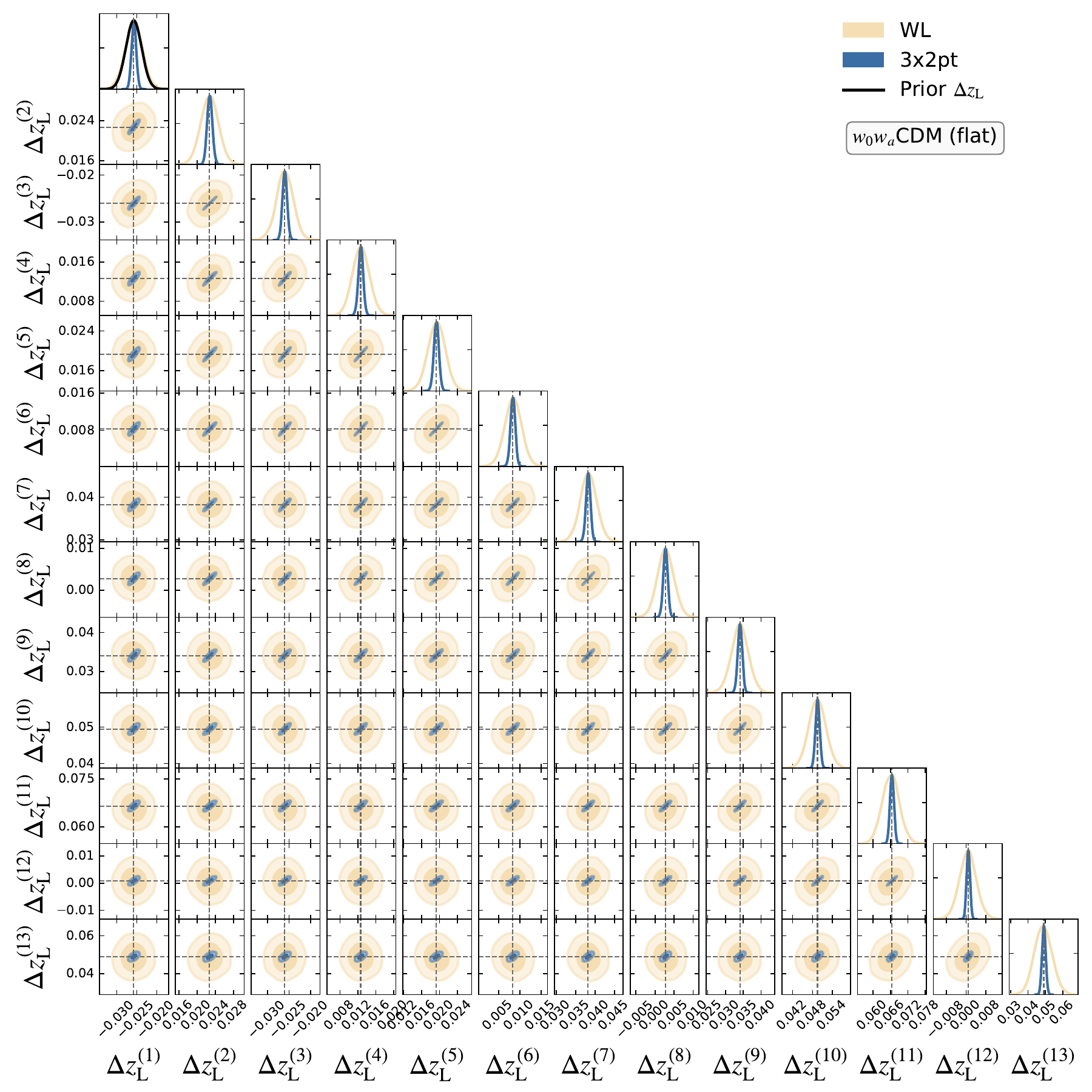}
\caption{Forecast of the constraints for the photometric nuisance parameters of the \wowaCDM cosmological model (adopting a flat geometry) using \Euclid photometric probes: cosmic shear (WL), and cosmic shear, galaxy-galaxy lensing and angular clustering (3\texttimes2pt). For the photometric probes, we used $\ell_{\rm max} = 5000$ for cosmic shear and $\ell_{\rm max} = 3000$ for photometric angular clustering, and galaxy-galaxy lensing. The degeneracy between adjacent bins is evident.}
\label{fig:triangle_w0waCDM_nuisance3}
\end{figure*}

Although not explicitly shown for all parameters, the analysis of \wowaCDM, adopting a flat geometry and joint full \Euclid analysis, reveals that several nuisance parameters exhibit exceptionally strong mutual correlations, both positive and negative, which can have significant implications for parameter inference.  Notably, in \cref{fig:triangle_w0waCDM_nuisance1}, we can see that the polynomial photometric galaxy bias parameters and the polynomial magnification bias parameters show strong positive and negative correlations, although not among them. For example, all polynomial magnification bias parameters are heavily (anti)-correlated with absolute values up to $0.9$. The photometric galaxy bias parameters also display strong correlations, such as $b_{\mathrm{G, 1}}$ vs.\ $b_{\mathrm{G, 2}}$ ($-0.905$) and $b_{\mathrm{G, 1}}$ vs.\ $b_{\mathrm{G, 3}}$ ($0.860$). Overall, the nuisance parameters shown in \cref{fig:triangle_w0waCDM_nuisance1} are well constrained and remain within the specified prior ranges. The inclusion of the GCsp data does not lead to significantly tighter constraints on the photometric galaxy bias, magnification bias, or intrinsic alignment parameters. This is expected, as the 3\texttimes2pt combination already provides substantially stronger constraining power than GCsp alone, and the addition of GCsp does not break any further parameter degeneracies. This last behaviour is also observed on multiplicative biases shown in \cref{fig:triangle_w0waCDM_nuisance2}.

\begin{figure*}
\centering
\includegraphics[width=1.\textwidth]{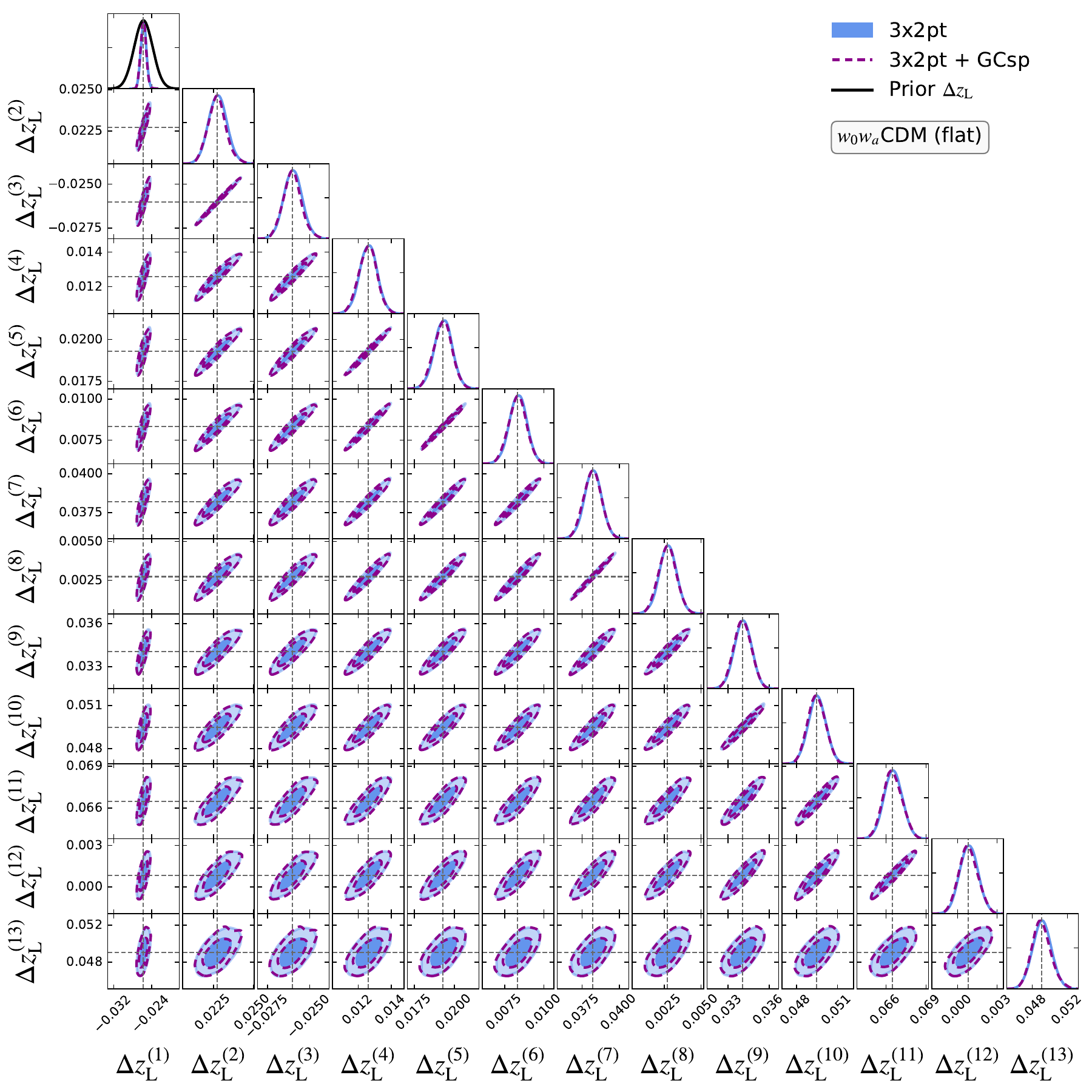}
\caption{Forecast of the constraints for the photometric nuisance parameters of the \wowaCDM cosmological model (adopting a flat geometry) using \Euclid photometric and spectroscopic probes: cosmic shear, galaxy-galaxy  lensing and angular clustering (3\texttimes2pt), and the combination of both of them as described in \cref{sec:results_full} (3\texttimes2pt + GCsp). For the photometric probes, we used $\ell_{\rm max} = 5000$ for cosmic shear and $\ell_{\rm max} = 3000$ for photometric angular clustering, and galaxy-galaxy lensing. The degeneracy between adjacent bins is evident.}
\label{fig:triangle_w0waCDM_nuisance4}
\end{figure*}

Multiplicative biases are also well constrained but remain prior-dominated, regardless of the data combination used. This is expected, as the priors imposed on these parameters (see, for example, the multiplicative bias prior associated with the first redshift bin in \cref{fig:triangle_w0waCDM_nuisance2}) are based on the assumption that the \Euclid\ science requirements are met -- requirements that are already highly constraining. Although not explicitly shown, the multiplicative biases exhibit mild degeneracies with $\sotto$ and $\lnAs$, as all three affect the amplitude of the power spectra.

The per-bin redshift shift nuisance parameters for the lensing sample, shown in \cref{fig:triangle_w0waCDM_nuisance3}, are well constrained but remain prior-dominated when using WL data alone. This is expected, given the stringent \Euclid\ science requirements. However, when the full photometric 3\texttimes2pt combination is considered, these parameters are no longer dominated by the prior. For instance, the posterior for the first redshift bin shift parameter, shown in \cref{fig:triangle_w0waCDM_nuisance4}, illustrates this transition (may the reader note that other plotting scale is applied to the rest of the two-dimensional distributions to highlight the results when GCsp is also incorporated). This improvement is anticipated, as these redshift shift parameters are strongly correlated with $\Ob$ and $\Oc$, which are poorly constrained in the WL-only case but better determined in the 3\texttimes2pt analysis. This enhanced constraining power helps to break degeneracies and allows the redshift nuisance parameters to be more tightly constrained by the data. Similarly to the case of photometric galaxy and magnification biases, GCsp data do not further constrain per-bin redshift shift nuisance parameters (see purple contours in \cref{fig:triangle_w0waCDM_nuisance4}).

\begin{figure*}
\centering
\includegraphics[width=1.\textwidth]{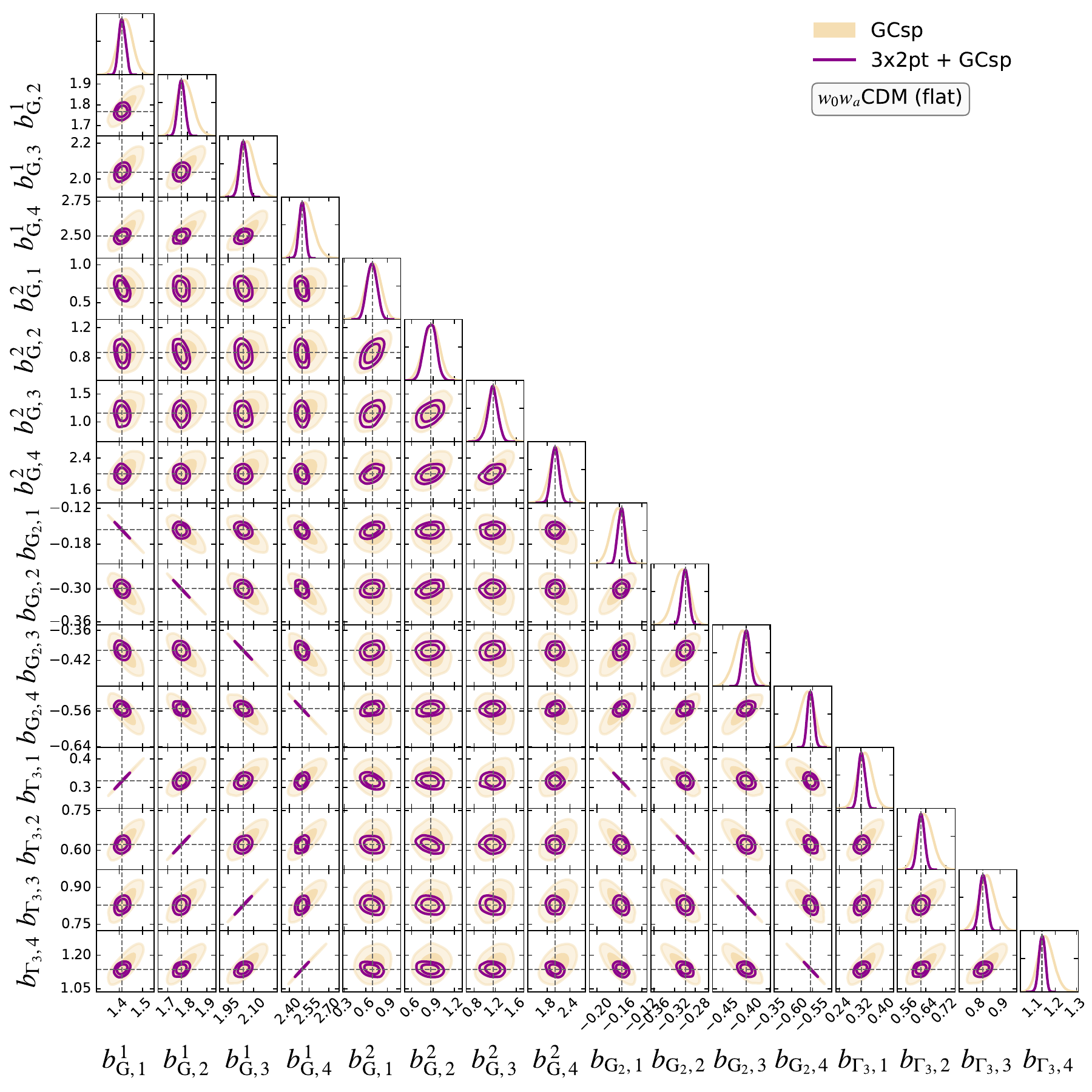}
\caption{Forecast of the constraints for the spectroscopic nuisance parameters of the \wowaCDM cosmological model (adopting a flat geometry) using \Euclid photometric and spectroscopic probes: spectroscopic galaxy clustering only (GCsp), and the combination with3\texttimes2ptas described in \cref{sec:results_full} (3\texttimes2pt + GCsp).}
\label{fig:triangle_w0waCDM_nuisance5}
\end{figure*}

On the other hand, the spectroscopic galaxy bias nuisance parameters, while well constrained within their priors, exhibit non-Gaussian posteriors when sampled in isolation using the GCsp probe only. This behaviour is significantly mitigated when the full photometric probe is included in the analysis, leading to a notable reduction in the uncertainties of their best-fit values (see \cref{fig:triangle_w0waCDM_nuisance5} and \cref{tab:CL_nuis_spectro}). This is due to strong correlations between these parameters and \lnAs, \Ob, and \Oc. The figure also includes derived parameters -- those showing strong correlations -- computed according to the prescriptions outlined in \cref{tab:fiducial_model}. 

\begin{figure*}
\centering
\includegraphics[width=1.\textwidth]{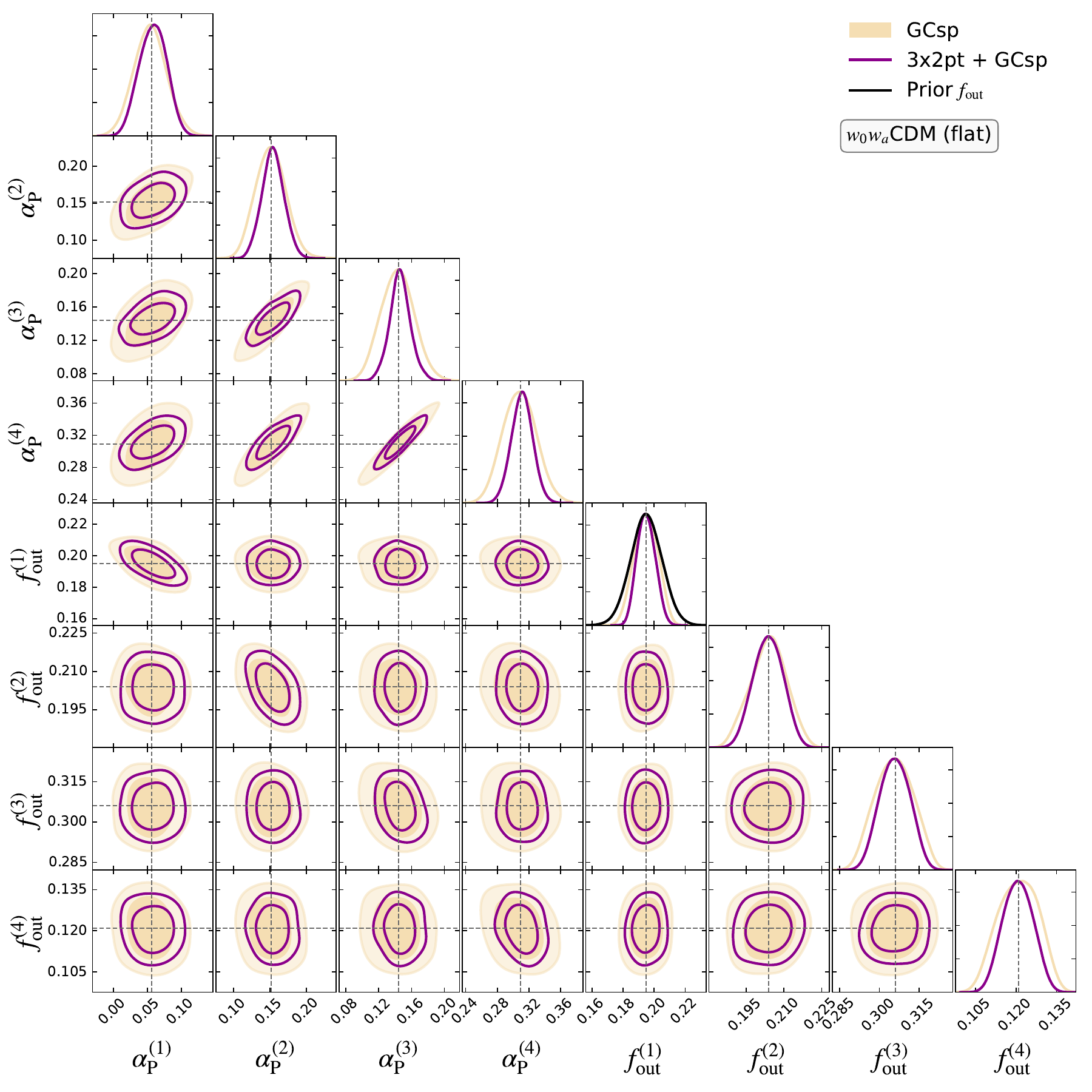}
\caption{Forecast of the constraints for the spectroscopic nuisance parameters of the \wowaCDM cosmological model (adopting a flat geometry) using \Euclid photometric and spectroscopic probes: spectroscopic galaxy clustering only (GCsp), and the combination with 3\texttimes2pt as described in \cref{sec:results_full} (3\texttimes2pt + GCsp).}
\label{fig:triangle_w0waCDM_nuisance6}
\end{figure*}

Further improvements in the constraints of spectroscopic nuisance parameters are observed in the calibration and purity parameters when 3\texttimes2pt data are included (see \cref{fig:triangle_w0waCDM_nuisance6}). In particular, the spectroscopic calibration parameters $\alpha_{\rm P}^{2}$, $\alpha_{\rm P}^{3}$, and $\alpha_{\rm P}^{4}$ exhibit strong mutual correlations, with correlation coefficients of $0.763$, $0.790$, and $0.909$, respectively. The calibration parameters are heavily correlated with \ns. The purity parameters, which are all sampled from Gaussian priors (see an example of the Gaussian prior distribution used in the purity parameter associated to the first redshift bin in \cref{fig:triangle_w0waCDM_nuisance6}) are mildly correlated with the cosmological parameters.

In conclusion, the observed correlations among nuisance parameters underscore the importance of careful modelling and regularisation in cosmological analyses, as they can propagate into degeneracies and uncertainties in the cosmological parameter estimates (see illustrative examples in \cref{sec:results_3x2pt}).

\newpage

\section{Confidence intervals of nuisance parameters for the \wowaCDM model}\label{app:CL}
In this section, we list in Tables \ref{tab:CL_cosmo}, \ref{tab:CL_nuis_spectro}, and \ref{tab:CL_nuis_photo} the 68\% confidence limits of the sampled parameters corresponding to both the cosmological and the nuisance parameters for WL, 3\texttimes2pt, and GCsp for the \wowaCDM cosmological model for single probes, combination 3\texttimes2pt, and joint analysis 3\texttimes2pt + GCsp. The definition of each parameter can be found in \cref{tab:fiducial_model}.

\begin{table}[h!]
\centering
    \caption{Same as \cref{tab:CL_cosmo}, but for the photometric nuisance parameters.}
    \renewcommand{\arraystretch}{1.4}
\begin{tabular} { l  c c c}
\noalign{\vskip 3pt}\hline\noalign{\vskip 1.5pt}\hline\noalign{\vskip 5pt}
 \multicolumn{1}{c}{\bf } &  \multicolumn{1}{c}{\bf WL} &  \multicolumn{1}{c}{\bf 3\texttimes2pt} &  \multicolumn{1}{c}{\bf 3\texttimes2pt + GCsp}\\
\noalign{\vskip 3pt}\cline{1-4}\noalign{\vskip 3pt}

% Parameter &  68\% limits &  68\% limits &  68\% limits\\
%\hline
{$\logten({T_{\rm AGN}}/$K)} & $7.743\pm 0.067            $ & $7.750\pm 0.015            $ & $7.749^{+0.014}_{-0.012}   $\\

{$\mathcal{A}_{\rm IA}$} & $0.163^{+0.024}_{-0.029}   $ & $0.1605\pm 0.0063          $ & $0.1609\pm 0.0064          $\\

{$\eta_{\rm IA}  $} & $1.62\pm 0.33              $ & $1.658\pm 0.051            $ & $1.654\pm 0.050            $\\

{$\Delta z_{\rm L}^{(1)}$} & $-0.0257\pm 0.0022         $ & $-0.02571\pm 0.00058       $ & $-0.02575\pm 0.00058       $\\

{$\Delta z_{\rm L}^{(2)}$} & $0.0227\pm 0.0020          $ & $0.02276\pm 0.00058        $ & $0.02271\pm 0.00057        $\\

{$\Delta z_{\rm L}^{(3)}$} & $-0.0260\pm 0.0020         $ & $-0.02599\pm 0.00055       $ & $-0.02603\pm 0.00054       $\\

{$\Delta z_{\rm L}^{(4)}$} & $0.0127\pm 0.0019          $ & $0.01262\pm 0.00054        $ & $0.01259\pm 0.00054        $\\

{$\Delta z_{\rm L}^{(5)}$} & $0.0194\pm 0.0020          $ & $0.01931\pm 0.00057        $ & $0.01928\pm 0.00056        $\\

{$\Delta z^{(6)}_{\rm L}$} & $0.0084\pm 0.0020          $ & $0.00835\pm 0.00057        $ & $0.00831\pm 0.00056        $\\

{$\Delta z_{\rm L}^{(7)}$} & $0.0383\pm 0.0022          $ & $0.03823\pm 0.00059        $ & $0.03820\pm 0.00059        $\\

{$\Delta z_{\rm L}^{(8)}$} & $0.0028\pm 0.0022          $ & $0.00276\pm 0.00060        $ & $0.00272\pm 0.00060        $\\

{$\Delta z_{\rm L}^{(9)}$} & $0.0341\pm 0.0024          $ & $0.03409\pm 0.00064        $ & $0.03405\pm 0.00063        $\\

{$\Delta z_{\rm L}^{(10)}$} & $0.0494\pm 0.0026          $ & $0.04952\pm 0.00065        $ & $0.04948\pm 0.00063        $\\

{$\Delta z_{\rm L}^{(11)}$} & $0.0661\pm 0.0030          $ & $0.06653\pm 0.00069        $ & $0.06649\pm 0.00070        $\\

{$\Delta z_{\rm L}^{(12)}$} & $0.0007\pm 0.0035          $ & $0.00086\pm 0.00070        $ & $0.00081\pm 0.00071        $\\

{$\Delta z_{\rm L}^{(13)}$} & $0.0487\pm 0.0051          $ & $0.0491\pm 0.0011          $ & $0.04904^{+0.00097}_{-0.0011}$\\

{$m_{\rm L}^{(1)}$} & $0.00000\pm 0.00050        $ & $0.00001\pm 0.00049        $ & $0.00002\pm 0.00051        $\\

{$m_{\rm L}^{(2)}$} & $0.00000\pm 0.00050        $ & $0.00000\pm 0.00050        $ & $-0.00003\pm 0.00050       $\\

{$m_{\rm L}^{(3)}$} & $0.00000\pm 0.00050        $ & $0.00001\pm 0.00050        $ & $-0.00002\pm 0.00050       $\\

{$m_{\rm L}^{(4)}$} & $0.00000\pm 0.00050        $ & $-0.00001\pm 0.00050       $ & $0.00000\pm 0.00051        $\\

{$m_{\rm L}^{(5)}$} & $0.00000\pm 0.00050        $ & $0.00001\pm 0.00050        $ & $0.00001\pm 0.00051        $\\

{$m_{\rm L}^{(6)}$} & $0.00001\pm 0.00050        $ & $0.00000\pm 0.00049        $ & $0.00002\pm 0.00046        $\\

{$m_{\rm L}^{(7)}$} & $0.00000\pm 0.00049        $ & $0.00000\pm 0.00049        $ & $-0.00003\pm 0.00052       $\\

{$m_{\rm L}^{(8)}$} & $0.00000\pm 0.00050        $ & $0.00002\pm 0.00049        $ & $-0.00002\pm 0.00049       $\\

{$m_{\rm L}^{(9)}$} & $0.00001\pm 0.00050        $ & $-0.00001\pm 0.00050       $ & $0.00004\pm 0.00048        $\\

{$m_{\rm L}^{(10)}$} & $0.00001\pm 0.00050        $ & $0.00000\pm 0.00049        $ & $0.00000\pm 0.00050        $\\

{$m_{\rm L}^{(11)}$} & $-0.00001\pm 0.00050       $ & $0.00000\pm 0.00048        $ & $-0.00001\pm 0.00044       $\\

{$m_{\rm L}^{(12)}$} & $-0.00001\pm 0.00048       $ & $0.00001\pm 0.00047        $ & $0.00004\pm 0.00049        $\\

{$m_{\rm L}^{(13)}$} & $-0.00001\pm 0.00050       $ & $0.00000\pm 0.00048        $ & $0.00002\pm 0.00047        $\\

{$b_{{\rm G},\,0}  $} & & $1.3327\pm 0.0036          $ & $1.3324\pm 0.0034          $\\

{$b_{{\rm G},\,1}  $} & & $-0.723\pm 0.014           $ & $-0.722\pm 0.013           $\\

{$b_{{\rm G},\,2}  $} & & $1.017\pm 0.015            $ & $1.017\pm 0.013            $\\

{$b_{{\rm G},\,3}  $} & & $-0.1489\pm 0.0041         $ & $-0.1487\pm 0.0037         $\\

{$b_{{\rm mag},\,0}$} & & $-1.51\pm 0.24             $ & $-1.51\pm 0.24             $\\

{$b_{{\rm mag},\,1}$} & & $1.35\pm 0.58              $ & $1.35\pm 0.58              $\\

{$b_{{\rm mag},\,2}$} & & $0.08\pm 0.44              $ & $0.08\pm 0.44              $\\

{$b_{{\rm mag},\,3}$} & & $0.04\pm 0.10              $ & $0.04\pm 0.10              $\\
\hline
\end{tabular}
    \label{tab:CL_nuis_photo}
\end{table}
\newpage

\vspace{-10pt}
\begin{table}[h!]
    \caption{Same as \cref{tab:CL_cosmo}, but for the spectroscopic nuisance parameters.}
    \centering
    \renewcommand{\arraystretch}{1.3}
\begin{tabular} { l  c c}
\noalign{\vskip 3pt}\hline\noalign{\vskip 1.5pt}\hline\noalign{\vskip 5pt}
 \multicolumn{1}{c}{\bf } &  \multicolumn{1}{c}{\bf GCsp} &  \multicolumn{1}{c}{\bf 3\texttimes2pt + GCsp}\\
\noalign{\vskip 3pt}\cline{1-3}\noalign{\vskip 3pt}
% Parameter &  68\% limits &  68\% limits\\
%\hline
{$\alpha_{\rm P}^{(1)}$} & $0.054\pm 0.024            $ & $0.058\pm 0.020            $\\

{$\alpha_{\rm P}^{(2)}$} & $0.151\pm 0.021            $ & $0.154\pm 0.015            $\\

{$\alpha_{\rm P}^{(3)}$} & $0.143\pm 0.020            $ & $0.146\pm 0.013            $\\

{$\alpha_{\rm P}^{(4)}$} & $0.308\pm 0.021            $ & $0.311\pm 0.013            $\\

{$b^1_{\rm G,\,1}$} & $1.426\pm 0.031            $ & $1.412\pm 0.014            $\\

{$b^1_{\rm G,\,2}$} & $1.789^{+0.037}_{-0.043}   $ & $1.769\pm 0.017            $\\

{$b^1_{\rm G,\,3}$} & $2.062\pm 0.048            $ & $2.039\pm 0.021            $\\

{$b^1_{\rm G,\,4}$} & $2.527^{+0.053}_{-0.064}   $ & $2.496\pm 0.022            $\\

{$b^2_{\rm G,\,1}$} & $0.70\pm 0.11              $ & $0.686\pm 0.068            $\\

{$b^2_{\rm G,\,2}$} & $0.88\pm 0.11              $ & $0.860\pm 0.081            $\\

{$b^2_{\rm G,\,3}$} & $1.19\pm 0.15              $ & $1.151\pm 0.095            $\\

{$b^2_{\rm G,\,4}$} & $2.08\pm 0.20              $ & $2.005\pm 0.099            $\\

{$f_{\rm out}^{(1)}$} & $0.1948\pm 0.0075          $ & $0.1949\pm 0.0057          $\\

{$f_{\rm out}^{(2)}$} & $0.2039^{+0.0077}_{-0.0069}$ & $0.2038\pm 0.0058          $\\

{$f_{\rm out}^{(3)}$} & $0.3056\pm 0.0071          $ & $0.3058\pm 0.0056          $\\

{$f_{\rm out}^{(4)}$} & $0.1209\pm 0.0073          $ & $0.1208\pm 0.0055          $\\

$b_{{\rm G_2},\,1}$ & $-0.163\pm 0.013           $ & $-0.1567\pm 0.0058         $\\

$b_{{\rm G_2},\,2}$ & $-0.307^{+0.016}_{-0.014}  $ & $-0.2997\pm 0.0067         $\\

$b_{{\rm G_2},\,3}$ & $-0.408\pm 0.017           $ & $-0.4002\pm 0.0077         $\\

$b_{{\rm G_2},\,4}$ & $-0.564^{+0.020}_{-0.017}  $ & $-0.5547\pm 0.0071         $\\

{$b_{{\Gamma_3},\,1}$} & $0.335\pm 0.027            $ & $0.323\pm 0.012            $\\

{$b_{{\Gamma_3},\,2}$} & $0.637^{+0.029}_{-0.034}   $ & $0.621\pm 0.014            $\\

{$b_{{\Gamma_3},\,3}$} & $0.844\pm 0.035            $ & $0.827\pm 0.016            $\\

{$b_{{\Gamma_3},\,4}$} & $1.156^{+0.034}_{-0.039}   $ & $1.137\pm 0.014            $\\
\hline
\end{tabular}
    \label{tab:CL_nuis_spectro}
\end{table}

The Gaussian priors set for the baryonic feedback parameter, multiplicative bias, photometric per-bin redshift bins and spectroscopic purity nuisance parameters establish baseline $\sigma$-uncertainties of $0.18$, $0.0005$, $0.01$, and approximately $0.002$, respectively. Comparing these prior widths to the best-fit posterior constraints presented in \cref{tab:CL_nuis_photo} and \cref{tab:CL_nuis_spectro} reveals a significant tightening of the parameter estimates. To quantify the improvement in constraint precision, we define the \textit{tightening factor} as the ratio between the prior standard deviation and the posterior standard deviation, i.e., \( T \equiv \sigma_{\mathrm{prior}} / \sigma_{\mathrm{posterior}} \). Applying this definition, we find tightening factors of approximately $2.3$ for the multiplicative bias, $1.6$ for spectroscopic purity parameters, $3.3$ for the per-bin redshift shifts and $4.7$ for the baryonic feedback amplitude. These results demonstrate that the data significantly sharpen our knowledge of these nuisance parameters, reducing their uncertainties well below the levels assumed a priori and thus improving the robustness of the cosmological inference both at theoretical and observational level.

\end{appendix}
\end{document}